\documentclass[article]{aa}
\usepackage{natbib}         
\usepackage{graphicx}
\usepackage{txfonts}
\usepackage[colorlinks=true, linkcolor=blue, citecolor=blue, urlcolor=blue]{hyperref}
\usepackage{nameref}
\bibpunct{(}{)}{;}{a}{}{,} 

\newcommand{\gaia}{\it Gaia}
\newcommand{\kepler}{\it Kepler}
\newcommand{\hipparcos}{\it Hipparcos}

\begin{document}

\title{{\kepler} meets {\gaia} DR3: homogeneous extinction-corrected color-magnitude diagram and binary classification}

\titlerunning{{\kepler} field characterization with {\gaia} DR3}

\authorrunning{Godoy-Rivera et al.}

\author{D. Godoy-Rivera\inst{1,2}
          \and
          S. Mathur\inst{1,2}
          \and
          R. A. Garc\'ia\inst{3}          
          \and
          M. H. Pinsonneault\inst{4}
          \and
          \^A. R. G. Santos\inst{5} 
          \and
          P. G. Beck\inst{1,2,6}
          \and \\
          D. H. Grossmann\inst{1,2,6}
          \and
          L. Schimak\inst{7} 
          \and
          M. Bedell\inst{8}          
          \and
          J. Merc\inst{1,9}
          \and
          A. Escorza\inst{1,2}
          }

\institute{
            Instituto de Astrofísica de Canarias (IAC), E-38205 La Laguna, Tenerife, Spain; \email{godoyrivera.astro@gmail.com}
         \and
             Universidad de La Laguna (ULL), Departamento de Astrofísica, E-38206 La Laguna, Tenerife, Spain   
         \and
             Universit\'e Paris-Saclay, Universit\'e Paris Cité, CEA, CNRS, AIM, 91191, Gif-sur-Yvette, France   
         \and
             Department of Astronomy, The Ohio State University, 140 West 18th Avenue, Columbus, OH 43210, USA
         \and
             Instituto de Astrof\'isica e Ci\^encias do Espa\c{c}o, Universidade do Porto, CAUP, Rua das Estrelas, PT4150-762 Porto, Portugal             
         \and
             Institut für Physik, Karl-Franzens Universität Graz, Universitätsplatz 5/II, NAWI Graz, 8010 Graz, Austria   
         \and
             Sydney Institute for Astronomy (SIfA), School of Physics, University of Sydney, NSW 2006, Australia   
         \and
             Center for Computational Astrophysics, Flatiron Institute, 162 5th Avenue, Manhattan, NY, USA
         \and
             Astronomical Institute, Faculty of Mathematics and Physics, Charles University, V Holešovičkách 2, 180000, Prague, Czechia
             }

\abstract{The original {\kepler} mission has delivered unprecedented high-quality photometry. These data have impacted numerous research fields (e.g., asteroseismology and exoplanets), and continue to be an astrophysical goldmine. Because of this, thorough investigations of the $\sim$ 200,000 stars observed by {\kepler} remain of paramount importance. In this paper, we present a state-of-the-art characterization of the {\kepler} targets based on {\gaia} DR3 data. We place the stars on the color-magnitude diagram (CMD), account for the effects of interstellar extinction, and classify targets into several CMD categories (dwarfs, subgiants, red giants, photometric binaries, and others). Additionally, we report various categories of candidate binary systems spanning a range of detection methods, such as Renormalised Unit Weight Error (RUWE), radial velocity variables, {\gaia} non-single stars (NSS), {\kepler} and {\gaia} eclipsing binaries from the literature, among others. First and foremost, our work can assist in the selection of stellar and exoplanet host samples regarding CMD and binary populations. We further complement our catalog by quantifying the impact that astrometric differences between {\gaia} data releases have on CMD location, assessing the contamination in asteroseismic targets with properties at odds with {\gaia}, and identifying stars flagged as photometrically variable by {\gaia}. We make our catalog publicly available as a resource to the community when researching the stars observed by {\kepler}.}


\maketitle
\section{Introduction} 
\label{sec:introduction}

Since its launch 15 years ago, the {\kepler} space telescope \citep{borucki10} has deeply revolutionized astrophysics. The impact of its high-precision photometry has allowed unprecedented discoveries, with contributions spanning the fields of asteroseismology, stellar rotation, stellar activity, exoplanets, and Galactic archaeology, among others (e.g., \citealt{bedding11,beck12,howard12,mosser12,miglio13,batalha13,mcquillan14,vansaders16,silvaaguirre17}). Although the mission was retired in 2018, after being refurbished into K2 for the second part of its life \citep{howell14}, the community is still analyzing its data and producing novel scientific studies (e.g., \citealt{santos21,santos23,li22,mathur22,mathur23,vrard22,long23,martinezpalomera23,reinhold23,bhalotia24,kamai25}). Hence, detailed characterizations of the stars observed by {\kepler} remain highly relevant. 

In this context, the {\gaia} mission \citep{gaia16,gaia18} has provided exquisite all-sky data that allow for unprecedented characterizations of stars (e.g.,  \citealt{berger18,brandt18,cantatgaudin18,godoyrivera21a,kuhn19}). The Early Data Release 3 (EDR3) reported the latest astrometric and photometric data \citep{gaia21,lindegren21a,riello21}, and the more recent DR3 delivered several lists of binary systems, as well as radial velocity measurements, extinction values, and other stellar parameters \citep{gaia23a,gaia23b,creevey23,fouesneau23,halbwachs23,katz23}. 

The goal of this paper is to leverage the latest {\gaia} data and perform a thorough characterization of the {\kepler} stars, specifically regarding target selections of singles vs. binaries and main sequence (MS) vs. post-MS phases. For instance, when focused on single stars, removing stars with binary companions is relevant as they may pollute the signal or influence the evolution given their proximity \citep{curtis19,lu22,santos24}. Conversely, studies dedicated to binary stars require a comprehensive selection of these systems, spanning a range of properties and detection methods \citep{gaulme16,ball23,grossmann25}. Similarly, post-MS evolution can produce variations in stellar properties driven by the structural changes experienced by stars \citep{garcia14,santos19,hall21,gehan24}. Some of the analyses that would benefit from such a characterization regarding {\kepler} include the aforementioned fields of stellar, exoplanetary, and Galactic astrophysics.

This paper is structured as follows. In Sect.~\ref{sec:data} we present the {\kepler} sample and the {\gaia} data we use to characterize it. In Sect.~\ref{sec:characterization_CMD} we investigate the color-magnitude diagram (CMD) and separate the sample into different CMD categories. In Sect.~\ref{sec:characterization_binaries} we use a variety of detection methods to identify several categories of binary systems. In Sect.~\ref{sec:differenceDR2vsDR3} we study the impact that revised DR3 distances have on CMD location. In Sect.~\ref{sec:misclassifiedstars} we illustrate how the {\gaia} data can shed light on puzzling asteroseismic targets. In Sect.~\ref{sec:variability} we examine the {\gaia} variability classification on the {\kepler} targets. We conclude in Sect.~\ref{sec:conclusions}.
\section{Data} 
\label{sec:data}

\begin{figure*}
\centering
\includegraphics[width=8.0cm]{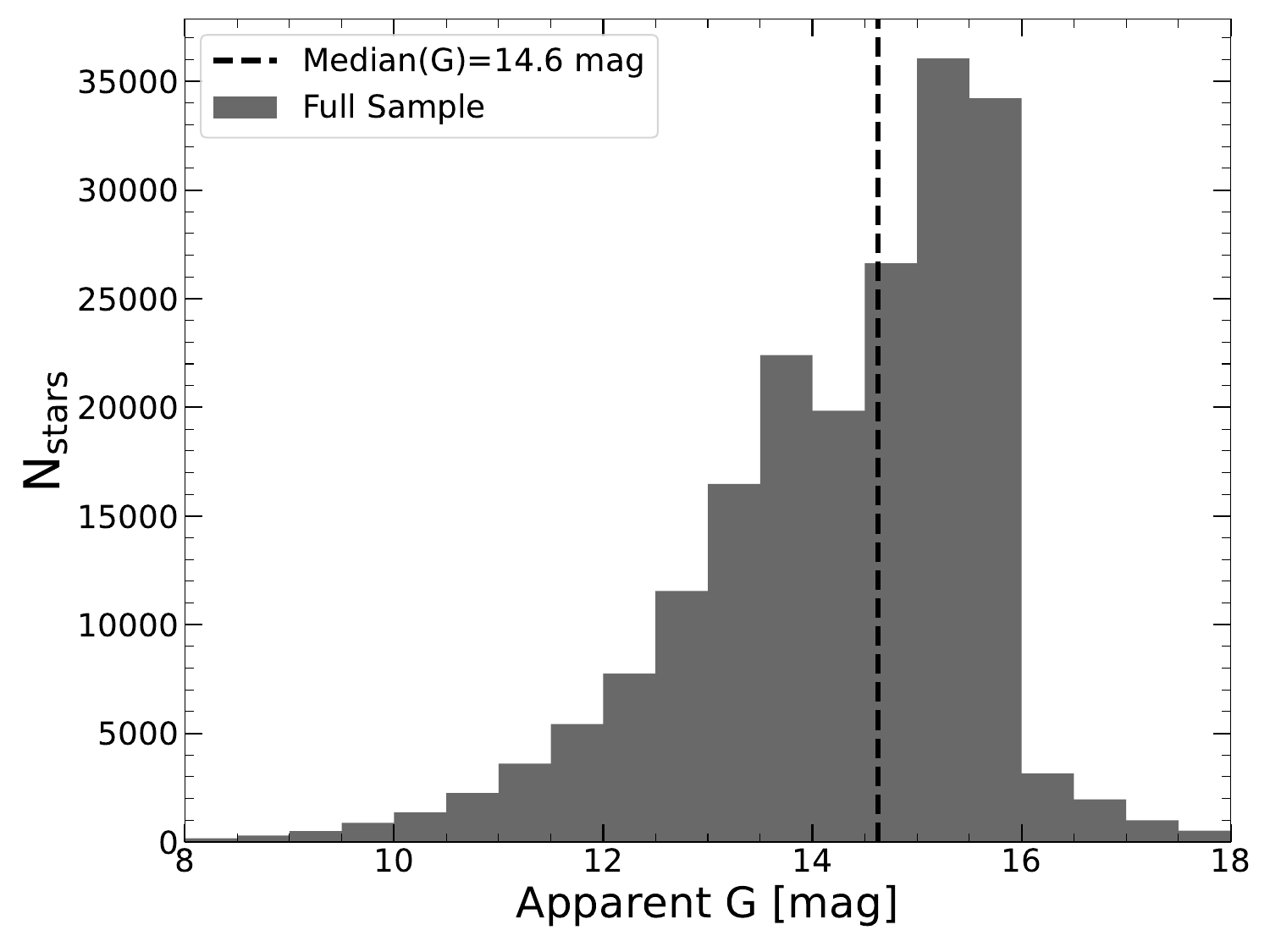}
\includegraphics[width=8.0cm]{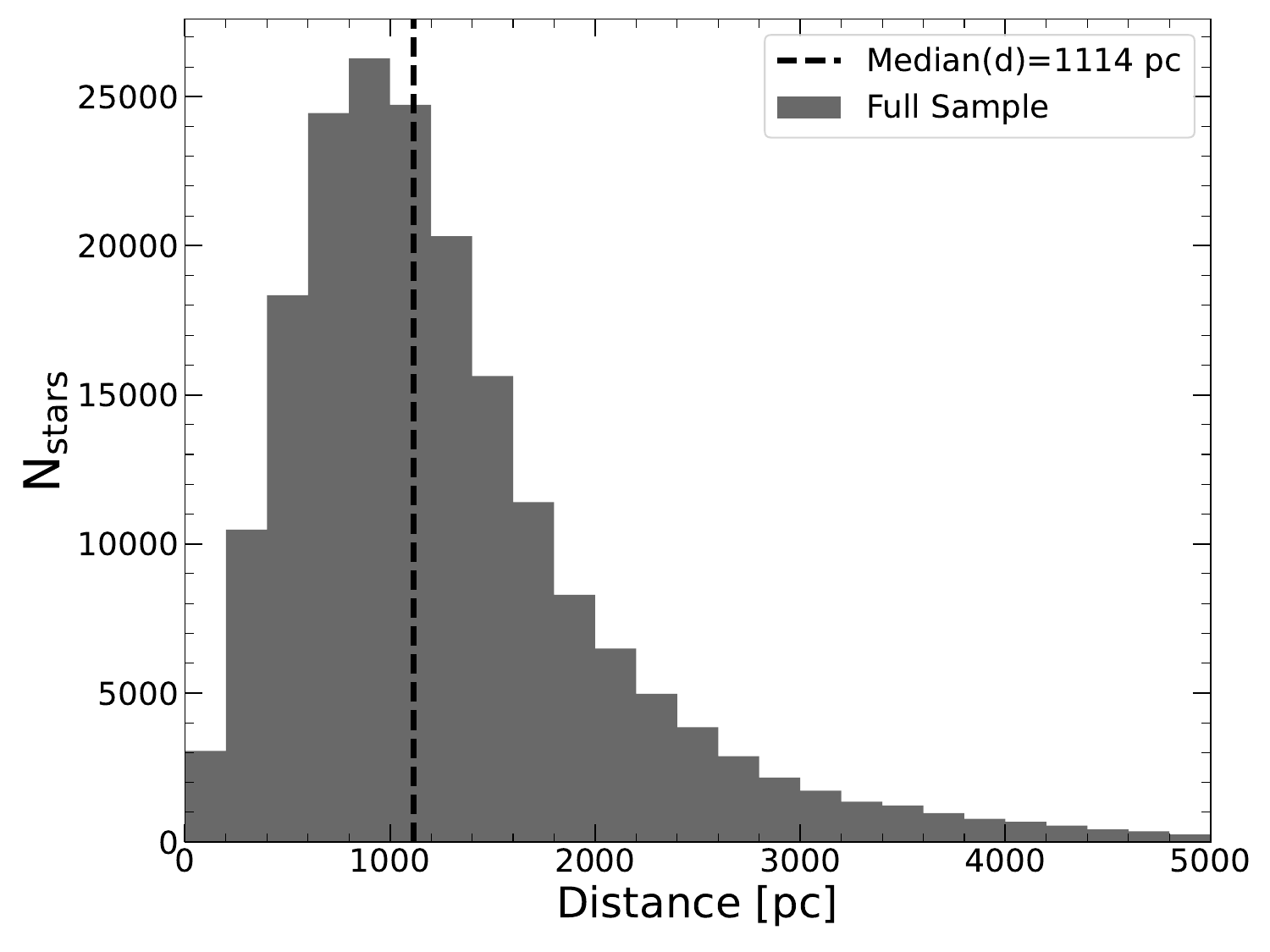}\\
\includegraphics[width=8.0cm]{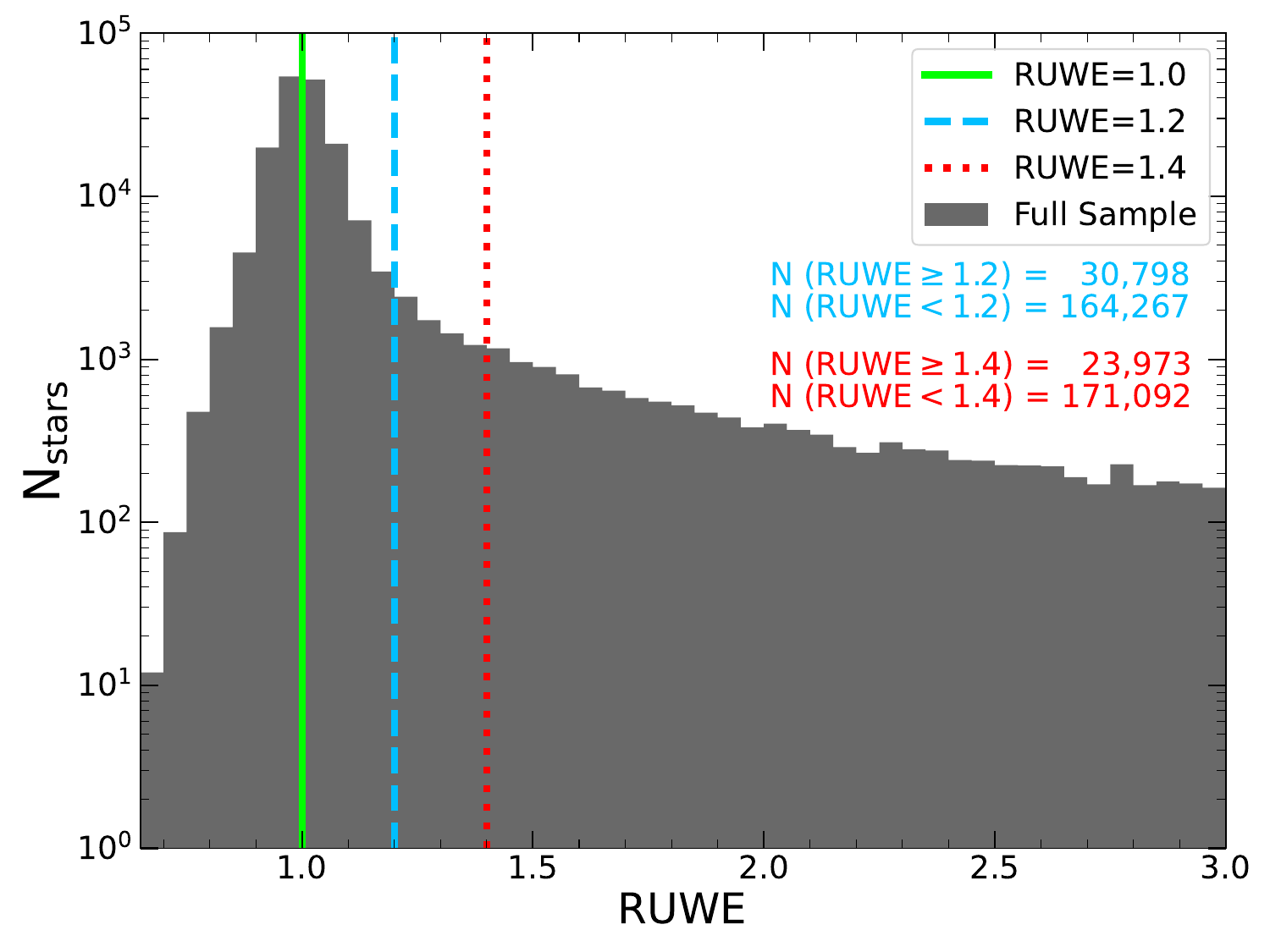}
\includegraphics[width=8.0cm]{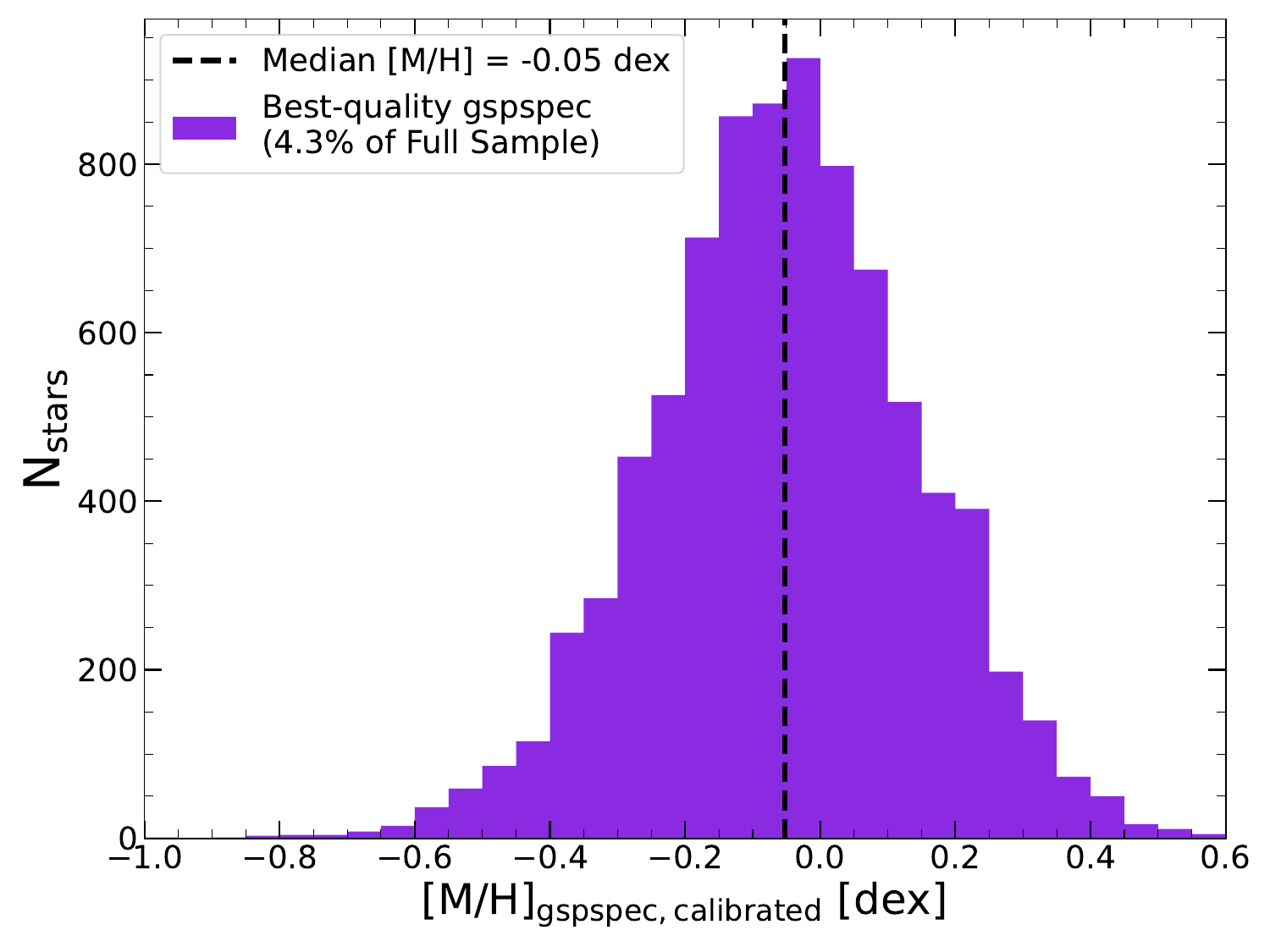}\\
\caption{Characterization of the {\kepler} targets. Top-left: Distribution of apparent {\gaia} $G$-band magnitudes. The targets are mostly concentrated in the $10 < G < 16$ mag range. Top-right: Distribution of {\gaia} distances. The distribution peaks around $\sim$ 1 kpc. Bottom-left: Distribution of RUWE values, with the vertical lines indicating RUWE = 1.0 (green), 1.2 (cyan), and 1.4 (red). From this, RUWE binaries are later identified in Sect.~\ref{sec:characterization_binaries}. Bottom-right: Distribution of calibrated metallicity values, for the subsample with best-quality \texttt{gspspec} data. The distribution peaks around [M/H]=0 dex, in agreement with the literature.}
\label{fig:characterization_Gmag_distance_RUWE_MHgspspec}
\end{figure*}

We define the target sample as the stars observed by the {\kepler} mission reported in the \texttt{one-to-one} \texttt{Gaia-Kepler.fun}\footnote{\url{https://gaia-kepler.fun/}. Download date = 13/06/2024.} crossmatch. This amounts to a list of 196,762 stars with {\kepler} Input Catalog (KIC) identifiers (hereafter KIC IDs; \citealt{brown11}) that also have {\gaia} DR3 IDs.

Given the much higher angular resolution of the {\gaia} mission compared to {\kepler}, for some stars the crossmatching may not be straightforward (e.g., a given {\kepler} target may be matched to more than one {\gaia} source). For this work, we focus on the \texttt{one-to-one match} table, i.e., the subset of {\kepler} targets that can be confidently matched to exactly one {\gaia} ID and vice versa. We note that this crossmatch is conservative to some extent, as for a given {\kepler} star, it was calculated by enforcing three conditions: 1) that within 4{\arcsec}, the nearest {\gaia} match in angular separation is also the nearest match by magnitude difference between {\gaia} $G$ and {\kepler} $K_p$; 2) that the angular separation is less than 1{\arcsec} after proper motion correction; and 3) that the $G$ and $K_p$ brightnesses are within 2 magnitudes.

\begin{figure*}
\centering
\includegraphics[width=8.0cm]{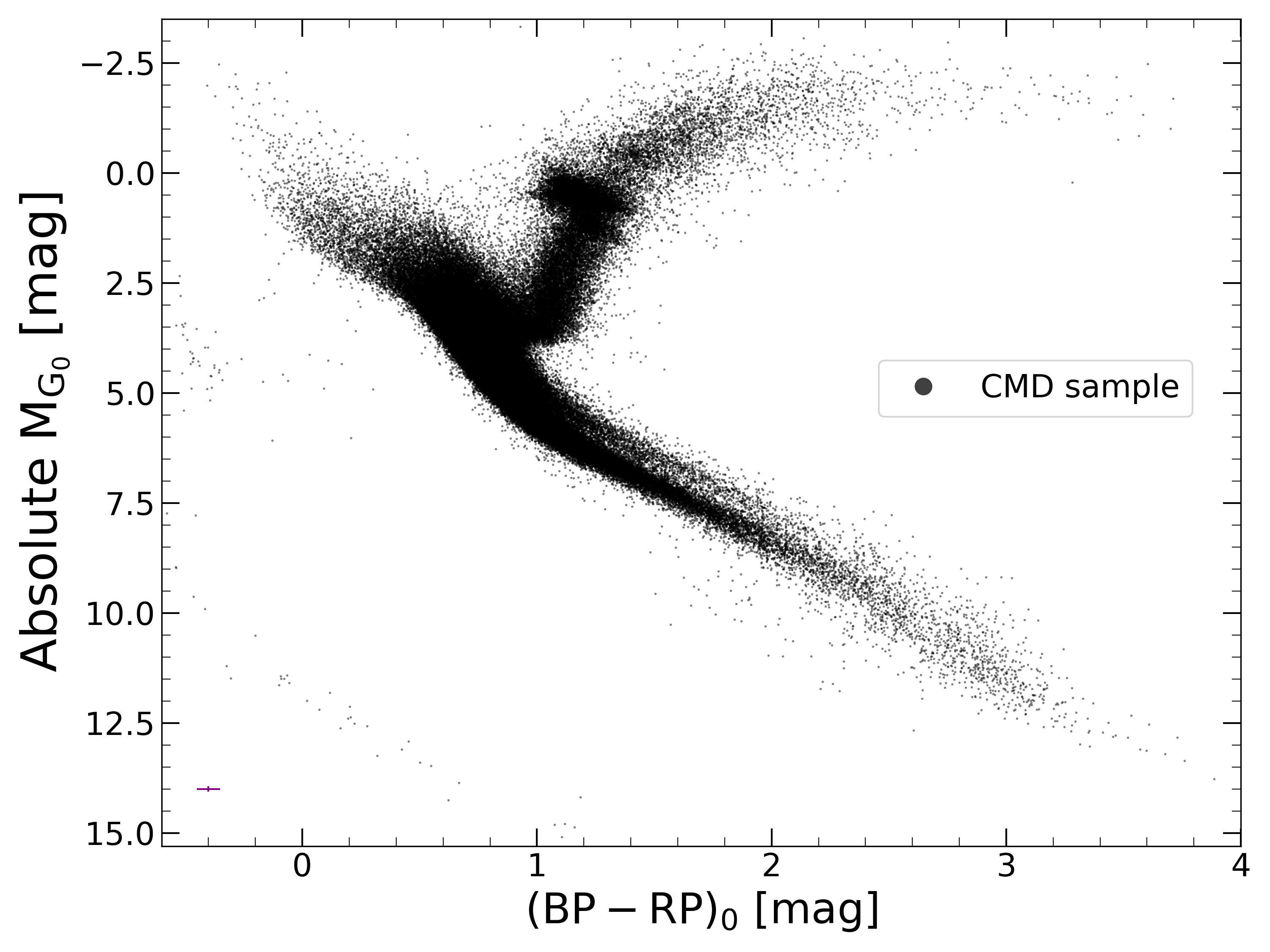}
\includegraphics[width=8.0cm]{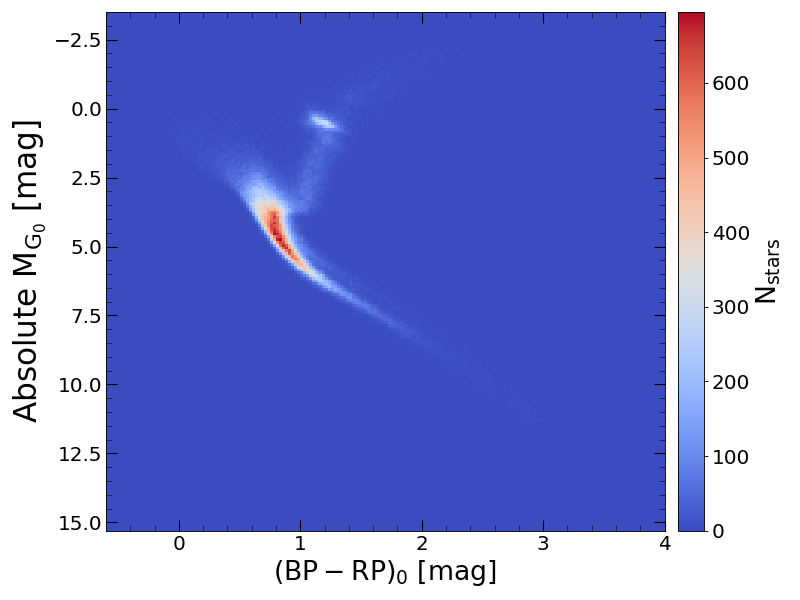}\\
\includegraphics[width=16.0cm]{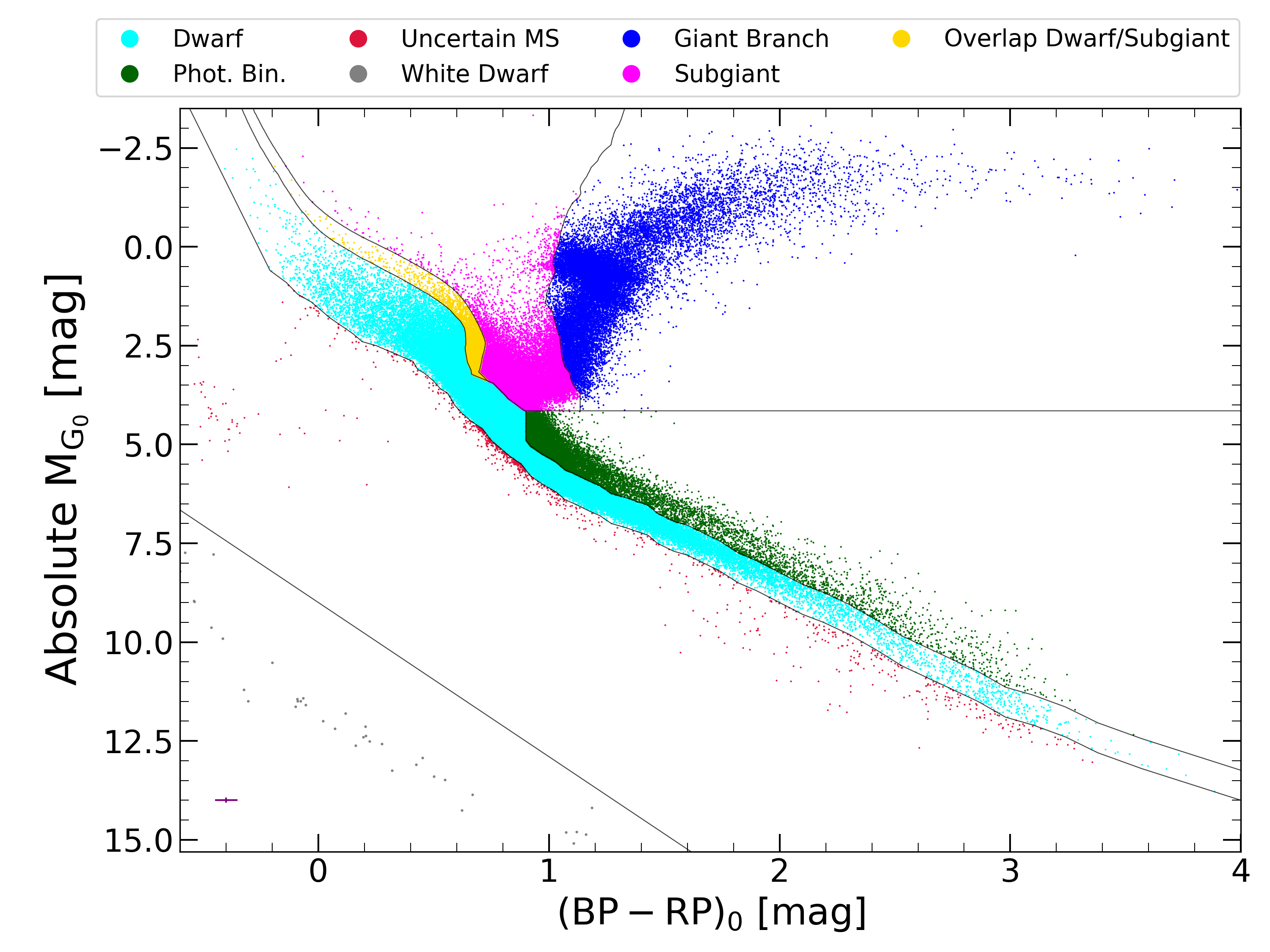}\\
\caption{Absolute and de-reddened {\gaia} CMD of the {\kepler} targets. Top-left: CMD sample described in Sect.~\ref{sec:characterization_CMD}. The purple marker illustrates the median error bars. Top-right: Hess diagram of the CMD sample. Bottom: CMD sample, with the stars color-coded according to the CMD categories we define in Sect.~\ref{sec:characterization_CMD}. The black lines illustrate the borders of the CMD regions.}
\label{fig:characterization_CMD}
\end{figure*}

While not fully complete, this crossmatch includes the vast majority of the stars observed by {\kepler}. For instance, our target sample includes 99.1\% (195,288 out of 197,096 stars) of the {\kepler} DR25 catalog by \citet{mathur17}. We leave the remaining 0.9\% (1,808 stars) to be further examined in future studies. 

Using these {\gaia} DR3 IDs, we obtain the {\gaia} magnitudes, parallaxes, and Renormalised Unit Weight Error (RUWE) values by querying the \texttt{gaiadr3.gaia\_source} table. We correct the parallaxes by subtracting the zero-point values\footnote{Queried with the \texttt{gaiadr3\_zeropoint} tool hosted at \url{https://gitlab.com/icc-ub/public/gaiadr3_zeropoint}.} from \citet{lindegren21b,lindegren21a}, and the parallax errors by considering the inflation factors from \citet{elbadry21}. We calculate distances by inverting the (corrected) parallax values. Given the quality cuts we introduce in Sect.~\ref{sec:characterization_CMD}, the parallax inverse is a safe estimate for our targets \citep{bailerjones21}. To confirm this we query the \texttt{gaiaedr3\_distance} table from \citet{bailerjones21} and find that our distances agree well with their geometric distances (e.g., both estimates agree at the 95\% level or better for 97.4\% of cases). Additionally, we query the \texttt{gaiadr3.astrophysical\_parameters} table for spectroscopic metallicities ($[\text{M/H}]$), surface gravities ($\log(g)$), and effective temperatures ($T_{\text{eff}}$). We find 12.1\% of the sample (23,894 stars) prior to any quality cuts. We discuss the calibrations applied to the $[\text{M/H}]$ and $\log(g)$ values, as well as the selection of targets with high-quality spectroscopic parameters, in Appendix \ref{sec:app_calibration_quality_gspspec}.

In Table~\ref{tab:table_catalog}, we report the KIC IDs, {\gaia} DR3 IDs, and main astrometric, photometric, and spectroscopic {\gaia} DR3 data for the sample. To further enhance the value of our catalog, in Table~\ref{tab:table_catalog} we also include the TESS Input Catalog (TIC; \citealt{stassun18,stassun19}) and Two Micron All Sky Survey (2MASS; \citealt{cutri03,skrutskie06}) ID of the targets. These were obtained from the \texttt{KIC-to-TIC}\footnote{\url{https://github.com/jradavenport/kic2tic}} and \texttt{Gaia-Kepler.fun} tables, and are available for 99.9\% of the sample (196,532 and 196,535 stars, respectively).

To characterize the targets, we show the distributions of their apparent $G$-band magnitudes, distances, RUWE, and metallicity values in Figure~\ref{fig:characterization_Gmag_distance_RUWE_MHgspspec}. Regarding the apparent magnitudes, most of the targets are in the range of $12 < G < 16$ mag (median of $\approx$ 14.6 mag). In terms of distances, the distribution peaks around $\sim$ 1 kpc (median of $\approx$ 1.11 kpc), with a tail extending up to several kpc. Regarding RUWE, this parameter quantifies how appropriate the {\gaia} single-star astrometric solution is, and thus helps in the identification of binary and non-single star candidates (see Sect.~\ref{sec:characterization_binaries} for details). The RUWE distribution is strongly peaked around values of 1.0, with a tail extending to higher values. For metallicity, the \texttt{gspspec} distribution is centered around solar, with a median value of $\approx$ -0.05 dex ($1\sigma$ range of -0.25 to +0.14 dex) for the subset with best-quality spectroscopic parameters (see Appendix \ref{sec:app_calibration_quality_gspspec}). This is in agreement with other spectroscopic studies \citep{dong14,frasca16}.
\section{Color-magnitude diagram characterization} 
\label{sec:characterization_CMD}

In this section, we perform a thorough CMD characterization of the {\kepler} targets. We define several CMD categories that can be identified by the `Flag CMD' column in Table~\ref{tab:table_catalog}. We show these in Figure~\ref{fig:characterization_CMD}, and summarize the number of targets in each category in Table~\ref{tab:summary_table}.
\subsection{Quality cuts} 
\label{subsec:characterization_CMD_qualitycuts}
 
Before constructing the CMD of the {\kepler} sample, we apply a series of quality cuts to ensure a reliable CMD placement. First, we do not consider stars that are missing either of the following parameters in Table~\ref{tab:table_catalog}: $G$-band magnitude, $BP-RP$ color, parallax, or distance (see Sect.~\ref{sec:data}). Second, we discard all the stars that have a corrected parallax signal-to-noise-ratio (SNR) of $\varpi_{\text{Corr}}/\sigma_{\varpi,\text{Corr}} \leq 10$, or a flux SNR of $f/\sigma_{f} \leq 50$ in the $G$-band or $f/\sigma_{f} \leq 20$ in the $BP$- and $RP$-bands \citep{gaia18b}. Third, to avoid overestimated $BP$-band magnitudes, we impose a maximum value of apparent $BP < 20.3$ mag \citep{riello21}. Fourth and last, to manage background contamination in the $BP$ and $RP$ photometry, we discard all stars with corrected $BP$ and $RP$ flux excess factors located outside the 3$\sigma$ scatter level \citep{riello21}. Further details on the quality cuts are described in Appendix \ref{sec:app_specifics_quality_cuts}. The fraction of targets that pass all these quality cuts corresponds to 91.1\% of the sample (179,295 stars).
\subsection{Extinction} 
\label{subsec:characterization_CMD_extinction}

To accurately place the targets on the CMD, the observed photometry needs to be corrected for the effect of interstellar extinction. After considering and comparing several extinction maps from the literature, for the remainder of this paper, we adopt the values from \citet{vergely22}. The choice of this map is motivated by its good agreement with other extinction references, and by its high completeness (e.g., it provides extinctions for every {\kepler} star with a distance value). A comparison of literature extinction maps is shown in Appendix~\ref{sec:app_comparing_extinction_maps}. Further details on the adopted extinctions (and their uncertainties) are discussed in Appendix~\ref{sec:app_characterization_extinction_values}.
\subsection{CMD sample} 
\label{subsec:characterization_CMD_CMDsample}

The stars that pass all the quality cuts from Sect.~\ref{subsec:characterization_CMD_qualitycuts} can be reliably placed on the CMD, hence, for the remainder of this paper we refer to them as the `CMD sample'. We calculate de-reddened colors as 
\begin{equation}
(BP-RP)_0 = (BP-A_{BP}) - (RP - A_{RP}) ,
\end{equation}
and absolute magnitudes as 
\begin{equation}
M_{G_0} = (G - A_G) + 5 - 5 \log_{10}(d) , 
\end{equation}
where $A_{BP}$, $A_{RP}$, and $A_{G}$ are the extinction values in the corresponding bands as obtained in Appendix~\ref{sec:app_characterization_extinction_values}, and $d$ is the {\gaia} distance in pc. Their uncertainties are calculated from error propagation as 
\begin{equation}
\sigma_{(BP-RP)_0}=\sqrt{ \sigma_{BP}^2+ \sigma_{A_{BP}}^2+ \sigma_{RP}^2 + \sigma_{A_{RP}}^2}
\end{equation}
and
\begin{equation}
\sigma_{M_{G_0}} = \sqrt{\sigma_G^2 + \sigma_{A_{G}}^2 + \left(\frac{5}{\ln(10)} \frac{\sigma_{d}}{d}\right)^2}.
\end{equation}
The median errors are $\sigma_{(BP-RP)_0}=0.049$ mag and $\sigma_{M_{G_0}} = 0.063$ mag. The color errors are heavily dominated by the extinction uncertainties (particularly $\sigma_{A_{BP}}$), while for the magnitude errors both distance and extinction uncertainties contribute almost equally (with the former being slightly more prominent). The $(BP-RP)_0$, $\sigma_{(BP-RP)_0}$, $M_{G_0}$, and $\sigma_{M_{G_0}}$ values are reported in Table~\ref{tab:table_catalog}.

We show the CMD and Hess diagram of this sample in Figure~\ref{fig:characterization_CMD}, with the median error bars shown as the purple marker for reference. The stars observed by the {\kepler} mission span a range of evolutionary stages, with the targets being mostly concentrated towards MS solar-like stars \citep{huber14,berger20,wolniewicz21}. 

The stars excluded by the quality cuts of Sect.~\ref{subsec:characterization_CMD_qualitycuts} are not shown in the CMD. They are identified as such in Table~\ref{tab:table_catalog} by having `Flag CMD = \texttt{notinCMDsample}', and amount to 8.9\% of the {\kepler} sample (17,467 stars). We note that some of these do have measured magnitudes and distances, and hence could have been plotted on the CMD. We have chosen to restrict the `CMD sample' to a more high-quality data set, but we nonetheless report all the relevant data in Table~\ref{tab:table_catalog} for interested readers.
\subsection{CMD categories} 
\label{subsec:characterization_CMD_categories}

We now classify the {\kepler} stars into different categories based on their locations on the CMD. The results of this classification are shown in the bottom panel of Figure~\ref{fig:characterization_CMD}, and reported in Table~\ref{tab:table_catalog} by the `Flag CMD' column.

\begin{figure*}
\centering
\includegraphics[width=8.0cm]{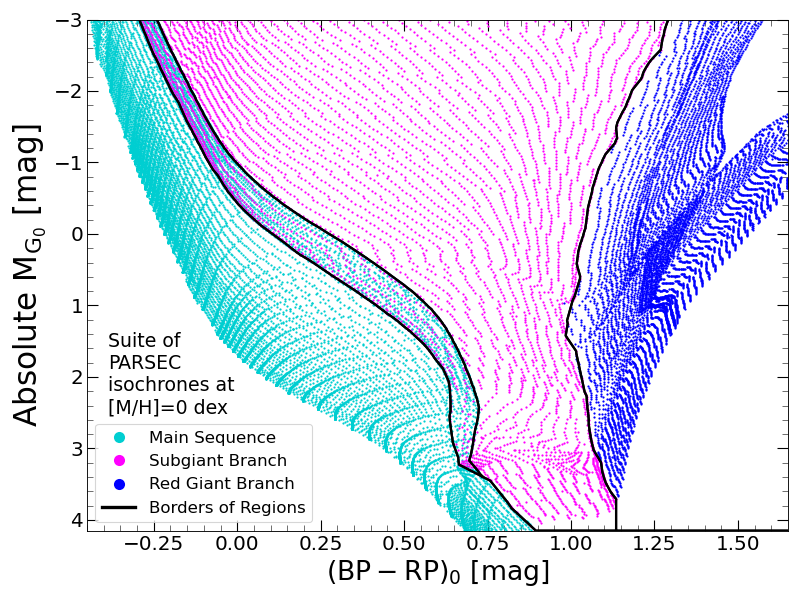}
\includegraphics[width=8.0cm]{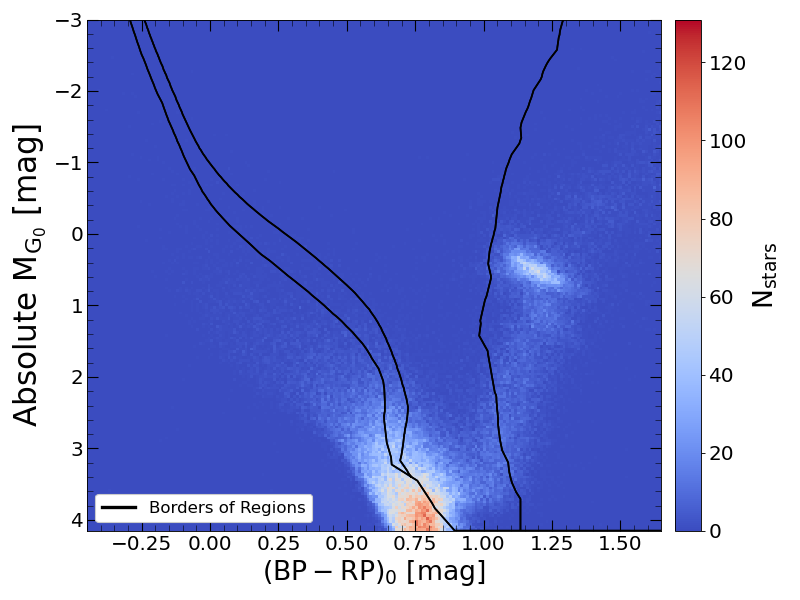}\\
\includegraphics[width=8.0cm]{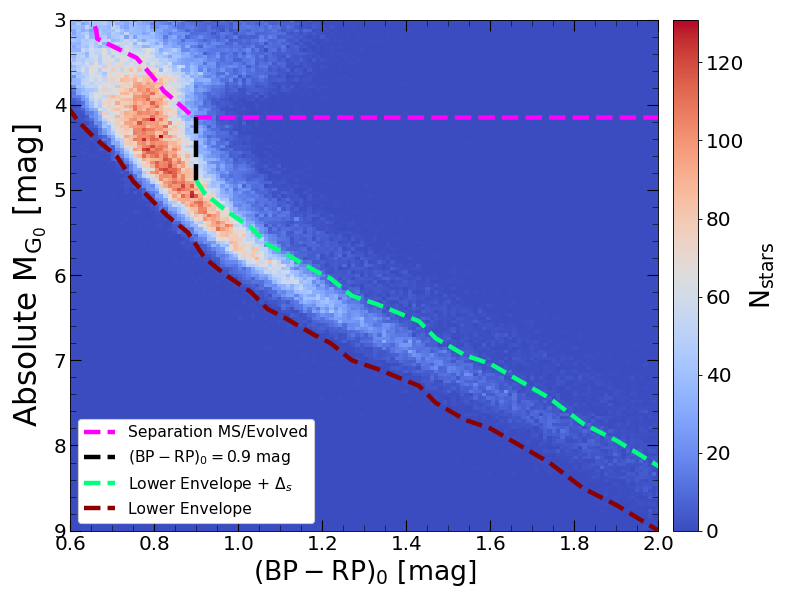}
\includegraphics[width=8.0cm]{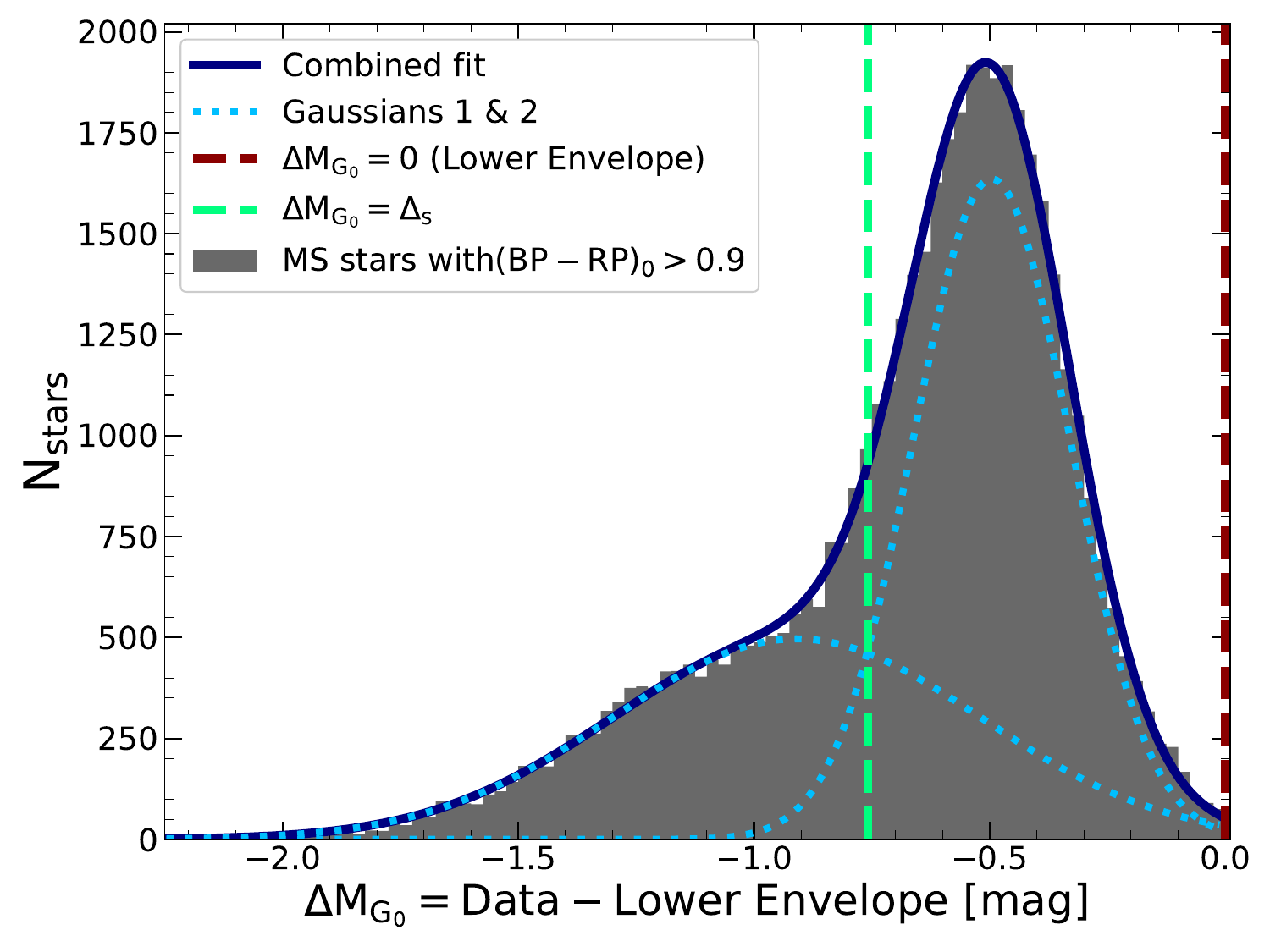}\\
\caption{Characterization of the CMD regions presented in Sect.~\ref{sec:characterization_CMD}. The top-left panel shows the suite of PARSEC models we use to define the upper-CMD regions, and the projections of these onto the Hess diagram are displayed in the top-right panel. The bottom-left panel shows the Hess diagram and borders of the lower-CMD regions. The bottom-right panel shows the distribution of $\Delta M_{G_0}$ values we use to define the \texttt{`Photometric Binary'} region.}
\label{fig:characterization_CMDregions_upper_lower}
\end{figure*}

We begin by broadly defining the CMD regions occupied by some of the main evolutionary stages, namely MS, subgiant branch (SGB), and red giant branch (RGB) (e.g., \citealt{donada23}). For this, we choose to use PARSEC\footnote{\url{http://stev.oapd.inaf.it/cgi-bin/cmd}} (PAdova and tRieste Stellar Evolutionary Code; \citealt{bressan12,chen14,nguyen22}), as this family of models reports the SGB separated from the MS and RGB phases\footnote{The different evolutionary phases in the PARSEC models are identified by combining interior physics and Hertzsprung–Russell diagram (HRD) morphology \citep{bressan12}. They are specified via the \texttt{label} column ($1=$ MS, $2=$ SGB, $3=$ RGB, among others).}. We download a suite of PARSEC isochrones, in the range of $6.60<\log_{10}(\text{age/yr})<10.13$ (in steps of 0.01 dex), with $\text{[M/H]}=0$ dex (see Sect.~\ref{subsec:characterization_CMD_validation} for a discussion on the metallicity-dependence). We display their CMD in the top-left panel of Figure~\ref{fig:characterization_CMDregions_upper_lower}, and color-code the points according to their evolutionary stage (MS in cyan, SGB in pink, and RGB in blue). From this, we use \texttt{Python}'s \texttt{alphashape} package to delineate the borders that separate the regions, and use them on the CMD of the {\kepler} sample. 

We find the PARSEC SGB and RGB regions to be clearly separated in the top-left panel of Figure~\ref{fig:characterization_CMDregions_upper_lower}, with no evolved star populating absolute magnitudes $M_{G_0}\gtrsim4$ mag. We use the borders of these regions on the CMD sample and classify the stars that fall to the right of the PARSEC SGB/RGB limit as \texttt{`Giant Branch'}. This region, which includes stars on the RGB, red clump (RC), and asymptotic giant branch (AGB), is not further subdivided as performing a detailed distinction between them is beyond the scope of this paper. These correspond to 11.2\% of the {\kepler} sample (22,098 stars), and we show them as blue points in Figure~\ref{fig:characterization_CMD}.

The separation between the PARSEC MS and SGB, on the other hand, is not as straightforward. As shown in the top-left panel of Figure~\ref{fig:characterization_CMDregions_upper_lower}, some overlap exists in the region of the blue loop of stars (where the cyan and pink points coincide). As stars in this CMD location cannot be cleanly classified as either MS or SGB in a global sense (rather, the specific classification depends on the evolutionary stage of each star), we adopt a conservative approach and define this overlap region as its own category. We define this region as \texttt{`Overlap Dwarf/Subgiant'}, corresponding to 2.8\% of the sample (5,578 stars), and targets that fall inside it are shown as yellow points in Figure~\ref{fig:characterization_CMD}. Again based on the top-left panel of Figure~\ref{fig:characterization_CMDregions_upper_lower}, the stars located in between the \texttt{`Overlap Dwarf/Subgiant'} and \texttt{`Giant Branch'} regions are given the \texttt{`Subgiant'} classification. These correspond to 13.2\% of the sample (25,901 stars), and are shown by the fuchsia points in Figure~\ref{fig:characterization_CMD}. 

We note that some RC stars appear to scatter to the \texttt{`Subgiant'} region (fuchsia points with $M_{G_0} \lesssim 0.84$ mag in Figure~\ref{fig:characterization_CMD}). We examine their \texttt{gspspec} metallicities and find that they correspond to subsolar-metallicity stars for which their bluer colors place them to the left of the solar-metallicity SGB/RGB limit. These nonetheless amount to a small number of targets, as illustrated on the zoomed-in Hess diagram in the top-right panel of Figure~\ref{fig:characterization_CMDregions_upper_lower}. We assess the reliability of our CMD classification in Sect.~\ref{subsec:characterization_CMD_validation}, where the impact of metallicity is discussed, and targets with extreme [M/H] values are flagged.

We also note the presence of 38 stars (0.02\% of the sample) with CMD locations that coincide with the white dwarf sequence (e.g., \citealt{gaia18b}). A Simbad\footnote{\url{https://simbad.cds.unistra.fr/simbad/}} search returns that all of them have been previously identified as either confirmed or candidate white dwarfs in the literature \citep{maoz15,doyle17,gentilefusillo21}. To avoid confusion with the lower-CMD categories we define below, we delineate the \texttt{`White Dwarf'} region as that with $M_{G_0} > 3.9 (BP-RP)_0 + 9$, with the targets appearing as the grey points in Figure~\ref{fig:characterization_CMD}. We note that our \texttt{`White Dwarf'} region is similar to that of \citet{elbadry21} (see also \citealt{gaia18b} and \citealt{gentilefusillo21}).

Proceeding with the rest of the CMD, except for the aforementioned white dwarfs, the stars that fall underneath the border defined by the regions \texttt{`Overlap Dwarf/Subgiant'}, \texttt{`Subgiant'}, and \texttt{`Giant Branch'}, may all be technically classified as MS stars. Nonetheless, we aim for a more detailed classification motivated by the presence of two interesting populations. First, a small but noticeable fraction of the targets scatter below the MS. Assuming their CMD locations are correct, these targets likely correspond to a mix of hot subdwarfs, metal-poor MS stars, and cataclysmic variables (CVs), i.e., interacting binaries consisting of an MS star and a white dwarf (e.g., see \citealt{abril20} and \citealt{dubus24} for illustrations of their {\gaia} CMD). Second, the top panels of Figure~\ref{fig:characterization_CMD} show an overdensity of MS stars with $(BP-RP)_0\gtrsim 0.9$ mag that are slightly more luminous than the bulk population. These systems likely correspond to unresolved binaries (e.g., \citealt{hurley98,lewis22}), with their exact CMD location being determined by their mass-ratio, age, and metallicity values. Although finding multiple systems in this way is more straightforward when done for star clusters (given the common age and metallicity of their members; \citealt{li20,pang23}), we attempt a global CMD identification of photometric binaries in the {\kepler} field (see also \citealt{gordon21,messias22,cantomartins23}). 

To rigorously define these two under- and over-luminous CMD populations, we proceed as follows. We first determine the lower envelope of the MS CMD distribution. This is calculated as a smoothed version of the running 99.5th percentile of $M_{G_0}$ values as a function of $(BP-RP)_0$ color. This lower envelope is illustrated as the red dashed line in the bottom-left panel of Figure~\ref{fig:characterization_CMDregions_upper_lower}. We define the CMD region below the lower envelope as \texttt{`Uncertain~MS'} (i.e., towards larger $M_{G_0}$ values). This region corresponds to 0.4\% of the full {\kepler} sample (828 stars), and targets that fall inside it are shown as the red points in Figure~\ref{fig:characterization_CMD}.

Regarding the photometric binary stars, we first define the bluest color of our search for these targets as $(BP-RP)_0 = 0.9$ mag, corresponding to $T_{\text{eff}} \sim 5400$ K (e.g., \citealt{pecaut13}). This limit is set to avoid misclassifying the stars that are leaving the MS and populating the turn-off region as photometric binaries (e.g., \citealt{simonian19,simonian20}). Then, for each MS star with $(BP-RP)_0 > 0.9$ mag, we calculate the difference between its absolute magnitude and that of the lower envelope evaluated at its $(BP-RP)_0$ color. We name this quantity $\Delta M_{G_0}$ and report it in Table~\ref{tab:table_catalog}. The distribution of $\Delta M_{G_0}$ values is shown in the bottom-right panel of Figure~\ref{fig:characterization_CMDregions_upper_lower}. We fit the $\Delta M_{G_0}$ distribution as the sum of two Gaussians. One of these Gaussians represents the population of photometric binaries (with best-fit parameters $\mu_{\text{Phot.Bin.}}=-0.910$, $\sigma_{\text{Phot.Bin.}}=0.391$, and amplitude$_{\text{Phot.Bin.}}=497$), and the other represents the population of single dwarfs (with best-fit parameters $\mu_{\text{Single}}=-0.496$, $\sigma_{\text{Single}}=0.166$, and amplitude$_{\text{Single}}=1635$). We then find the value at which both Gaussians contribute equally, $\Delta M_{G_0} \equiv \Delta_s \approx -0.758$ mag, and take this as the border between the regions (vertical green line in the bottom-right panel of Figure~\ref{fig:characterization_CMDregions_upper_lower}). This corresponds to a truncated version of the lower envelope shifted by $\Delta_s$ (i.e., towards higher luminosities), as shown by the green dashed line in the bottom-left panel of Figure~\ref{fig:characterization_CMDregions_upper_lower}. We define the region of MS stars more luminous than this as \texttt{`Photometric Binary'}, corresponding to 7.2\% of the sample (14,117 stars), with targets that fall inside it shown as the green points in Figure~\ref{fig:characterization_CMD}.

The double-Gaussian fit is performed in the entire $(BP-RP)_0 > 0.9$ color range, and therefore the results are valid in a global sense. We nonetheless remark that the density of stars decreases towards redder colors, and the MS gets wider and more diffuse beyond $(BP-RP)_0 \gtrsim 2$ (see Figure~\ref{fig:characterization_CMD}), making the separation between photometric binaries and the MS more uncertain. We also note that the value of $\Delta_s \approx -0.758$ mag is close to the theoretically expected magnitude shift between an MS star and an unresolved equal-mass photometric binary of the same color, $\Delta m = -2.5\log_{10}(2) = -0.753$ mag \citep{hurley98}. For this comparison, we highlight the importance of using an appropriate distance indicator. Had we used the \citet{bailerjones21} photogeometric distances instead of the parallax inverse, an inaccurate $\Delta_s$, and therefore \texttt{`Photometric Binary'} CMD region, would have been obtained. This is because the photogeometric \citet{bailerjones21} CMD prior did not include a population of photometric binaries (it rather assumed all stars to be single sources), hence causing a systematic shift in their distances (see their Section 5.5).

Finally, the remaining region of MS stars, bounded by the evolved and \texttt{`Photometric Binary'} regions at higher luminosities (lower $M_{G_0}$ values), and by the \texttt{`Uncertain~MS'} region at lower luminosities (higher $M_{G_0}$ values), is defined as the \texttt{`Dwarf'} region. This corresponds to 56.3\% of the {\kepler} sample (110,735 stars), and is shown as the cyan points in Figure~\ref{fig:characterization_CMD}. 

The sample sizes of all the CMD categories are summarized in Table~\ref{tab:summary_table}. We display the full extent of the CMD regions, and overlay them on the Hess diagram, in Appendix~\ref{sec:app_CMD_categories}. While not perfect, our approach allows a straightforward classification of the {\kepler} stars based on their CMD locations, which can be further characterized in follow-up studies.
\subsection{Impact of metallicity}
\label{subsec:characterization_CMD_metallicity_impact}

The CMD regions of Sect.~\ref{subsec:characterization_CMD_categories} were derived from a suite of PARSEC models at solar-metallicity for the evolved regions, and a sample of mostly solar-metallicity stars for the MS regions (see Figure \ref{fig:characterization_Gmag_distance_RUWE_MHgspspec}). Although spectroscopic metallicities are not available for most stars in our sample, and performing a star-by-star classification is beyond the scope of this paper, we can assess the impact of global metallicity changes. We test the robustness of our CMD categories against the effects of metallicity in three steps: assuming a global solar-metallicity (Sect.~\ref{subsubsec:characterization_CMD_validation_montecarlo}), assuming a moderate metallicity change given the typical scatter (Sect.~\ref{subsubsec:characterization_CMD_validation_modelsuites}), and analyzing the metal-poor and metal-rich tails of the distribution (Sect.~\ref{subsubsec:characterization_CMD_validation_metallicity_outliers}).
\subsubsection{Monte Carlo sampling assuming solar metallicity}
\label{subsubsec:characterization_CMD_validation_montecarlo}

\begin{figure}
\centering
\includegraphics[width=8.0cm]{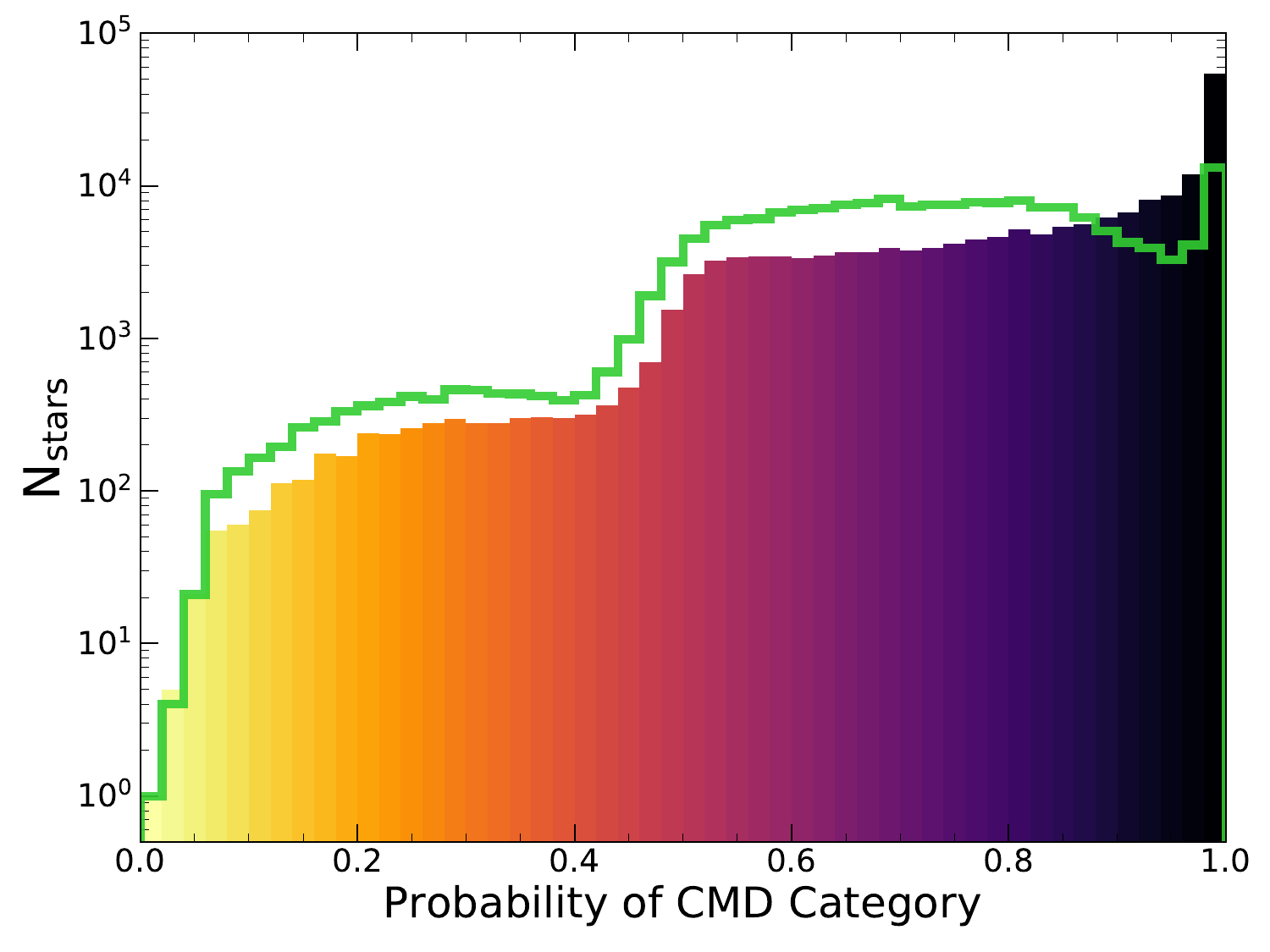}\\
\includegraphics[width=8.0cm]{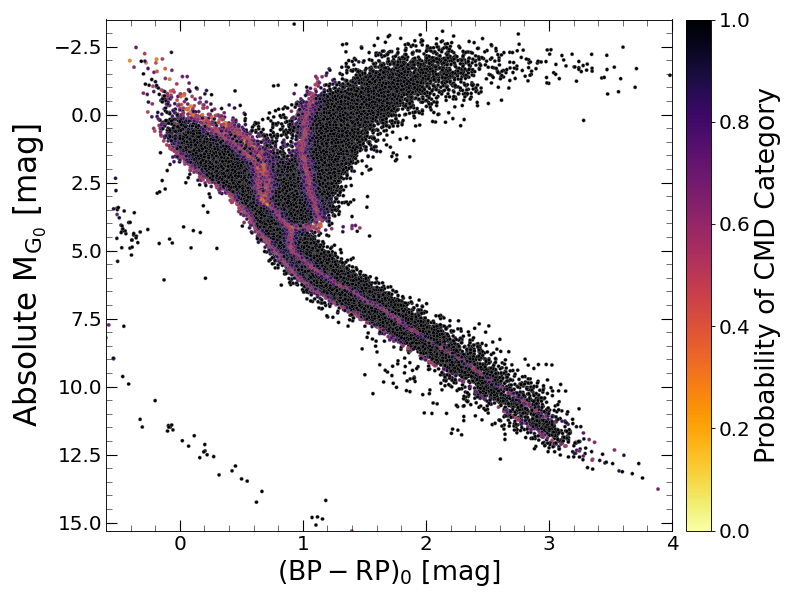}
\caption{Validation of the CMD categories via the Monte Carlo method presented in Sect.~\ref{subsec:characterization_CMD_validation}. Top: Logarithmic distribution of the `Probability of CMD Category' parameter, $P_{\text{CMD}}$. The filled histogram represents the fiducial simulation (Sect.~\ref{subsubsec:characterization_CMD_validation_montecarlo}), while the open histogram represents the simulation that includes metallicity effects (Sect.~\ref{subsubsec:characterization_CMD_validation_modelsuites}). Bottom: CMD projection color-coded by the fiducial $P_{\text{CMD}}$ values. Most of the stars have high $P_{\text{CMD}}$ values, indicating that their assigned CMD categories are reliable. Targets with low $P_{\text{CMD}}$ values are located near the boundaries of the CMD regions (see bottom panel of Figure~\ref{fig:characterization_CMD}), and their CMD categories are consequently less reliable.}
\label{fig:characterization_CMD_MonteCarlo}
\end{figure}

We first examine the confidence of the aforementioned CMD categories ignoring any direct metallicity effects. We quantify this by calculating the probability that a given star recovers its assigned CMD category from Sect.~\ref{subsec:characterization_CMD_categories} (i.e., the `Flag CMD' column from Table~\ref{tab:table_catalog}), given the uncertainties on its CMD position. We perform a Monte Carlo simulation and sample the CMD location of each star $N_{\text{s}}=1,000$ times following Gaussian distributions $\mathcal{N}((BP-RP)_0,\sigma_{(BP-RP)_0})$ for color and  $\mathcal{N}(M_{G_0},\sigma_{M_{G_0}})$ for magnitude. For each sampling, we infer the CMD category given the regions in the bottom panel of Figure~\ref{fig:characterization_CMD}. Then, we calculate the number of times that the assigned CMD category is recovered and divide it by $N_{\text{s}}$. We call this the `Probability of CMD Category' ($P_{\text{CMD}}$), which ranges from 0 to 1. For instance, a star where the assigned CMD category from Sect.~\ref{subsec:characterization_CMD_categories} is found in 750 of the 1,000 samplings will have a value of $P_{\text{CMD}}=0.75$. We report the $P_{\text{CMD}}$ values in Table~\ref{tab:table_catalog}.

We show the distribution of the $P_{\text{CMD}}$ values as the filled histogram in the top panel of Figure~\ref{fig:characterization_CMD_MonteCarlo}. The distribution is heavily concentrated at values of $P_{\text{CMD}}\approx 1$ (note the logarithmic $y$-axis), with a median value of $P_{\text{CMD}} = 0.90$. More specifically, 96\% of the CMD sample has $P_{\text{CMD}}\geq 0.50$, and 77\% of it has $P_{\text{CMD}}\geq 0.70$. In the bottom panel of Figure~\ref{fig:characterization_CMD_MonteCarlo}, we show the CMD projection color-coded by the $P_{\text{CMD}}$ values (in the same color scheme as the top panel). The diagram illustrates that the targets with low $P_{\text{CMD}}$ values correspond to stars located near the boundaries of the CMD regions, i.e., stars more prone to scattering to different CMD regions given their measurement uncertainties. Accordingly, given the smaller area it covers on the CMD, the \texttt{`Overlap Dwarf/Subgiant'} region has the highest fraction of stars with lower $P_{\text{CMD}}$ values. Overall, however, the fact that most of the stars have high $P_{\text{CMD}}$ values demonstrates the robustness of our CMD categories for solar-metallicity targets.
\subsubsection{Monte Carlo sampling given typical metallicity scatter}
\label{subsubsec:characterization_CMD_validation_modelsuites}

We now test the impact of the solar-metallicity assumption by comparing the regions that the evolved populations occupy in suites of metal-poor and metal-rich models. We illustrate these in Appendix~\ref{sec:app_PARSEC_metallicity}, where we overlay the borders of the CMD regions defined at [M/H]=0, with PARSEC suites at [M/H]=-0.3 dex and [M/H]=+0.3 dex. The comparison reveals that, at a global level, the regions shift by $\frac{\text{d}(BP-RP)_0}{\text{d}\text{[M/H]}} \sim 0.3$ mag/dex and $\frac{\text{d}M_{G_0}}{\text{d}\text{[M/H]}} \sim 0.9$ mag/dex (i.e., getting redder and less luminous with increasing metallicity).

To examine the impact that moderate metallicity shifts have on our CMD classification, we repeat the Monte Carlo sampling from Sect.~\ref{subsubsec:characterization_CMD_validation_montecarlo} and incorporate the effects of the unknown spectroscopic metallicity for most targets. We take these global color- and magnitude-shifts due to metallicity and multiply them by the standard deviation of the \texttt{gspspec} metallicity distribution (bottom-right panel of Figure~\ref{fig:characterization_Gmag_distance_RUWE_MHgspspec}), $\sigma_{\text{[M/H],gspspec}} = 0.2$ dex. Thus, at the 1$\sigma$ level, the global shifts in color and magnitude due to the unknown metallicity are 0.06 and 0.18 mag, respectively. We take these shifts as systematic errors and add them in quadrature with the nominal errors from Sect.~\ref{subsec:characterization_CMD_CMDsample}. This increases the median errors to $\sigma_{(BP-RP)_0}$ = 0.077 mag and $\sigma_{M_{G_0}}=0.191$ mag. 

We re-run the simulation from Sect.~\ref{subsubsec:characterization_CMD_validation_montecarlo} with these inflated errors, and show the resulting $P_{\text{CMD}}$ distribution as the green line in the top panel of Figure~\ref{fig:characterization_CMD_MonteCarlo}. As could be expected due to the larger error bars, the fraction of targets with very high probability decreases (i.e., they are more likely to scatter to other CMD regions). Nevertheless, the median $P_{\text{CMD}}$ value is only moderately reduced (0.90 before vs. 0.73 now), and the fraction of the sample with $P_{\text{CMD}} \geq 0.50$ remains high (96\% before vs. 93\% now). Thus, we conclude that our overall CMD classification is robust against moderate ($1\sigma$, or $\approx$ 0.2 dex) metallicity changes. We report these revised probability values in Table~\ref{tab:table_catalog} as $P_{\text{CMD,[M/H]}}$.
\subsubsection{Misclassification at the tails of the metallicity distribution}
\label{subsubsec:characterization_CMD_validation_metallicity_outliers}

Naturally, our CMD classification loses accuracy towards more extreme (non-solar) metallicity values. Although we lack spectroscopic metallicities for most targets, complementary data sets that leverage the lower resolution {\gaia} spectrophotometry can provide useful estimates. With this aim, we use the metallicities from \citet{andrae23b}, which are by construction on the APOGEE DR17 scale \citep{abdurrouf22}. Although subject to larger uncertainties compared to higher resolution spectroscopy, the \citet{andrae23b} values allow us to test the effects of metallicities that deviate considerably from solar on the CMD classification.

\begin{figure}
\centering
\includegraphics[width=8.0cm]{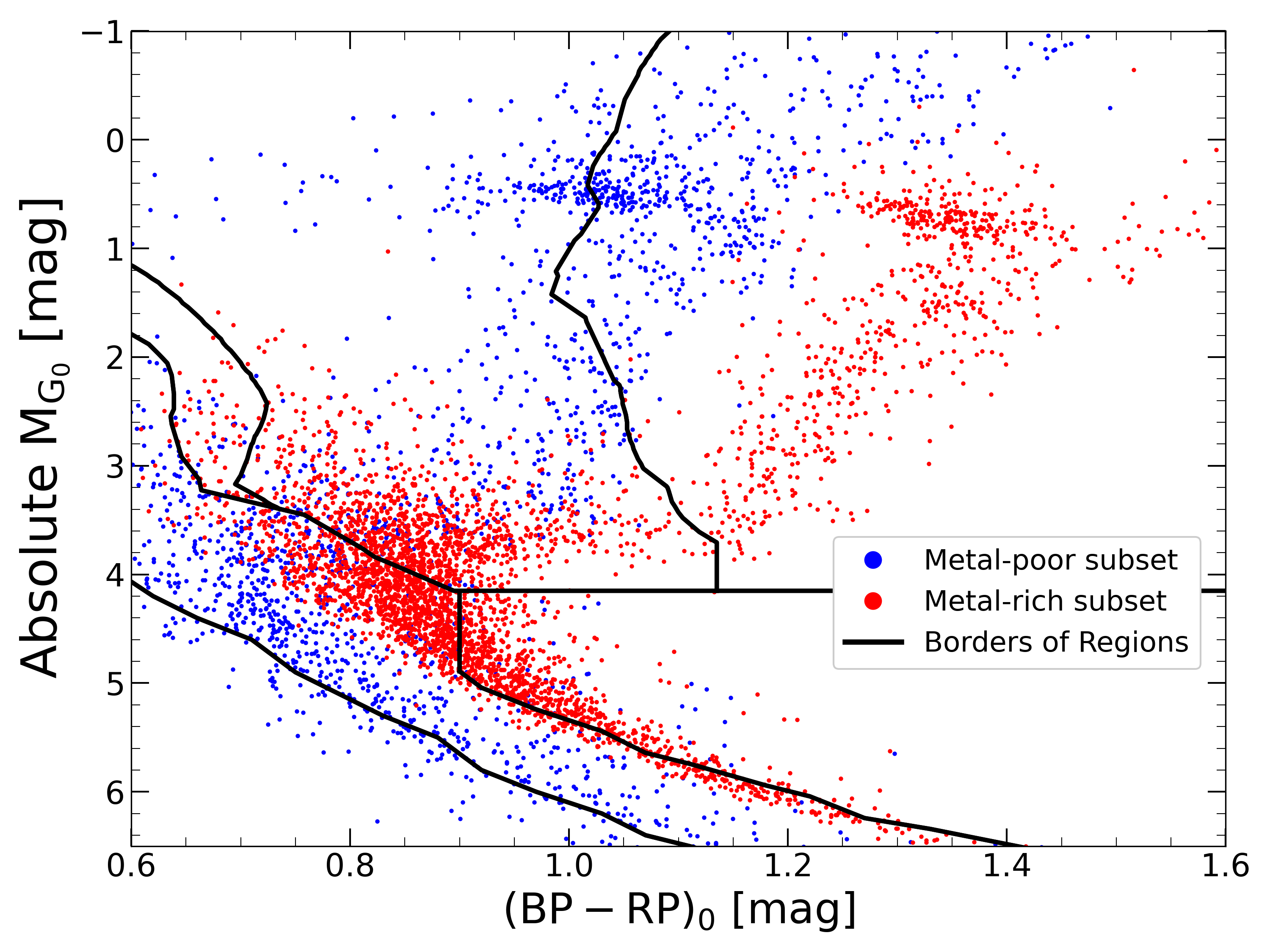}
\caption{Metallicity impact on the CMD classification. The blue and red points show, respectively, the metal-poor and metal-rich tails (beyond $2\sigma$) of the \citet{andrae23b} metallicity distribution. Our CMD classification loses accuracy towards extreme-metallicity values, and this effect is enhanced in certain regions of the CMD.}
\label{fig:characterization_CMD_metallicity_outliers}
\end{figure}

For this purpose, we select stars with \citet{andrae23b} metallicities that are beyond the 2$\sigma$ limits of the distribution for the {\kepler} targets, namely those with [M/H] values below the 2.3$^{\text{rd}}$ and above the 97.7$^{\text{th}}$ percentiles (which translate to [M/H] $< -0.63$ dex and [M/H] $> +0.29$ dex, respectively). These correspond to 7,672 out of the 168,635 targets found by crossmatching our sample with \citet{andrae23b}. Figure~\ref{fig:characterization_CMD_metallicity_outliers} illustrates these metal-poor and metal-rich targets as the blue and red points, respectively. The metallicity effects are readily apparent, with the metal-poor (metal-rich) subset being displaced from the fiducial solar-metallicity CMD regions towards bluer (redder) colors and brighter (fainter) magnitudes, respectively. This demonstrates that our CMD classification is not accurate for such metallicity-extreme targets. These deviations affect some regions of the CMD more heavily, of which we highlight three: metal-poor MS stars shifting to the \texttt{`Uncertain MS'} region, metal-rich MS stars shifting to the \texttt{`Photometric Binary'} region, and metal-poor RC stars shifting to the \texttt{`Subgiant'} region.

We flag these targets via the `Flag Metal-Poor Tail' and `Flag Metal-Rich Tail' columns in Table~\ref{tab:table_catalog}. As demonstrated in Figure~\ref{fig:characterization_CMD_metallicity_outliers}, these stars are more prone to having inaccurate CMD classifications, and thus their `Flag CMD' labels should be used with extreme caution (especially for the three aforementioned cases). Nonetheless, we note that even though their CMD categories might be mistaken, their CMD locations are accurate and reliable.

\subsection{Validation} 
\label{subsec:characterization_CMD_validation}

\subsubsection{Comparison with asteroseismology}
\label{subsubsec:characterization_CMD_validation_asteroseismology}

To validate the CMD categories with an external reference, we now perform a comparison with the asteroseismic catalogs of evolved stars from \citet{yu18} and \citet{yu20}. Combining both data sets, they report a list of 18,604 unique KIC IDs. 92.1\% of these (17,133 stars) pass our CMD quality cuts, and we examine the classification flag we assign to them. In our catalog, most of their stars are classified as \texttt{`Giant Branch'} (97.7\%), followed by the \texttt{`Subgiant'} category (2.2\%). Combining these, they add up to 99.8\% (17,105 stars) of the asteroseismic sample being in evolved CMD categories, in excellent agreement with expectations. The remaining 0.2\% of the sample belongs to the other CMD categories, with 24 out of 28 stars being found in the \texttt{`Dwarf'} region. This likely hints at targets with contamination in their {\kepler} light curves due to bright neighboring stars (see Sect.~\ref{sec:misclassifiedstars}). 

We perform an analogous comparison with APOKASC-3 \citep{pinsonneault24}, which combines {\kepler} asteroseismology with APOGEE spectroscopy \citep{abdurrouf22} to derive stellar parameters. Importantly for our purposes, they derive evolutionary states, thus providing a comparison point for our CMD classification. The APOKASC-3 catalog reports a list of 15,808 KIC IDs, 88.4\% of which (13,977 stars) pass our CMD quality cuts. Restricting the comparison to 13,626 targets classified as `RGB', `RC', or `RC/RGB' in APOKASC-3, we classify 96.6\% of them (13,167 stars) as CMD \texttt{`Giant Branch'} and 3.2\% of them (441 stars) as CMD \texttt{`Subgiant'}, with the remaining 0.1\% (18 stars) being classified in the other CMD categories. The excellent agreements with these independent catalogs further validate our CMD classification scheme.
\subsubsection{Comparison with {\gaia} DR3 evolutionary stages}
\label{subsubsec:characterization_CMD_validation_gaiadr3}

As part of the Astrophysical Parameters table, {\gaia} DR3 reports the analysis of the Final Luminosity Age Mass Estimator (FLAME) module \citep{creevey23,fouesneau23}. In brief, this module takes as input stellar parameters from \texttt{gspphot} and/or \texttt{gspspec}, extinction from \texttt{gspphot}, and distance estimates (parallax inverse or \texttt{gspphot} distance) to produce stellar mass as well as evolutionary parameters by comparing with BASTI solar-metallicity models \citep{hidalgo18}. One of the derived properties is the \texttt{evolstage\_flame} ($\epsilon$), which quantifies the evolutionary stage of stars. We compare this parameter with our CMD categories.

To facilitate the analysis, we translate the $\epsilon$ values to the three labels defined in Section 3.3.2 of \citet{fouesneau23}, namely $100 \leq \epsilon \leq 420$ as MS, $420 < \epsilon \leq 490$ as SGB, and $490 < \epsilon$ as RGB. For the subset of stars in common, we compare these labels with our CMD flags\footnote{For practicality, when comparing with the FLAME labels, we group our CMD categories as main sequence = (\texttt{`Dwarf'}+\texttt{`Photometric Binary'}+\texttt{`Uncertain MS'}), subgiant = (\texttt{`Subgiant'}+\texttt{`Overlap Dwarf/Subgiant'}), and red giants = (\texttt{`Giant Branch'}).} from Sect.~\ref{subsec:characterization_CMD_categories}. We find agreements of 93.9\% for MS stars, 68.0\% for SGB stars, and 81.6\% for RGB stars.

Regarding their overall comparison, despite differences in the specific modeling (PARSEC vs. BASTI) and input data choices (particularly extinctions and distances), the good global agreement with FLAME validates our CMD classification. In absolute terms, relative to the full {\kepler} sample (196,762 stars), our CMD classification is available for an extra 8\% of targets than the FLAME \texttt{evolstage\_flame} parameter (179,295 vs. 163,314, respectively). Moreover, our analysis identifies additional categories of interesting populations that are relevant for complementary studies (e.g., photometric binaries and their role in stellar rotation; \citealt{stauffer18}).
\subsection{Caveats} 
\label{subsec:characterization_CMD_caveats}

The CMD classification presented throughout Sect.~\ref{sec:characterization_CMD} will allow searches of interesting populations for follow-up studies (see Sect.~\ref{sec:differenceDR2vsDR3}, Sect.~\ref{sec:misclassifiedstars}, and Sect.~\ref{sec:variability}).
Naturally, however, the procedure we have adopted carries assumptions and simplifications that must be considered when using it. We now discuss the caveats of the method and potential improvements for future works:

\begin{itemize}
\item In this paper we have only analyzed the {\gaia} DR3 $M_{G_0}$ vs. $(BP-RP)_0$ CMD. This could be complemented by also leveraging photometry from other surveys (e.g., 2MASS, ALLWISE, Pan-STARRS; \citealt{skrutskie06,wright10,chambers16}). Moreover, the classification is based on photometric and astrometric data, but the use of spectroscopic ($\log(g)$, $T_{\text{eff}}$, [M/H]) and asteroseismic ($\Delta \nu$, $\nu_{\text{max}}$) parameters could help improve the methodology (e.g., \citealt{pinsonneault14,pinsonneault18,pinsonneault24,serenelli17}).
\item The photometric analysis has been based on the mean {\gaia} DR3 magnitudes (i.e., \texttt{phot\_g\_mean\_mag}, \texttt{phot\_bp\_mean\_mag}, and \texttt{phot\_rp\_mean\_mag}). Thus, the current CMD classification and quality cuts might not be fully accurate or appropriate for stars that exhibit photometric variability. Indeed, some of these targets have time-dependent CMD locations (e.g., see Figure 11 of \citealt{gaia19}). We further discuss this in Sect.~\ref{sec:variability}.
\item We have only used the PARSEC models as a reference for the CMD classification. Thus, complementing the analysis with other families of stellar models (e.g., MIST, DSEP; \citealt{choi16,dotter08}) could further refine and validate the categories here defined.
\item For simplicity, the classification ignores the presence of certain evolutionary stages (e.g., pre-MS) and other types of systems (e.g., symbiotic stars with white dwarf companions, where the surrounding nebula and circumstellar material can influence the measured flux and increase the extinction; \citealt{munari19}). Hence, if present, these stars will be misclassified by the `Flag CMD' column in Table~\ref{tab:table_catalog}, as likely more than pure {\gaia} CMD data is required to reliably identify them (e.g., \citealt{merc20,merc21}). For instance, in the case of pre-MS stars, if present in the sample, they would likely be classified in the \texttt{`Photometric Binary'} region. We nonetheless note that young stars are not expected in the {\kepler} field given its Galactic latitude \citep{zwintz22}.
\item The {\kepler} field is a population of mixed stellar ages and metallicities. Therefore, although based on our own (Figure~\ref{fig:characterization_Gmag_distance_RUWE_MHgspspec}) as well as the literature \citep{ren16,zong18} metallicity distribution, our assumption of a global solar metallicity for the fiducial CMD analysis will introduce inaccuracies in the classification. These are quantified for metallicity values in the best (solar), typical ($1\sigma$ scatter), and extreme (beyond $2\sigma$) scenarios in Sect.~\ref{subsec:characterization_CMD_metallicity_impact}. Nevertheless, a future improvement would be to perform the CMD classification on a star-by-star basis with knowledge of the spectroscopic metallicities. At the moment, 12.2\% of the {\kepler} sample have \texttt{gspspec} metallicities (4.3\% of them being the most reliable; see Appendix \ref{sec:app_calibration_quality_gspspec}) from the {\gaia} DR3 RVS (radial velocity spectrometer; \citealt{recioblanco23}). As we have initially demonstrated in Sect.~\ref{subsubsec:characterization_CMD_validation_metallicity_outliers}, however, this fraction may be significantly expanded in future works using novel data sets (e.g., \citealt{andrae23b,zhang23,khalatyan24}) that leverage the lower-resolution XP spectra \citep{deangeli23,montegriffo23}.

\item Regarding the \texttt{`Photometric Binary'} category, given the aforementioned mix of stellar ages and metallicities present in the sample, our decisions regarding its color and magnitude extent might not fully suit all purposes. To aid in this, in Table~\ref{tab:table_catalog} we report the $\Delta M_{G_0}$ values, from which readers may design their customized selection cuts.
\item We have used the latest {\gaia} DR3 parallaxes and photometry, and attempted to ensure well-measured magnitudes, colors, and distances via the quality cuts applied in Sect.~\ref{subsec:characterization_CMD_qualitycuts}. In the future, thanks to upcoming {\gaia} data releases, more precise CMD placement will be possible, and a larger sample may be classified thanks to more stars passing the quality cuts.
\end{itemize}

\begin{table}
\caption{Categories identified in this work.}
\label{tab:summary_table}
\centering                        
    \begin{tabular}{lrr}
    	\hline
    	\hline
        Category & N$_{\text{stars}}$ & Percentage \\
        \hline
        Full {\kepler} Sample & 196,762 & 100\% \\
        \hline
        CMD Dwarf & 110,735 & 56.3\% \\
        CMD Subgiant & 25,901 & 13.2\% \\
        CMD Giant Branch & 22,098 & 11.2\% \\       
        CMD Photometric Binary & 14,117 & 7.2\% \\         
        CMD Overlap Dwarf/Subgiant & 5,578 & 2.8\% \\
        CMD Uncertain MS & 828 & 0.4\% \\
        CMD White Dwarf & 38 & 0.02\% \\        
        Not in CMD Sample & 17,467 & 8.9\% \\
        \hline
        Binary RUWE & 23,973 & 12.2\% \\
        Binary RV Variable & 4,072 & 2.1\% \\
        Binary NSS & 4,005 & 2.0\% \\
        Binary Eclipsing {\kepler} & 2,865 & 1.5\% \\
        Binary Eclipsing {\gaia} & 854 & 0.4\% \\
        Binary {\gaia} Variable & 775 & 0.4\% \\
        Binary SB9 & 49 & 0.02\% \\
        Binary or Multiple NEA & 71 & 0.04\% \\
        Binary HGCA & 97 & 0.05\% \\        
        Binary WDS & 2,829 & 1.4\% \\
        Binary Union & 31,334 & 15.9\% \\
       \hline
    \end{tabular}
\tablefoot{Summary of the sample classifications, with their respective numbers and percentages relative to the full {\kepler} sample shown in the first block. The second block refers to the CMD categories defined in Sect.~\ref{sec:characterization_CMD}. The third block refers to the binary categories defined in Sect.~\ref{sec:characterization_binaries}. Note that while the CMD categories are mutually exclusive (and add up to 100\% of the sample), the binary ones are not, and thus a given target can belong to one or more binary categories simultaneously (see Figure~\ref{fig:characterization_binaries}). We also highlight that the `Binary Union' flag counts stars in more than one binary category only once, and it is independent of the `Flag CMD' classification (hence does not account for the \texttt{`Photometric Binary'} category).}
\end{table}
\section{Binary characterization} 
\label{sec:characterization_binaries}

Binary (and higher-order multiple) systems are of crucial importance in astrophysics \citep{duquennoy91,raghavan10,torres10,duchene13,prsa18,serenelli21}. More specifically, binary stars observed by the {\kepler} mission are providing fundamental constraints to numerous problems (e.g., \citealt{hambleton13,hambleton16,hambleton18,beck14,beck22,sandquist16,lurie17,godoyrivera18,gehan22}).

In this section, we investigate our catalog in search of binary candidates using a range of detection methods (e.g., \citealt{gaia23c}). We do this by leveraging astrometric and radial velocity (RV) {\gaia} data (Sect.~\ref{subsec:characterization_binaries_ruwe} and Sect.~\ref{subsec:characterization_binaries_rvvariable}), as well as by crossmatching with binary tables published in {\gaia} DR3 and complementary literature databases (Sect.~\ref{subsec:characterization_binaries_nss} through Sect.~\ref{subsec:characterization_binaries_wds}). We generate a flag for each of these binary categories, and also merge them into a combined one for convenience to users (Sect.~\ref{subsec:characterization_binaries_union}). We report the binary flags in Table~\ref{tab:table_catalog}, and summarize the number of systems in each category in Table~\ref{tab:summary_table}. For illustration purposes, Figure~\ref{fig:characterization_binaries} shows the CMD projection of these populations.

We highlight that, contrary to the \texttt{`Photometric Binary'} category described in Sect.~\ref{sec:characterization_CMD}, the binary categories defined in this section are not subject to the CMD quality cuts from Sect.~\ref{subsec:characterization_CMD_qualitycuts}, and are completely independent of the `Flag CMD' classifications from Sect.~\ref{subsec:characterization_CMD_categories}. Thus, the systems classified as binary candidates in Sect.~\ref{sec:characterization_binaries} that lack CMD information, are absent from Figure~\ref{fig:characterization_binaries}. Analogously, while some overlap exists, the binary categories of this section are defined independently of the \texttt{`Photometric Binary'} CMD category from Sect.~\ref{sec:characterization_CMD}. 
\subsection{RUWE} 
\label{subsec:characterization_binaries_ruwe}

For a given star, the RUWE value is an astrometric parameter that characterizes how appropriate the {\gaia} single-star solution is \citep{gaia21,lindegren21a}. By construction, well-behaved single stars have RUWE values of around 1.0 \citep{lindegrenruwe18}, with larger values indicating binarity (\citealt{belokurov20,penoyre20}; see also \citealt{fitton22} for other applications). Since its introduction, it has been widely used in the literature to characterize binary candidates, with typical RUWE thresholds between 1.2 and 1.4 (e.g., \citealt{berger20,ziegler20,kervella22,penoyre22}).

\begin{figure*}
\centering
\includegraphics[width=8.0cm]{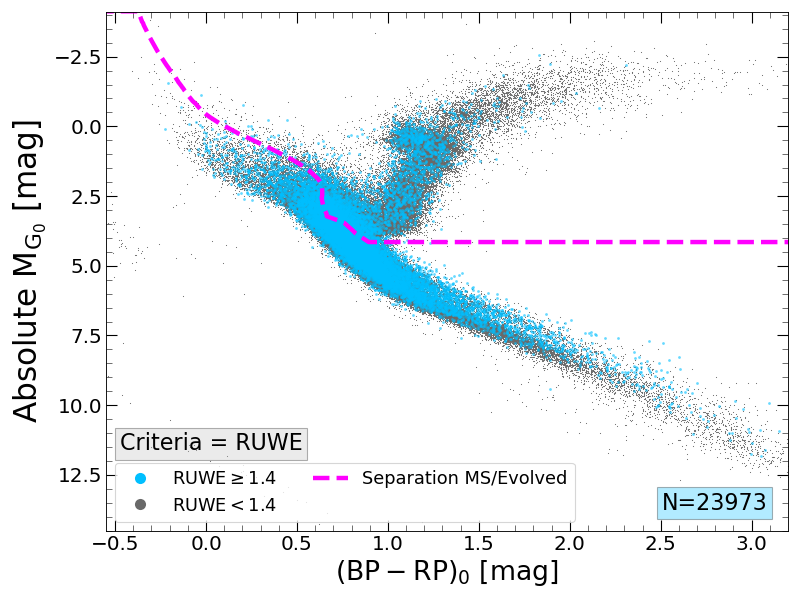}
\includegraphics[width=8.0cm]{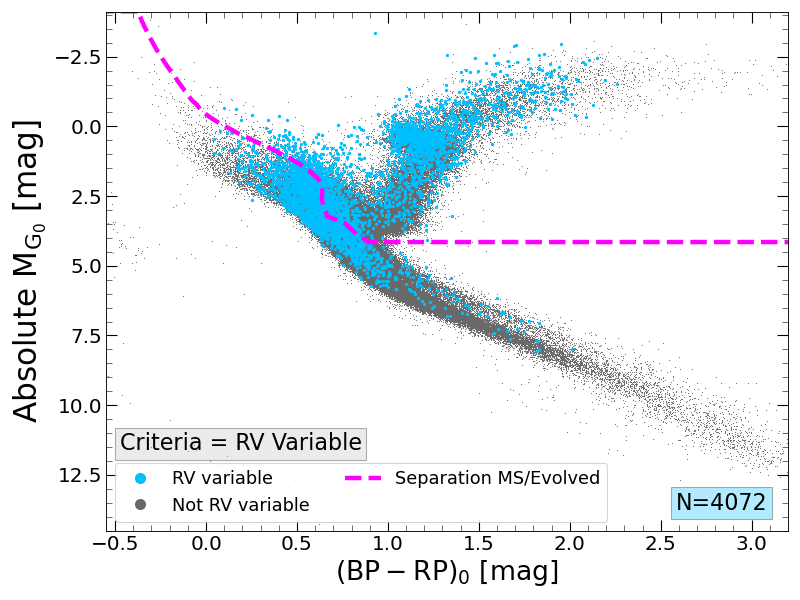}\\
\includegraphics[width=8.0cm]{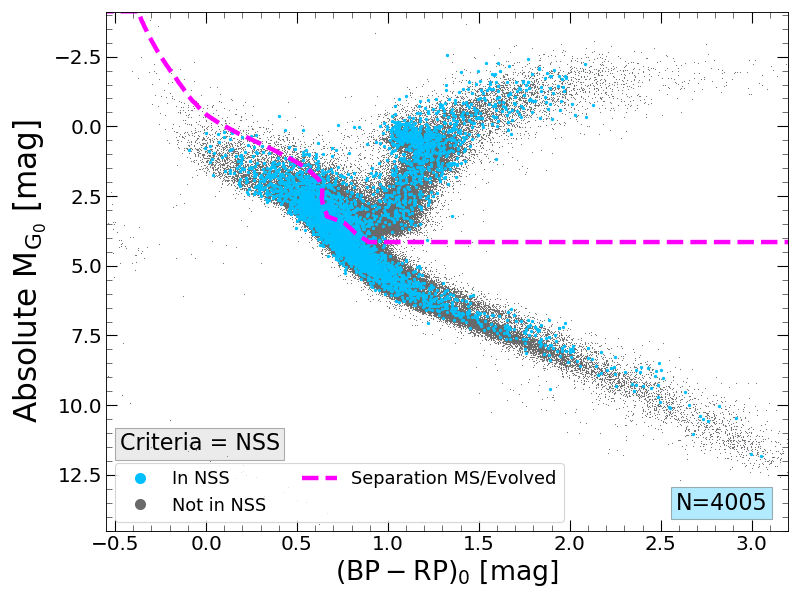}
\includegraphics[width=8.0cm]{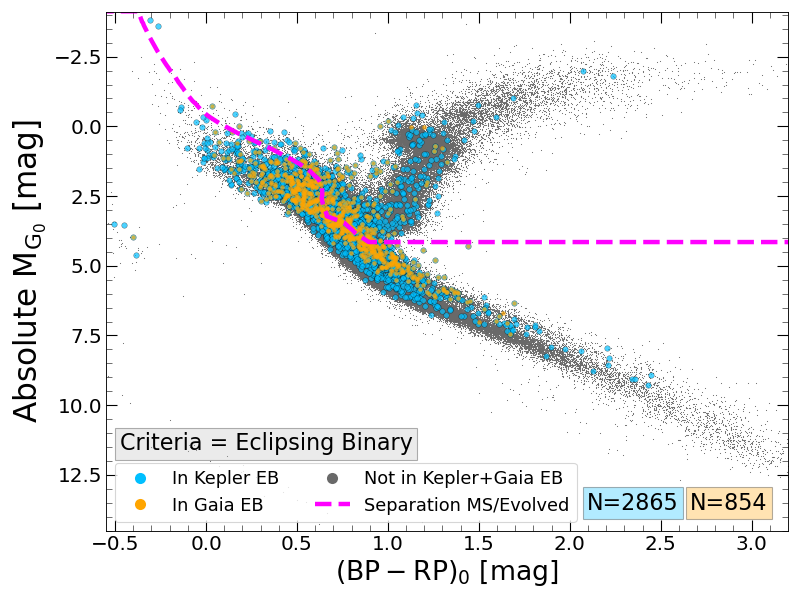}\\
\includegraphics[width=8.0cm]{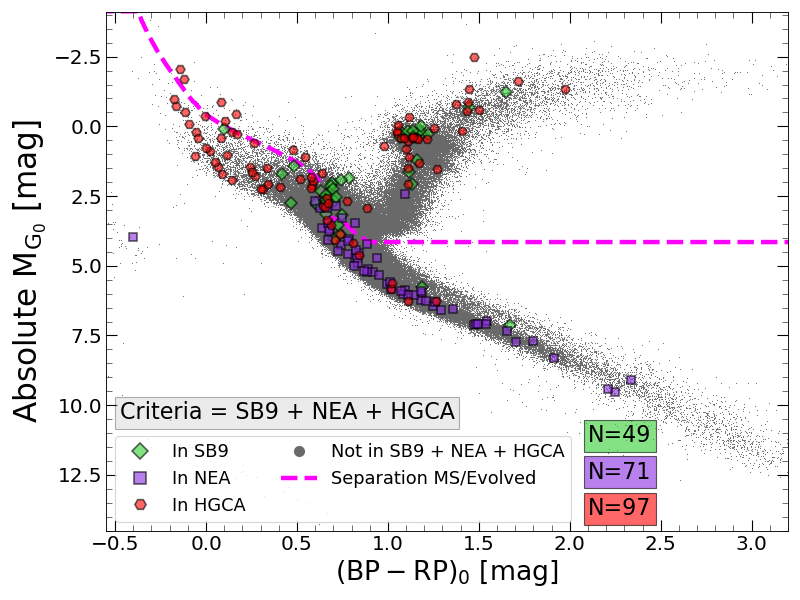}
\includegraphics[width=8.0cm]{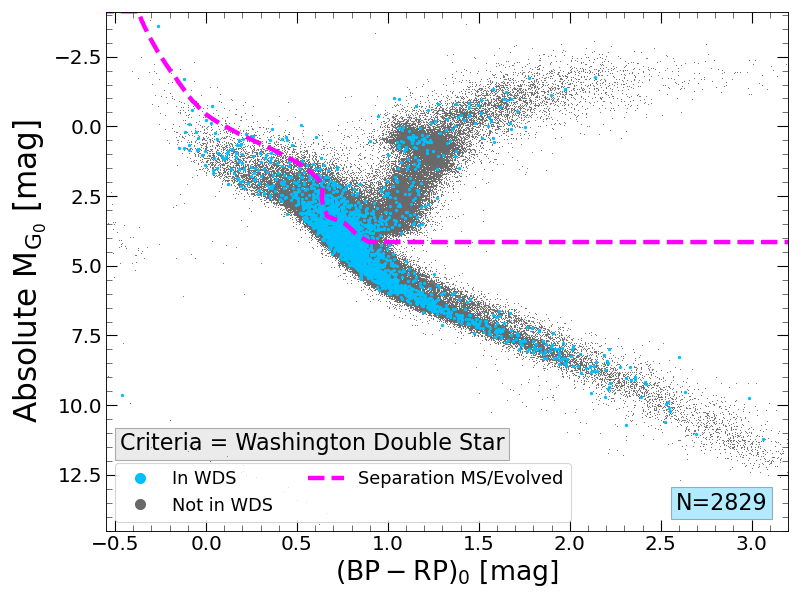}\\
\caption{Characterization of the binary categories we define in Sect.~\ref{sec:characterization_binaries}. The panels display the CMD projection of the RUWE, RV Variable, NSS, {\kepler} $+$ {\gaia} EB, SB9 $+$ NEA $+$ HGCA, and WDS binary samples, respectively. In all the CMDs, we show the separation between MS and evolved stars as guidance (dashed fuchsia line). The number of systems in each category is summarized in the bottom-right corner. The CMD projections help to illustrate the selection effects of each category (see Sect.~\ref{sec:characterization_binaries} for details).}
\label{fig:characterization_binaries}
\end{figure*}

The RUWE distribution for the targets is shown in the bottom-left panel of Figure~\ref{fig:characterization_Gmag_distance_RUWE_MHgspspec}. From this, we classify all the stars with RUWE~$\geq 1.4$ as binary candidates (dotted red line), which corresponds to 12.2\% of the {\kepler} sample (23,973 stars). The CMD projection of this subset is shown in the top-left panel of Figure~\ref{fig:characterization_binaries}, which illustrates that it spans the entirety of the CMD. Interestingly, this includes the \texttt{`Photometric Binary'} region from Sect.~\ref{sec:characterization_CMD}, thus providing further evidence of its binary nature. The RUWE binaries are identified as such in Table~\ref{tab:table_catalog} via the column `Flag RUWE'.

As this selection cut may be too stringent for certain purposes, and some users may wish to define their customized selections (e.g., \citealt{castroginard24}), we also report the RUWE values in Table~\ref{tab:table_catalog} for completeness. For instance, had we adopted a less stringent threshold of RUWE~$\geq 1.2$ (dashed cyan line in Figure~\ref{fig:characterization_Gmag_distance_RUWE_MHgspspec}), we would have classified 15.7\% of the {\kepler} sample (30,798 stars) as binary candidates.

Additionally, we note that the RUWE parameter does carry limitations for binary identification. Recently, \citet{beck24} showed that binary systems detected through other techniques may often still appear as having RUWE values below the binary threshold (e.g., for the literature catalog of spectroscopic binaries discussed in Sect.~\ref{subsec:characterization_binaries_sb9}, they found that $\sim40\%$ of systems have RUWE~$\leq1.4$). \citet{beck24} concluded that targets with high RUWE values are typically systems located close to Earth with longer orbital periods (see their Figure~4). Thus, these systems can more easily produce a clear astrometric signature of binary motion in {\gaia}. This highlights the importance of using complementary binary detection methods, as we do in the following subsections.
\subsection{RV variable stars} 
\label{subsec:characterization_binaries_rvvariable}

RV variations are a useful means of identifying binary systems. In {\gaia} DR3, mean RV values are available for 51.7\% of the {\kepler} sample (101,756 stars). Although the full epoch {\gaia} RV measurements will not be made available until DR4 (\citealt{gaia23a}; but see also \citealt{gaia23d}), the DR3 does contain RV variability indices. We use the prescription reported by \citet{katz23} to identify targets that can be classified as RV variables based on their RV scatter. In brief, the selection examines the consistency and noise of the RV time series\footnote{From the \texttt{gaiadr3.gaia\_source} table, we use the RV variability selection criteria $\texttt{rv\_nb\_transits}\geq10$, 3900 K $ \leq \texttt{rv\_template\_teff} \leq 8000$ K, $\texttt{rv\_chisq\_pvalue} \leq 0.01$, and $\texttt{rv\_renormalised\_gof} > 4$ (see Section 3.7 of \citealt{katz23}).}. While the \citet{katz23} criteria can be used for both binary systems and variable stars (e.g., Cepheids and RR Lyraes; see their Section 11), we find the overlap between both categories to be limited\footnote{Only 9.7\% (395 out of 4,072) of the RV variable targets from Sect.~\ref{sec:characterization_binaries} are also classified as photometrically variable in Sect.~\ref{sec:variability}.} for our target sample. Thus, we attribute the RV variability signal as coming predominantly from the presence of unresolved companions, and classify these systems as binary candidates (e.g., \citealt{cao22,patton24,silvabeyer23}).

By following this approach, we classify 4,072 targets as RV variables (2.1\% of the {\kepler} sample). The CMD projection of this subset is shown in the top-right panel of Figure~\ref{fig:characterization_binaries}, which illustrates that this criterion is biased towards more luminous targets, scarcely populating the region of $M_{G_0} \gtrsim 6$ mag. These RV variable binary candidates are identified as such in Table~\ref{tab:table_catalog} via the column `Flag RV Variable'.
\subsection{{\gaia} DR3 non-single stars} 
\label{subsec:characterization_binaries_nss}

{\gaia} DR3 published over 800,000 solutions for candidate non-single stars (NSS) systems \citep{gaia23a,gaia23c}. This sample is comprised of binaries flagged under different solution types and includes astrometric, spectroscopic, and eclipsing binaries (\citealt{halbwachs23,holl23,gosset25}).

We examine the \texttt{non\_single\_star} column in the \texttt{gaiadr3.gaia\_source} table and find 4,005 {\kepler} targets (2.0\% of the sample) flagged as NSS. The CMD projection of this subset is shown in the middle-left panel of Figure~\ref{fig:characterization_binaries}, which illustrates that it virtually spans the entirety of the CMD (see also Figure 4 of \citealt{gaia23c}). These NSS binaries are identified as such in Table~\ref{tab:table_catalog} via the column `Flag NSS'. We investigate them in more detail in Appendix \ref{sec:app_NSS_tables}, where the `NSS Type' and  `NSS Tables' columns of Table~\ref{tab:table_catalog} are generated.
\subsection{{\kepler} eclipsing binary stars} 
\label{subsec:characterization_binaries_eclipsing_kepler}

Numerous searches for eclipsing binary (EB) signatures have been performed in the {\kepler} light curves (e.g., \citealt{coughlin11,prsa11}). In particular, the Villanova catalog\footnote{Note that this catalog includes several classes of systems (e.g., with tertiary eclipses, with changing eclipse depths, and heartbeat systems, among others). We include the full table in our crossmatch. The catalog is hosted at \url{http://keplerebs.villanova.edu/}} of {\kepler} EBs provides a detailed inventory and characterization of such systems. For completeness, we download the most recent version\footnote{{\kepler} EB Catalog, Third Revision, version = 08/08/2019.} of the catalog reported by \citet{kirk16} and crossmatch it with our sample. We find 2,865 targets in the {\kepler} EB catalog (1.5\% of our target list). The CMD projection of this subset is shown in cyan in the middle-right panel of Figure~\ref{fig:characterization_binaries}, and is in good qualitative agreement with other literature studies (e.g., \citealt{mowlavi23}). These {\kepler} EBs are identified as such in Table~\ref{tab:table_catalog} via the column `Flag EB {\kepler}'.
\subsection{{\gaia} eclipsing binary stars} 
\label{subsec:characterization_binaries_eclipsing_gaia}

We complement the above with the {\gaia} DR3 catalog of EB candidates \citep{mowlavi23}. These systems were identified from photometric variability criteria applied on their $G$-band light curves, and a fraction of them also have NSS orbital solutions (see also Sect.~\ref{subsec:characterization_binaries_nss}). We crossmatch with the  \texttt{gaiadr3.vari\_eclipsing\_binary} table and find 854 targets (0.4\% of the sample). The CMD projection of this subset is shown in orange in the middle-right panel of Figure~\ref{fig:characterization_binaries}. These {\gaia} EBs are identified as such in Table~\ref{tab:table_catalog} via the column `Flag EB {\gaia}'.

In our catalog, both {\kepler} and {\gaia} EB flags are reported independently. We note, however, that a large overlap exists between them, in the sense that most of the {\gaia} EBs are contained within the {\kepler} EB sample. More specifically, 801 out of 854 {\gaia} EBs (94\%) are also flagged as {\kepler} EBs. Comparing the orbital periods in this overlap sample, we find 661 out of 801 systems (83\%) to have values along the 1:1 relation (fractional differences $<0.1\%$). This is in good agreement with the literature comparisons from \citet{mowlavi23}.
\subsection{Other {\gaia} variable binaries} 
\label{subsec:characterization_binaries_othervariable_gaia}

Beyond the aforementioned EBs (Sect.~\ref{subsec:characterization_binaries_eclipsing_gaia}), there are other categories of binary candidates that can be flagged based on their {\gaia} DR3 variability \citep{rimoldini23}. We elaborate on the {\gaia} variability classification for the {\kepler} targets in Sect.~\ref{sec:variability}. For the purpose of identifying binaries, however, we highlight the following four categories: cataclysmic variables (\texttt{CV}), ellipsoidal variables (\texttt{ELL}), RS Canum Venaticorum variables (\texttt{RS}), and symbiotic variables (\texttt{SYST}). We specifically look for stars classified in these categories, and find 15 as \texttt{CV}, 0 as \texttt{ELL}, 759 as \texttt{RS}, and 1 as \texttt{SYST}. Their CMD projection is shown later in Sect.~\ref{sec:variability}. For simplicity we group them into one category named `{\gaia} Variable Binaries', amounting to 775 targets (0.4\% of the sample). These systems are identified in Table~\ref{tab:table_catalog} via the column `Flag {\gaia} Variable Binary'.
\subsection{The $9^{\text{th}}$ catalog of spectroscopic binary orbits} 
\label{subsec:characterization_binaries_sb9}

We complement the characterization by searching for spectroscopic binaries reported in the literature. For this, we use the $9^{\text{th}}$ catalog of spectroscopic binary orbits (SB9\footnote{\url{https://sb9.astro.ulb.ac.be/}}) by \citet{pourbaix04}, which compiles up-to-date information for such targets and currently lists over 4,000 systems. In particular, we crossmatch with the latest SB9 version\footnote{SB9 catalog version = 22/04/2024}, following the curation procedure from \citet{beck24}, which accounts for multiple entries in the catalog and filters them accordingly (see their Appendix A.1). We find 49 targets in the SB9 catalog (0.02\% of the {\kepler} sample), with no triple systems being found. The CMD projection of this subset is shown as the green diamonds in the bottom-left panel of Figure~\ref{fig:characterization_binaries}, and illustrates a clear preference for early-type and more luminous stars. These spectroscopic binaries are identified as such in Table~\ref{tab:table_catalog} via the column `Flag SB9'.
\subsection{NASA exoplanet archive} 
\label{subsec:characterization_binaries_NEA}

As many of the {\kepler} stars have been studied in the context of exoplanet searches, for completeness we also query the NASA Exoplanet Archive (NEA)\footnote{\url{https://exoplanetarchive.ipac.caltech.edu/}}. Besides its function as an exoplanet database, NEA compiles relevant information for host star characterizations, including from ground-based spectroscopic follow-up. In particular, it includes a flag for binary or higher order systems in the \texttt{sy\_snum} value, i.e., the number of stars in the system. We find 1,979 of the {\kepler} stars in the NEA table\footnote{NEA confirmed exoplanets version = 01/06/2025.} of confirmed planets and their hosts. Of these, 71 targets (0.04\% of the {\kepler} sample) have values of \texttt{sy\_snum}$>1$ and are thus classified as multiple systems. More specifically, 66 targets have \texttt{sy\_snum}$=$2, 4 targets have \texttt{sy\_snum}$=$3 (KIC 4278221, KIC 6278762, KIC 9941662, and KIC 12069449), and 1 target has  \texttt{sy\_snum}$=$4 (KIC 4862625). The CMD projection of this subset is shown as the purple squares in the bottom-left panel of Figure~\ref{fig:characterization_binaries}, with almost all of the targets being located along the MS. These NEA multiple systems are identified as such in Table~\ref{tab:table_catalog} via the column `Flag NEA \texttt{sy\_snum}'. For completeness, we also include the \texttt{sy\_snum} values when available, which can be used to identify the confirmed exoplanet hosts (i.e., those targets with reported \texttt{sy\_snum} values).

While not explicitly included in Table~\ref{tab:table_catalog}, we also check for the presence of circumbinary planet hosts in the target sample (e.g., see \citealt{martin18}). We find 12 such systems, all classified as NEA binaries. These are: KIC 4862625, KIC 5095269, KIC 5473556, KIC 6504534, KIC 6762829, KIC 8572936, KIC 9472174, KIC 9632895, KIC 9837578, KIC 10020423, KIC 12351927, and KIC 12644769.
\subsection{{\hipparcos}-{\gaia} catalog of accelerations} 
\label{subsec:characterization_binaries_HGCA}

The {\gaia} astrometry can be combined with complementary databases, such as the {\hipparcos} mission \citep{esa97}. The comparison of precise proper motion measurements at different epochs allows the identification of accelerating stars due to wide companions (e.g., \citealt{brandt18, kervella22}). \citet{brandt21} published the {\hipparcos}-{\gaia} Catalog of Accelerations (HGCA), which presents a cross-calibration of the {\hipparcos} and {\gaia} EDR3 astrometry, and accounts for the different frames of reference. The HGCA quantifies the difference between the {\gaia} EDR3 and long-term proper motions using the $\chi^2$ parameter, where values of $\chi^2 \gtrsim 11.8$ correspond to targets with inconsistent constant proper motions at the $3\sigma$ level. The HGCA probes the bright limit of our sample, which has a median apparent $G \approx 14.6$ mag, as the {\hipparcos} catalog includes targets down to apparent $V \lesssim 12$ mag. In terms of orbital timescales, \citet{escorza23} used the HGCA on a sample of RV-confirmed binaries, and concluded that the $\chi^2 \gtrsim 11.8$ threshold is reliable in identifying binaries with periods $\gtrsim 10^3$ days.

We crossmatch our sample with the {\gaia} EDR3 HGCA, and find 275 entries in common. Of these, 97 targets (0.05\% of the sample) have values of $\chi^2 > 11.8$. We flag these as binary candidates, and show their CMD projection as the red hexagons in the bottom-left panel of Figure~\ref{fig:characterization_binaries}. The targets are heavily concentrated on the most luminous parts of the CMD, particularly the giant branch and the upper MS. These HGCA binary candidates are identified as such in Table~\ref{tab:table_catalog} via the column `Flag HGCA High $\chi^2$'.
\subsection{Washington double star catalog} 
\label{subsec:characterization_binaries_wds}

We supplement the binary search by crossmatching with the Washington Double Star (WDS) catalog \citep{mason01}. This catalog is a benchmark reference for multiple star systems, and currently lists over 150,000 binaries. We crossmatch with the latest WDS version\footnote{WDS catalog version = 02/12/2024.}, and find 2,829 {\kepler} targets (1.4\% of the sample). The CMD projection of this subset is shown in cyan in the bottom-right panel of Figure~\ref{fig:characterization_binaries}, which illustrates that it spans the entirety of the CMD. These WDS binaries are identified as such in Table~\ref{tab:table_catalog} via the column `Flag WDS'.

We note that the WDS binaries can be somewhat different from those of earlier sections, as in this case the components may have been resolved in the literature (i.e., visual binaries). This is the case for most of our crossmatch, which allows us to estimate some properties for them. The median angular separation of our WDS binaries is $\approx 3.2${\arcsec}, and using their {\gaia} distances we estimate a median (projected) physical separation of $\approx 2900$ AU. Their distribution of magnitude difference (in the sense of primary minus secondary) has 16th, 50th, and 84th percentiles of $\approx$ -0.9, -3.4, and -5.5 mag. To facilitate further analysis of these systems, in Table~\ref{tab:table_catalog} we also report the corresponding WDS names when appropriate.
\subsection{Union of binary categories} 
\label{subsec:characterization_binaries_union}

As some readers may want to identify all of the potential binary candidates in our catalog, regardless of their specific flag or detection method, Table~\ref{tab:table_catalog} also includes the `Flag Binary Union' column. This flag is the union of all the binary flags introduced in the previous subsections. This subset amounts to 31,334 targets (15.9\% of the sample), and it is heavily dominated by the RUWE binary candidates (see Table~\ref{tab:summary_table}). Note that, as introduced at the beginning of Sect.~\ref{sec:characterization_binaries}, this category is independent of the `Flag CMD' classification (and thus does not account for the \texttt{`Photometric Binary'} category from Sect.~\ref{sec:characterization_CMD}).

\begin{figure}
\centering
\includegraphics[width=8.0cm]{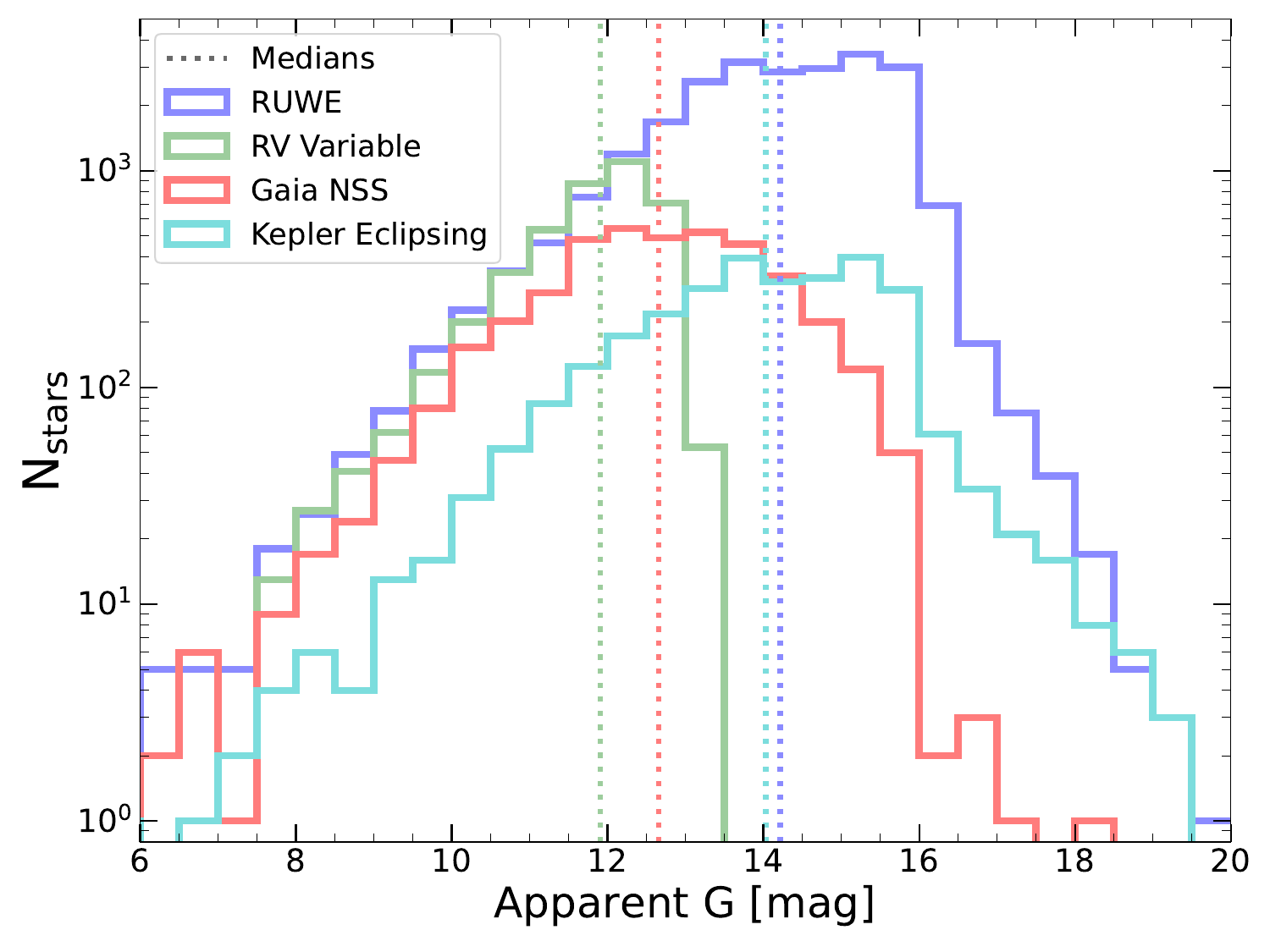}\\
\includegraphics[width=6.5cm]{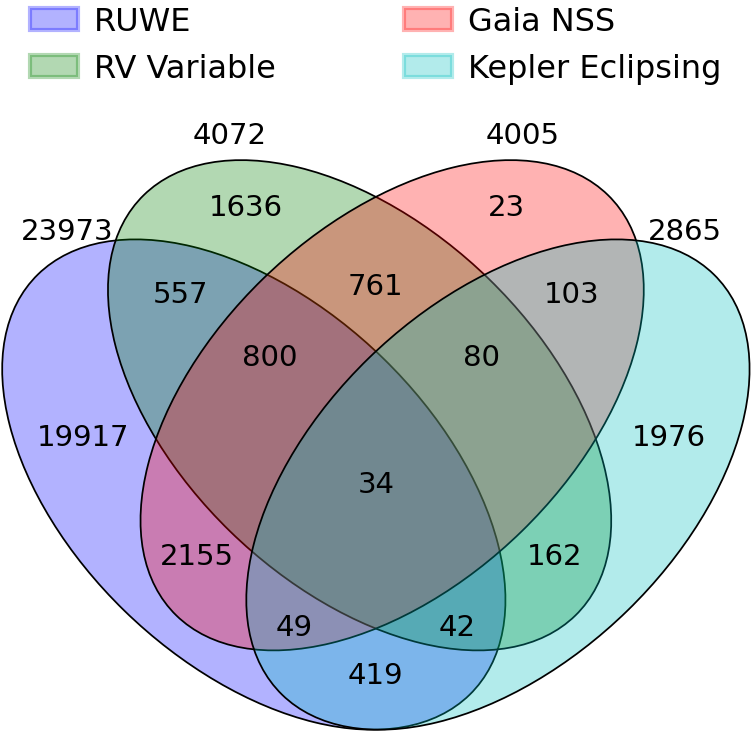}\\
\caption{Comparison of the four most numerous binary categories. Top: Distribution of apparent $G$-band magnitudes. The RV Variable and RUWE binary candidates are concentrated at the bright and faint limits of the {\kepler} sample, respectively. Bottom: Venn diagram. While some categories show a moderate degree of overlap (e.g., RV Variable and NSS), others do not (e.g., RUWE).}
\label{fig:characterization_union_binaries}
\end{figure}

To compare the different binary categories among each other, Figure~\ref{fig:characterization_union_binaries} shows the distributions of their apparent magnitudes, as well as a Venn diagram\footnote{Created with the \texttt{venny4py} tool.}. For practical reasons, we only include the four most numerous categories, namely RUWE (purple), RV Variable (green), NSS (red), and {\kepler} EB (cyan). The magnitude distribution illustrates that the selection of the RV Variable binaries is limited to apparent $G \lesssim 13.5$ mag (inherited from the {\gaia} DR3 RV availability), while the RUWE binaries span the entire magnitude range. The vertical dotted lines indicate the median of each distribution and correspond to 11.9 mag for the RV Variable, 12.7 mag for the NSS, 14.0 mag for the {\kepler} EB, and 14.2 mag for the RUWE binary candidates. For reference, the median value of the full {\kepler} sample is 14.6 mag (see Figure~\ref{fig:characterization_Gmag_distance_RUWE_MHgspspec}). 

Regarding the Venn diagram, for instance by comparing the NSS and RV Variable binaries, we find that they share $\sim$ 40\% of their samples (1,675 out of the 4,072 RV Variable and 4,005 NSS targets). More generally, an important finding of this comparison is that a significant fraction of the targets flagged by a given category are often not flagged by the rest. For instance, 81\% of the RUWE binary candidates are not found in other categories. Although the WDS binaries are not explicitly shown in Figure~\ref{fig:characterization_union_binaries} to avoid an excessively complicated diagram, these constitute the fifth most numerous category and 65\% of them are not found in the other binary samples. Interestingly, considering the criteria shown in Figure~\ref{fig:characterization_union_binaries}, 34 targets are classified as binaries by all four categories simultaneously. These findings highlight the power of integrating different binary criteria into one unique catalog.
\section{Astrometric differences between {\gaia} data releases and CMD implications}
\label{sec:differenceDR2vsDR3}

The {\gaia} mission has provided unprecedented astrometry for over a billion stars, becoming of paramount importance in stellar selections. Given its massive use across the literature, assessing the changes stars may have experienced from one data release to the next is highly relevant, as they could translate into differences in the derived stellar parameters. Global changes of the DR3 relative to the DR2, in terms of astrometric completeness and validation, have already been reported by \citet{fabricius21}. In this section, we focus on the main star-by-star astrometric differences for the DR3 versus DR2 \citep{lindegren18,lindegren21a}. Specifically for the {\kepler} targets, we study how these translate into changes to the CMD locations and classifications. For simplicity, we ignore magnitude and extinction effects, as the photometric systems between DR3 and DR2 are similar, albeit not identical \citep{riello21,maizapellaniz24}.

For every {\kepler} star, we look for potential DR2 crossmatches using the \texttt{gaiadr3.dr2\_neighbourhood} table in the {\gaia} archive, finding that most DR3 targets have one DR2 counterpart, but a fraction have multiple. The distributions of angular separations and magnitude differences between DR3 and DR2 are heavily concentrated towards $\Delta \theta < 0.1${\arcsec}  and $|\Delta G| < 0.05$ mag. We adopt these limits as the tolerances for a reliable crossmatch, prioritizing the nearest target (in terms of $\Delta \theta$) in case of multiple counterparts. 

\begin{figure}
\centering
\hspace*{0.7cm}\includegraphics[width=8.5cm]{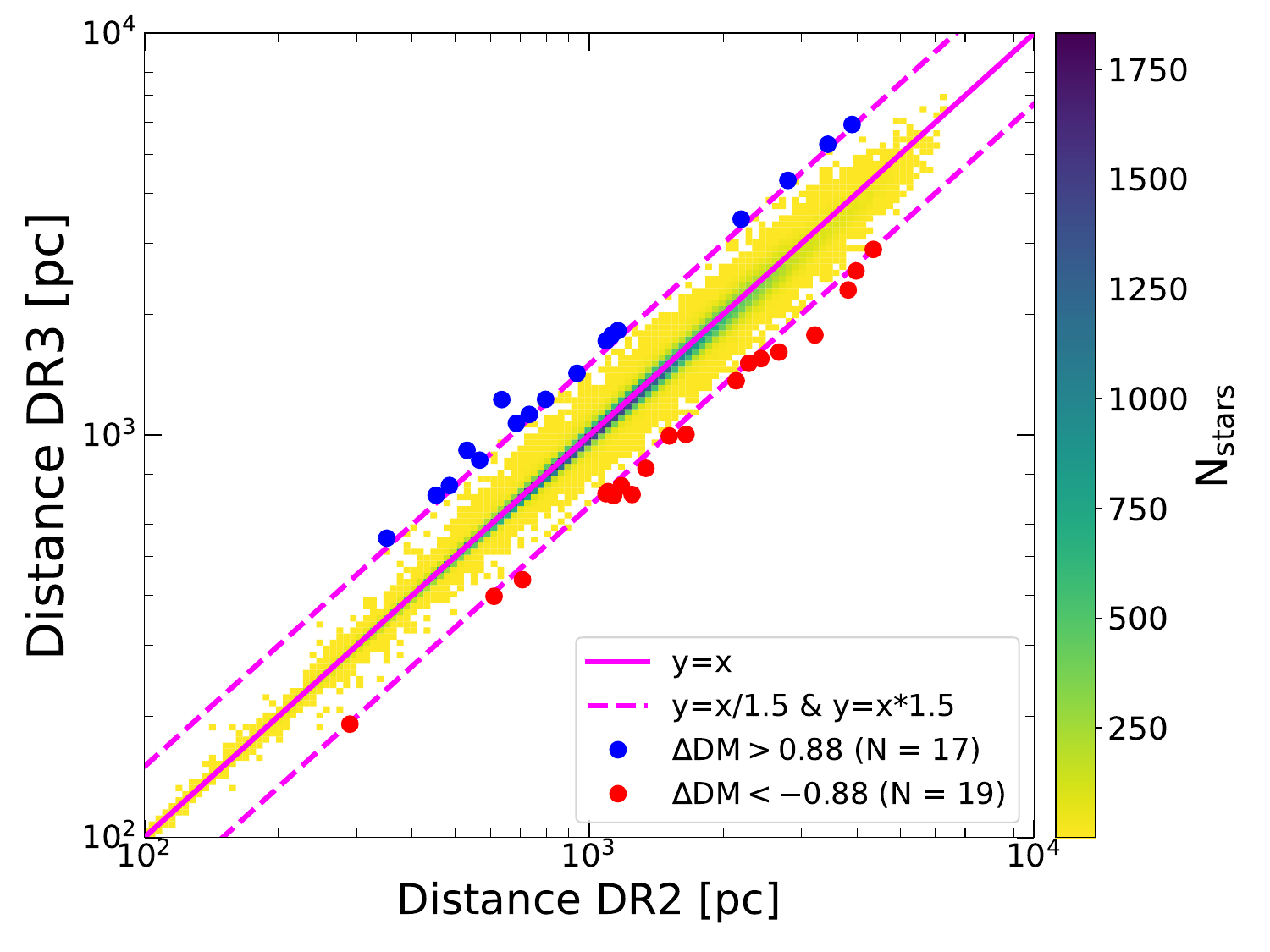}\\
\includegraphics[width=8.0cm]{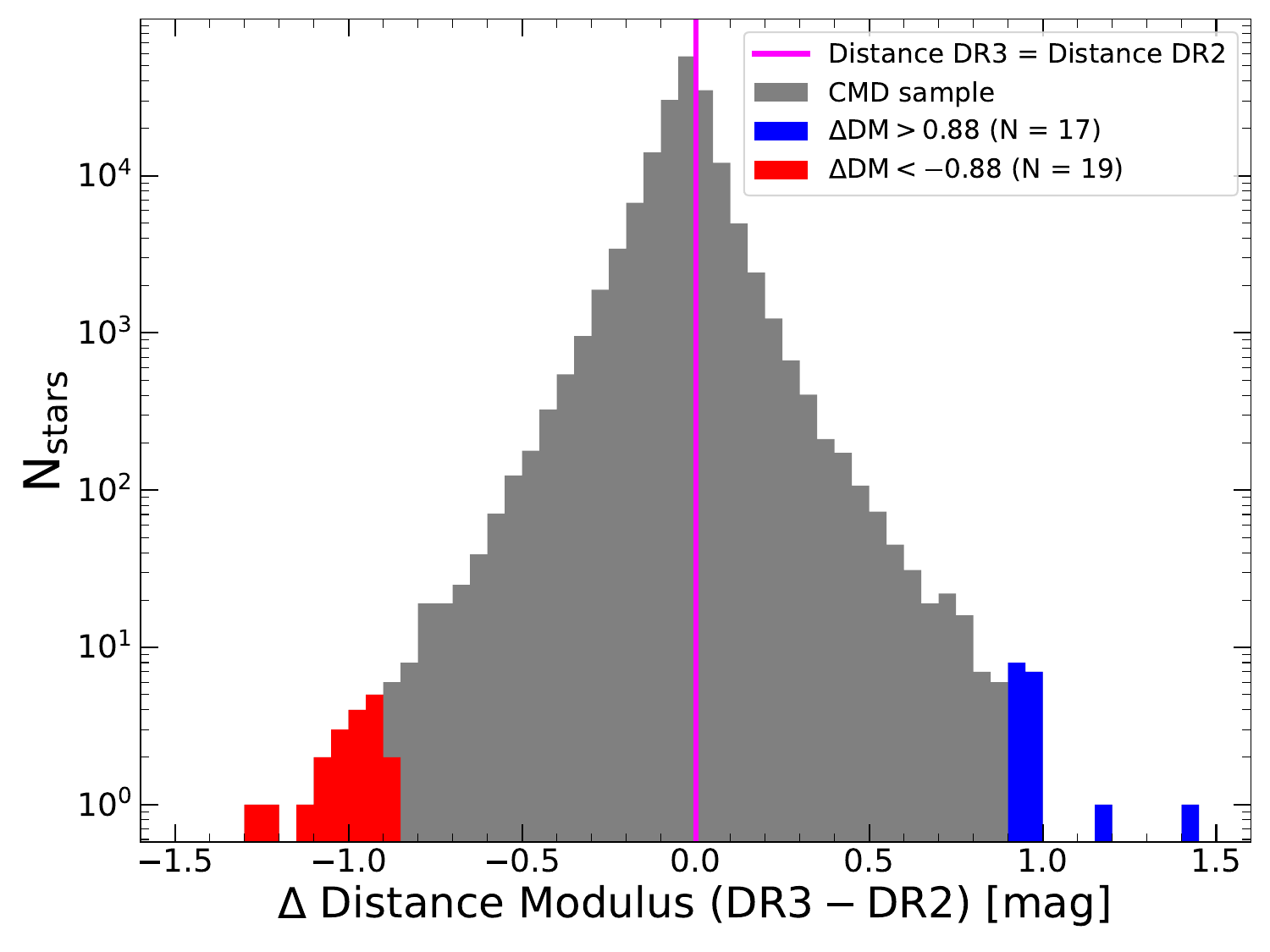}\\
\includegraphics[width=8.0cm]{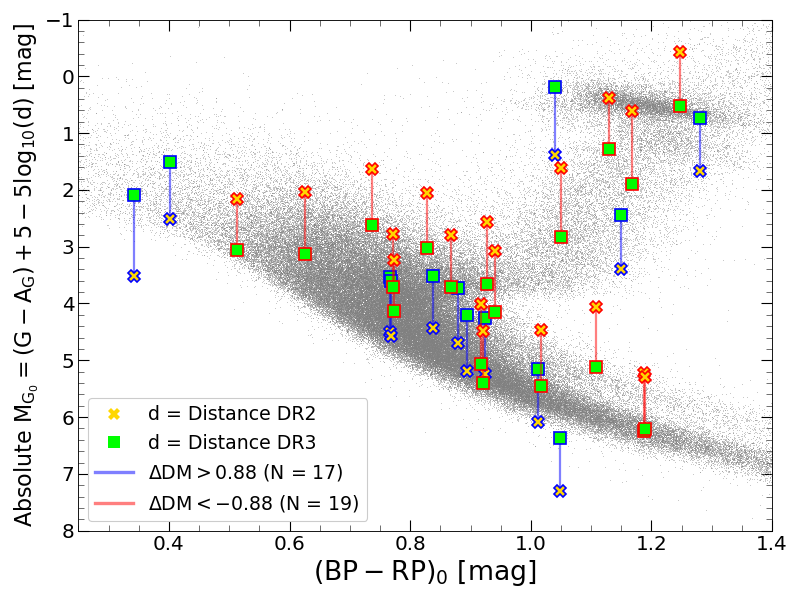}
\caption{Astrometric comparison between {\gaia} DR3 and DR2 from Sect.~\ref{sec:differenceDR2vsDR3}. Top: 2D histogram of the distance comparison. Most targets closely follow the 1:1 relation. Middle: Distribution of the distance modulus difference (in the sense of DR3 minus DR2). Stars are heavily centered around $\Delta \text{DM}= 0$, but outliers outside the 1.5:1 and 1:1.5 relations are selected for further inspection (blue and red samples). Bottom: CMD projection of the  $\Delta \text{DM}$ outliers, with their DR3 and DR2 positions shown in the green squares and yellow crosses, respectively. The astrometric differences can drive important CMD changes, thus modifying the derived stellar properties.}
\label{fig:comparisonDR3vsDR2}
\end{figure}

From these {\gaia} DR2 IDs, we query the tables \texttt{gaiadr2.gaia\_source} for parallax and \texttt{gaiadr2\_geometric\_distance} for distance information \citep{bailerjones18}. To ensure reliable astrometry, we limit the comparison to the stars in the CMD sample (Sect.~\ref{subsec:characterization_CMD_qualitycuts}), and also impose a DR2 parallax SNR of $\varpi/\sigma_{\varpi}>10$, corresponding to a subset of 173,612 stars (88.2\% of the {\kepler} sample). We illustrate the distance comparison between both data releases in the top panel of Figure~\ref{fig:comparisonDR3vsDR2}. The density map is heavily centered around the 1:1 line, albeit some scatter extends out to near the 1.5:1 and 1:1.5 ratios. While not explicitly shown, the analogous parallax comparison is equivalent (e.g., with $\approx$ 93\% and 98\% of the subset having parallaxes that agree at the $2\sigma$- and $3\sigma$-levels between DR3 and DR2, respectively).

We now examine the effects of the improved astrometry by comparing the changes in CMD positions due to the updated distance estimates. For this, we compute the difference in the distance modulus values from both data releases (in the sense of DR3 minus DR2), 
\begin{equation}
\Delta \text{DM}= (5\log_{10}(d_{\text{DR3}})-5) - (5\log_{10}(d_{\text{DR2}})-5),
\end{equation}
where $d_{\text{DR3}}$ and $d_{\text{DR2}}$ are the distances from DR3 and DR2 in pc. This new parameter allows us to easily re-compute absolute magnitudes using the DR2 distances, as $M_{G_{0},\text{DR2}} = M_{G_{0},\text{DR3}} + \Delta \text{DM}$. We report the $\Delta \text{DM}$ values in Table~\ref{tab:table_catalog}, and show its distribution in the middle panel of Figure~\ref{fig:comparisonDR3vsDR2}. The excellent overall distance agreement translates to a $\Delta \text{DM}$ distribution heavily centered at zero (vertical fuchsia line), with tails extending towards both positive and negative values.

At this point, we define two subsets of outliers from the $\Delta \text{DM}$ distribution. We select the stars that lie beyond the 1.5:1 and 1:1.5 distance relations (dashed lines in the top panel of Figure~\ref{fig:comparisonDR3vsDR2}), i.e., those with values of $\Delta \text{DM} >5\log_{10}(1.5/1) = 0.88$ mag and $\Delta \text{DM} < 5\log_{10}(1/1.5)= -0.88$ mag. These data sets contain 17 and 19 stars respectively, and we show them as the blue and red points in all panels of Figure~\ref{fig:comparisonDR3vsDR2}. In the bottom panel of Figure~\ref{fig:comparisonDR3vsDR2}, we show the DR3 CMD of the {\kepler} targets in the grey background, with the $\Delta \text{DM}$ outliers highlighted in colors. For these outliers, we show two sets of CMD positions, corresponding to the $M_{G_{0},\text{DR3}}$ (green squares) and $M_{G_{0},\text{DR2}}$ (yellow crosses) absolute magnitudes, and connect them with the corresponding blue or red lines.

For stars with large $\Delta \text{DM}$ values (either positive or negative), the CMD illustrates the importance of using the updated DR3 astrometry when deriving stellar properties. For instance, the blue outliers had under-luminous absolute magnitudes in DR2, and many of them go from lying underneath the MS and RC to being right on top of them with DR3. The analogous change in the opposite direction is true for the red outliers, with them going from over-luminous absolute magnitudes to (mostly) landing on higher-density regions. Upon further inspection of these 36 ($=17+19$) $\Delta \text{DM}$ outliers, we note that all except four are flagged as RUWE binaries in DR3, thus potentially explaining their astrometric discrepancies.

To generalize the analysis beyond these outliers, we replicate the CMD classification of Sect.~\ref{subsec:characterization_CMD_categories} to the 173,612 stars with $\Delta \text{DM}$ values. For each star, we infer the CMD category using its $M_{G_{0},\text{DR2}}$ absolute magnitude, and compare it with the original CMD classification obtained in Sect.~\ref{sec:characterization_CMD} using its $M_{G_{0},\text{DR3}}$ value. We find that 6,348 stars change CMD category between {\gaia} DR3 and DR2, and these are identified as such by the `Flag ${\Delta \text{DM}}$' column in Table~\ref{tab:table_catalog}. As may be expected, this sample is dominated by stars with low $P_{\text{CMD}}$ values, i.e., stars with unreliable CMD classifications (due to being close to the borders of the CMD regions; see Sect.~\ref{subsubsec:characterization_CMD_validation_montecarlo}). Restricting this to targets with higher probabilities, we find 971 stars with $P_{\text{CMD}}\geq 0.70$, and 171 stars with $P_{\text{CMD}}\geq 0.90$, that change CMD categories between {\gaia} DR2 and DR3.

Naturally, these changes hold important consequences for the properties of these stars, and that of their potential planets. For instance, among the stars with changing CMD categories, we find 52 exoplanet hosts according to NEA\footnote{These can be found by requiring `Flag ${\Delta \text{DM}}=$ \texttt{TRUE}' and measured values for `NEA \texttt{sy\_snum}'. Of the 52 targets, 46 have $P_{\text{CMD}}< 0.70$, and 6 have $P_{\text{CMD}}\geq 0.70$.}. Thus, readers are encouraged to check for potential CMD changes in their target samples using the `$\Delta \text{DM}$' and `Flag ${\Delta \text{DM}}$' columns in Table~\ref{tab:table_catalog}. All of the above highlights the importance of our catalog regarding astrometric differences between {\gaia} DR3 and DR2.
\section{Complementing {\kepler} seismic results with {\gaia}}
\label{sec:misclassifiedstars}

\begin{figure}
\centering
\includegraphics[width=8.0cm]{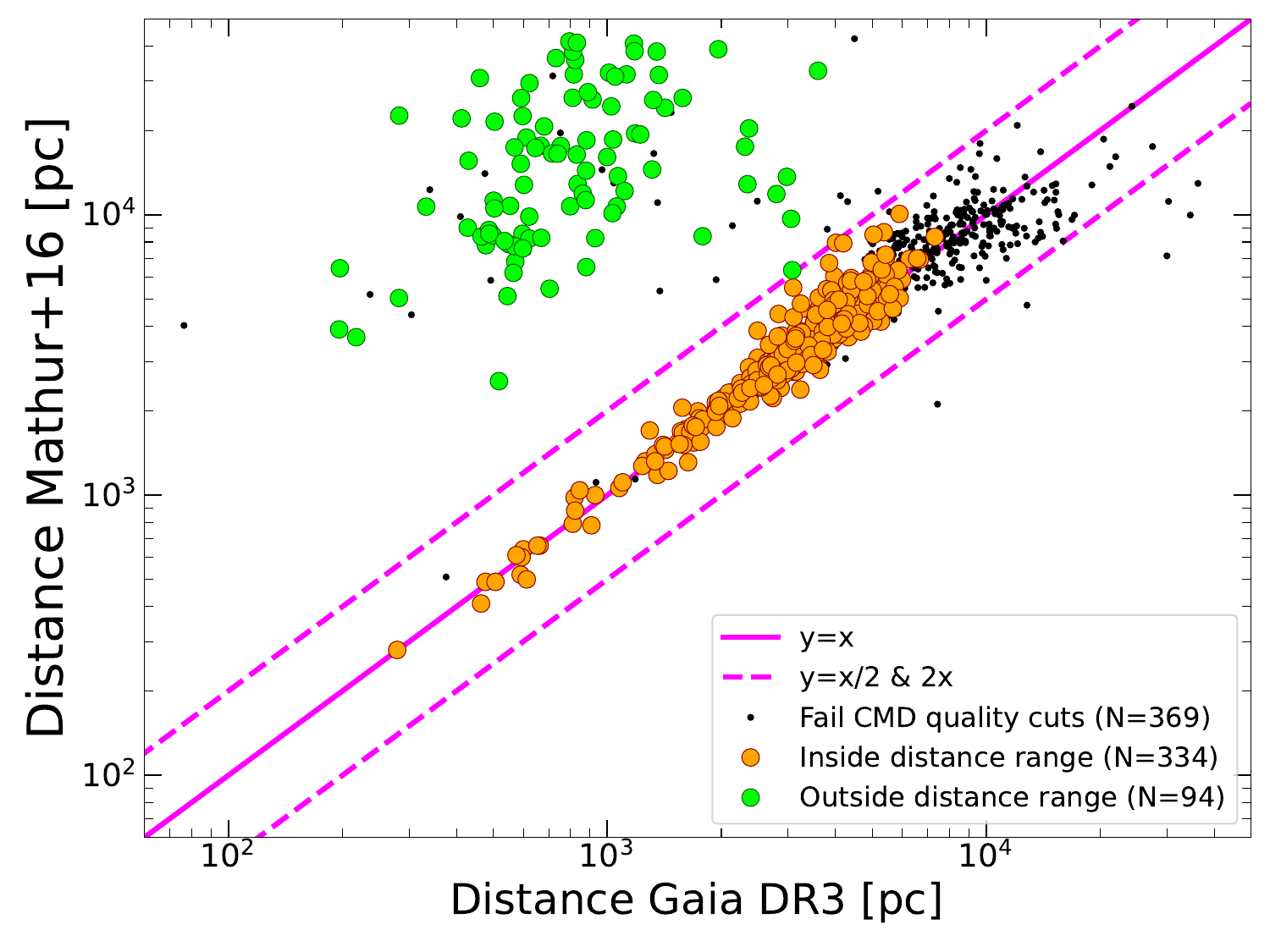}\\
\includegraphics[width=8.0cm]{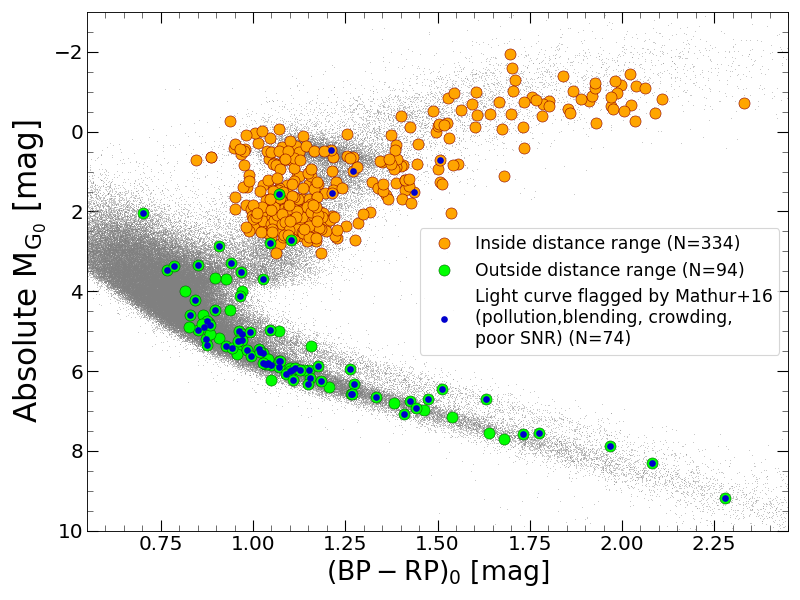}\\
\includegraphics[width=8.0cm]{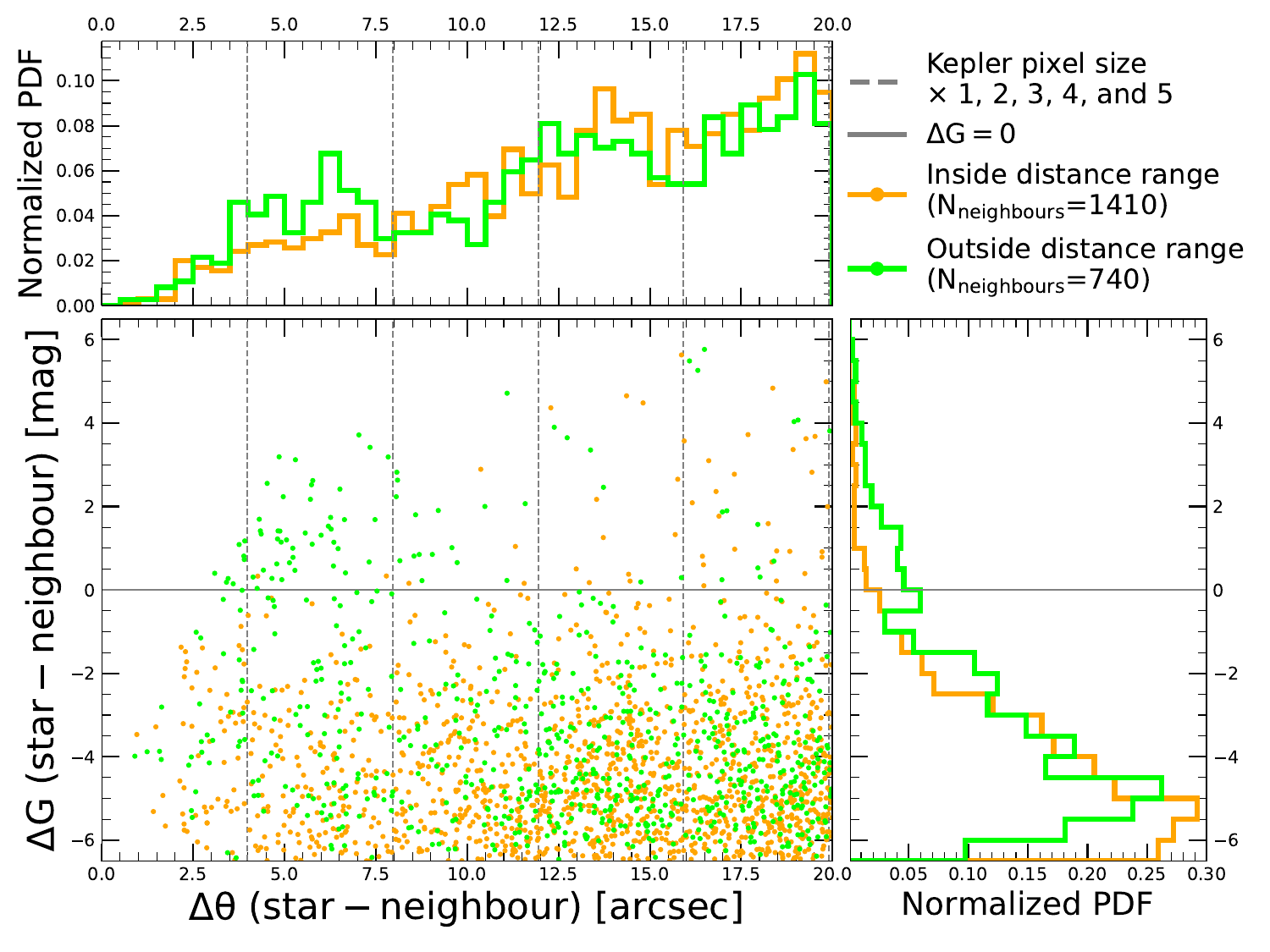}
\caption{{\gaia} DR3 analysis of the misclassified stars identified by \citet{mathur16} from Sect.~\ref{sec:misclassifiedstars}. Top: Distance comparison. The sample is split between targets inside (orange) and outside (green) the 2:1 and 1:2 lines. Middle: {\gaia} CMD projection. The targets with large distance discrepancies (green) appear predominantly as MS stars in {\gaia}, and most of them have crowding or blending flags (overplotted blue) in \citet{mathur16}. Bottom: $\Delta G$ vs. $\Delta \theta$ diagram (and marginalized distributions) of the stars' neighborhoods. The targets outside the distance range have more bright nearby neighbors than the targets inside the distance range, and thus the former are more prone to having contaminated {\kepler} light curves than the latter.}
\label{fig:misclassifiedstars}
\end{figure}

\citet{mathur16} identified over 800 targets with parameters consistent with dwarf stars according to the {\kepler} Star Properties Catalog (\citealt{huber14}; see also \citealt{brown11}), but where asteroseismic investigations of their light curves revealed oscillations corresponding to giants. They combined these asteroseismic analyses with broadband photometry and derived distances for the targets, finding most of them at several kpc. Furthermore, they reported a series of flags that identify targets with potential contamination in their light curves, namely pollution, blending, crowding, and poor SNRs. In this subsection, we revisit these targets in light of the {\gaia} DR3 data.

Starting with the 824 stars from \citet{mathur16}, we find that 27 are either missing from our catalog or lack {\gaia} distances, and are thus excluded from the following analysis. For the remaining 797 stars, we show the distance comparison between both catalogs in the top panel of Figure~\ref{fig:misclassifiedstars}. Of these, 369 fail the CMD quality cuts from Sect.~\ref{subsec:characterization_CMD_qualitycuts}, and we show them as black points. These stars are predominantly found $\gtrsim 5$ kpc away and are thus too distant to have well-measured parallaxes. 

We now focus on the subset of 428 stars that pass the CMD quality cuts. We split the sample between those that fall inside or outside the 2:1 and 1:2 lines in the distance comparison, hereafter referred to as the targets inside or outside the distance range, respectively. These are shown as the orange and green points in Figure~\ref{fig:misclassifiedstars}, with their sample sizes being 334 and 94 stars. We note that, for the targets outside the distance range, the \citet{mathur16} distances are always larger than the {\gaia} distances. 

In the middle panel of Figure~\ref{fig:misclassifiedstars}, we show the {\gaia} CMD projection of these targets. They occupy noticeably distinct CMD regions, with the subsets inside and outside the distance range appearing predominantly as giants and MS stars, respectively. We interpret this as follows. Those inside the distance range (orange) correspond to targets where the {\kepler} light curve is dominated by a giant star that matches the {\gaia} DR3 ID in our catalog (i.e., both {\kepler} light curve and {\gaia} DR3 ID recognize the same star, albeit the KIC ID in \citet{huber14} is that of a photometric dwarf). On the other hand, those outside the distance range (green) correspond to targets where the {\gaia} DR3 ID in our catalog is that of a dwarf, but where the {\kepler} light curve is likely contaminated by a nearby giant star, thus showing asteroseismic oscillations in \citet{mathur16} (i.e., both the KIC ID in \citet{huber14} and the {\gaia} DR3 ID recognize the same dwarf star, but the {\kepler} light curve is dominated by a bright giant in the neighborhood).

We further inspect these targets by including the light curve contamination flags from \citet{mathur16}. We show these as the smaller inset blue circles in the CMD of Figure~\ref{fig:misclassifiedstars}. Of the 74 flagged stars, 69 of them (93.2\%) correspond to targets outside the distance range, while 5 of them (6.8\%) correspond to targets inside the distance range. Consequently, there is a clear trend for most of the targets with contaminated light curves to coincide with those with large distance discrepancies, providing further evidence for our interpretation. For completeness, we also examine the fraction of targets classified as binaries, and find them to be comparable across both samples ($\approx 7\%$). Thus, binarity, in terms of the unresolved and resolved categories presented in Sect.~\ref{sec:characterization_binaries}, does not seem to be a significant factor in the above.

As a final piece of the analysis, we investigate the neighborhoods of these targets by querying the \texttt{gaiadr3.gaia\_source} table for other stars in their vicinity. We limit the search to neighbors within $\Delta \theta \leq 20${\arcsec} and $|\Delta G| \leq 6.5 $ mag. The $\Delta G$ vs. $\Delta \theta$ projection of this search is shown in the bottom panel of Figure~\ref{fig:misclassifiedstars}, with the marginalized distribution of each axis on the subpanels, and the number of neighbors (integrating over each subset) displayed in the top-right corner legend. For reference, we indicate the 3.98{\arcsec} size of the {\kepler} pixels and its multiples (up to $\times$5) as the vertical dashed lines. The results illustrate that the targets outside the distance range (green) have preferentially more bright ($\Delta G \approx -0.5$ to +3 mag), close ($\Delta \theta \approx$ 3{\arcsec} to 8{\arcsec}) neighbors than the targets inside the distance range (orange). These overdensities are indeed located within the size of a few {\kepler} pixels, thus explaining the aforementioned contamination in the light curves.

While other tools are better suited for the analysis of flux contamination in light curves \citep{schonhutstasik23}, our work illustrates the mismatches that such an effect can have on the CMD location of stars. All in all, the above highlights the power of the {\gaia} data in the study and interpretation of asteroseismic targets.
\section{{\gaia} variability classification for the {\kepler} stars}
\label{sec:variability}

In this section, we extend the analysis of the {\kepler} stars by leveraging complementary {\gaia} data products. In particular, we examine the photometric variability information of our targets as reported in {\gaia} DR3 \citep{eyer23}. Such variability analyses are allowed by the multiple-epoch observations of the mission, which span 34 months as of DR3 (e.g., \citealt{clementi23,distefano23,lebzelter23,marton23,mowlavi23,ripepi23})

We begin by studying the \texttt{phot\_variable\_flag} in the \texttt{gaiadr3.gaia\_source} table. This parameter identifies targets with variability in the photometric data (for the subset of sources that {\gaia} had processed by the release of DR3). We find 13,803 (7.0\%) of the {\kepler} targets classified as \texttt{VARIABLE}, and 182,959 (93.0\%) classified as \texttt{NOT AVAILABLE}. We further investigate the subset of \texttt{VARIABLE} targets by querying their variability classification as reported by the \texttt{best\_class\_name} in the \texttt{gaiadr3.vari\_classifier\_result} table \citep{rimoldini23}. This parameter identifies the best classification for variable sources obtained via statistical and machine learning methods trained on benchmark targets \citep{gavras23}. Of the 13,803 \texttt{VARIABLE} stars, we find 13,661 with variability classifications. We report both of these {\gaia} DR3 parameters in the `Flag Photometric Variability' and `Photometric Variability Class' columns in Table~\ref{tab:table_catalog}.

We find 19 different classes of variable sources in the sample. Their acronyms, taken verbatim from {\gaia} DR3, are as follows: \texttt{SOLAR\_LIKE} (solar-like star), \texttt{DSCT|GDOR|SXPHE} ($\delta$\,Scuti, $\gamma$\,Doradus, or SX\,Phoenicis star), \texttt{ECL} (eclipsing binary), \texttt{RS} (RS\,Canum Venaticorum variable), \texttt{LPV} (long-period variable), \texttt{S} (short-timescale object), \texttt{RR} (RR\,Lyrae star), \texttt{ACV|CP|MCP|ROAM|ROAP|SXARI} ($\alpha^2$\,Canum Venaticorum, or (magnetic) chemically peculiar, or rapidly oscillating Am-\,and\,Ap-type, or SX\,Arietis star), \texttt{AGN} (active galactic nucleus), \texttt{SPB} (slowly pulsating B-type variable), \texttt{CV} (cataclysmic variable), \texttt{EP} (star with exoplanet transits), \texttt{SDB} (subdwarf B), \texttt{BE|GCAS|SDOR|WR} (B-type emission line, $\gamma$Cassiopeiae, S\,Doradus, or Wolf-Rayet), \texttt{CEP} (Cepheid), \texttt{YSO} (young stellar object), \texttt{WD} (variable white dwarf), \texttt{SYST} (symbiotic variable star), and \texttt{BCEP} ($\beta$\,Cephei). Readers are referred to \citet{rimoldini23} for more detailed definitions.

\begin{figure}
\centering
\includegraphics[width=8.5cm]{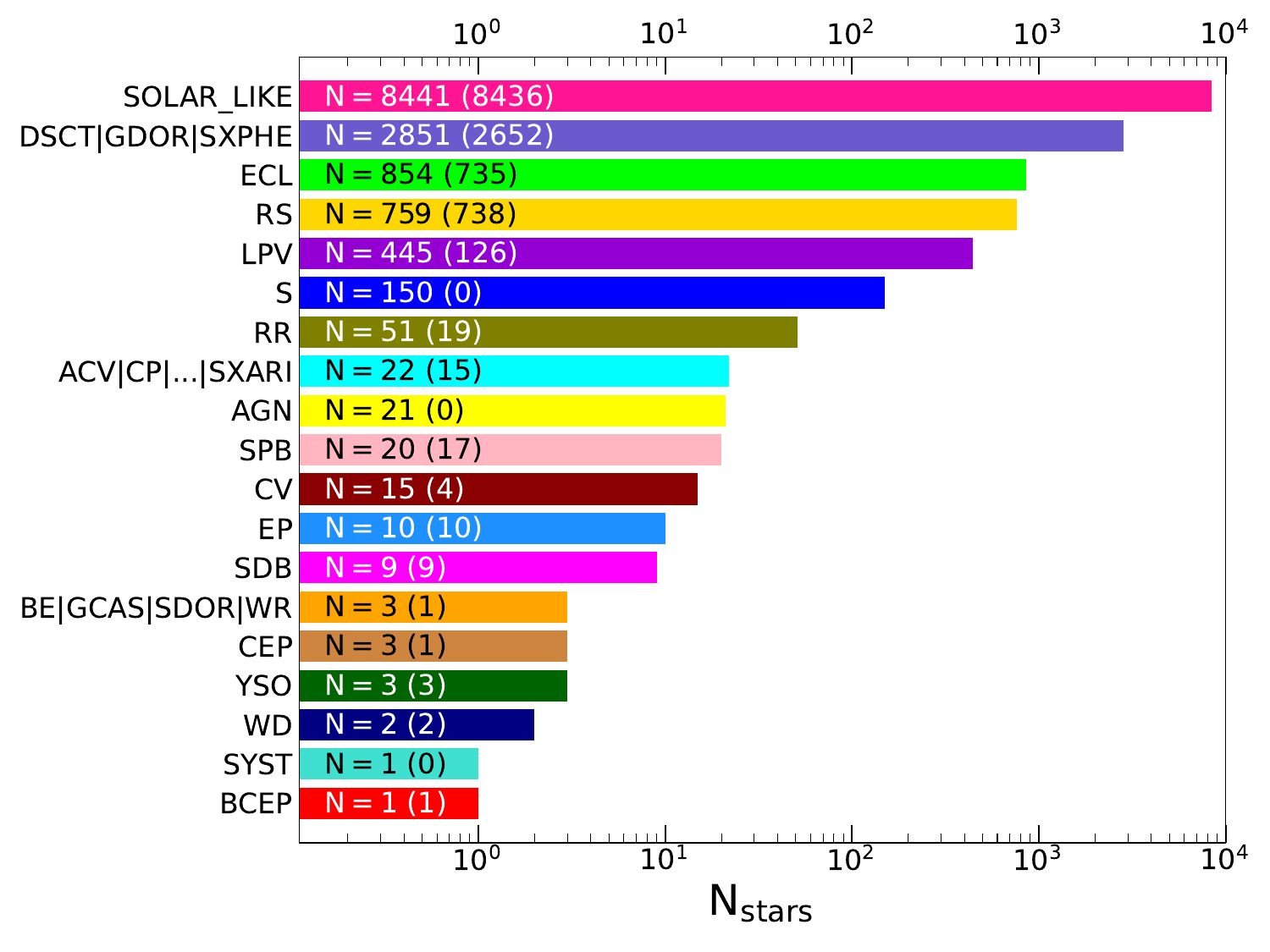}\\
\includegraphics[width=8.5cm]{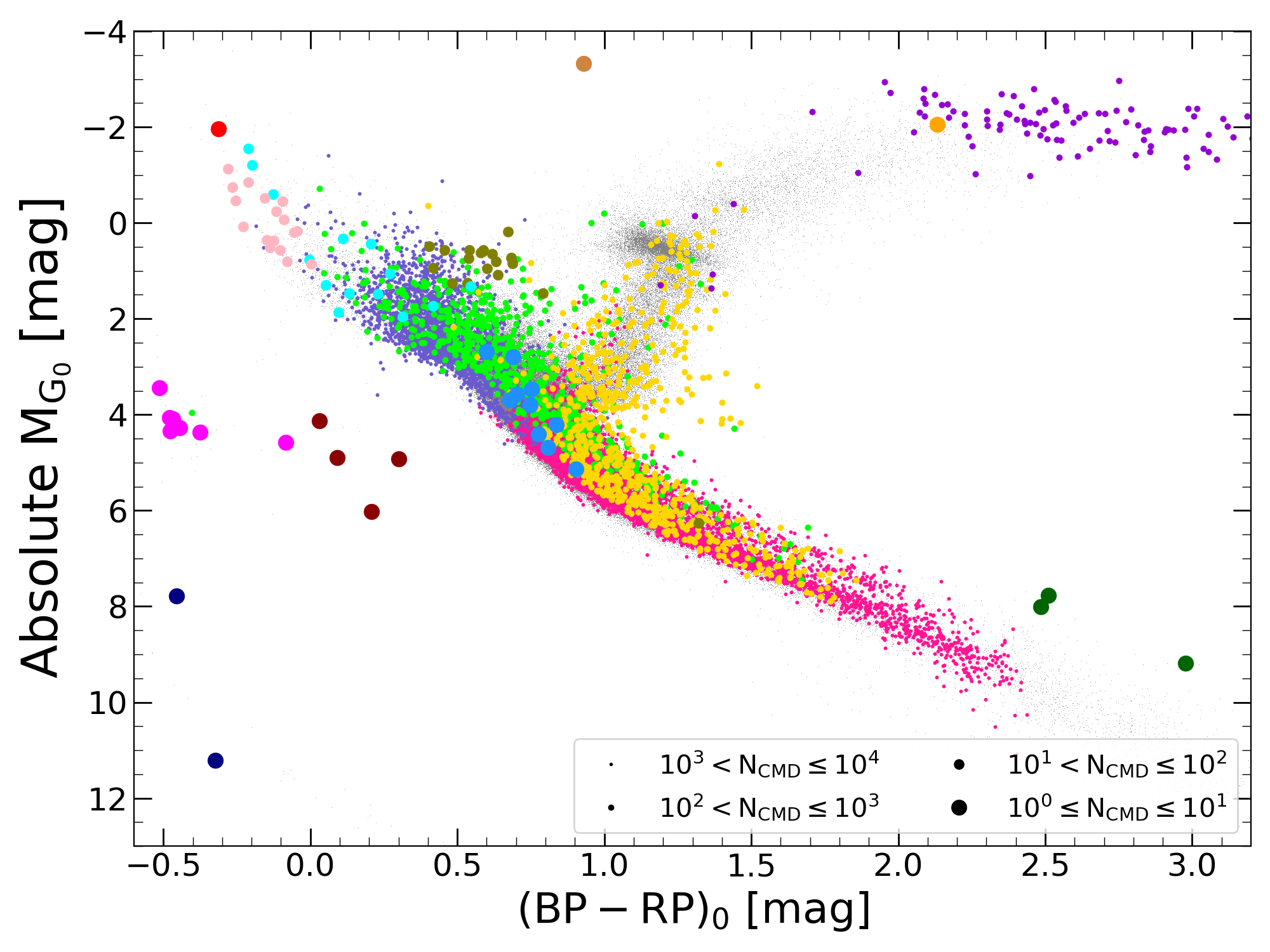}\\
\includegraphics[width=8.5cm]{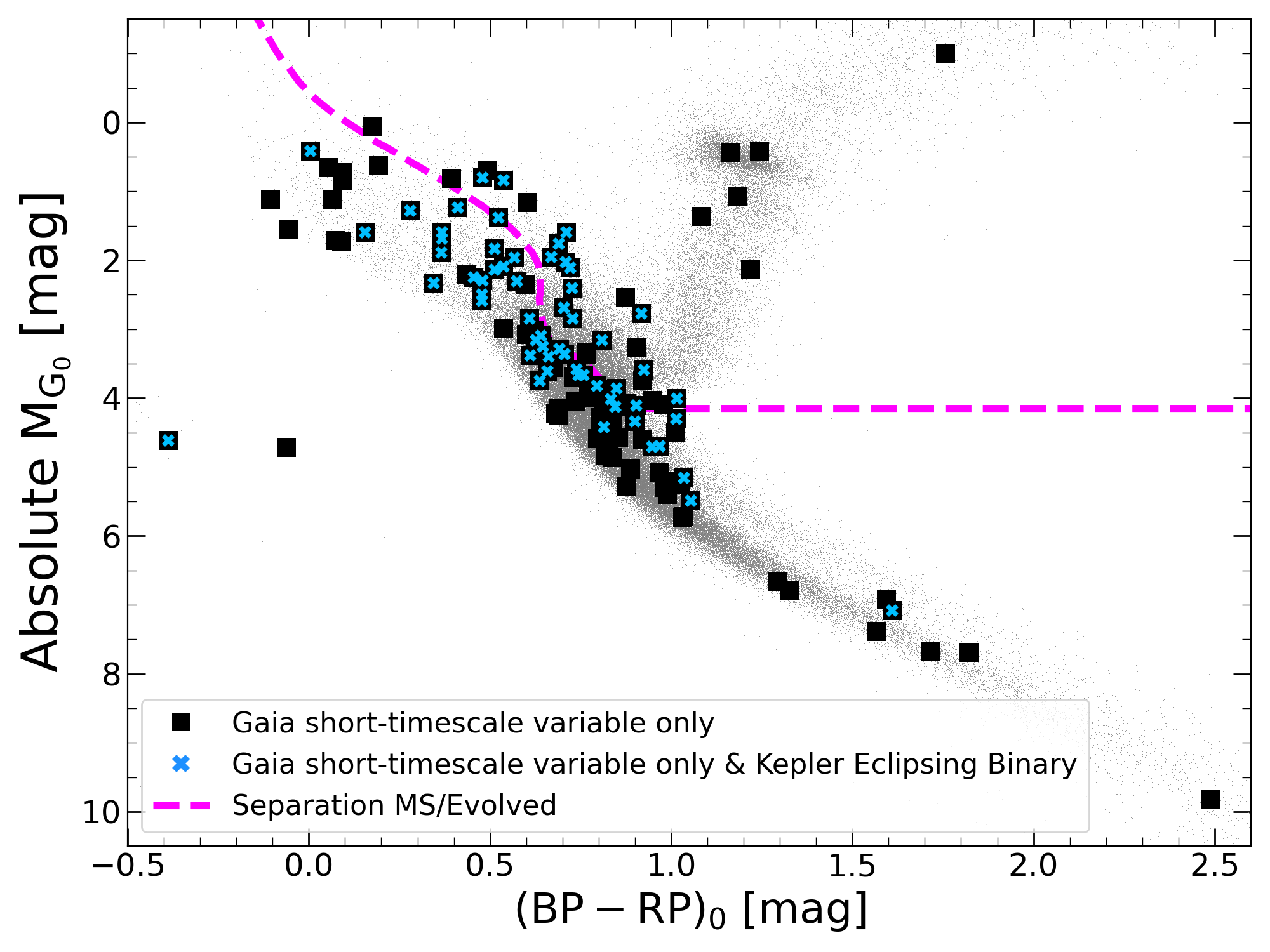}\\
\caption{Photometric variability analysis of the {\kepler} targets from Sect.~\ref{sec:variability}. Top: bar chart of the 19 classes of variable sources found in {\gaia} DR3. The text inside the bars lists the number of targets in each class (with the size of the CMD subset shown in parenthesis). Middle: CMD projection of the variable sources with classifications. Targets are color-coded following the bars in the top panel, and the symbol size increases for smaller samples. The full CMD sample (grey) is shown for reference in the background. The interpretations of the CMD locations are greatly benefited by the variability classifications in {\gaia} DR3. Bottom: CMD projection of the `{\gaia} short-timescale variable only' targets (black squares). A significant fraction of them are found to be {\kepler} eclipsing binaries (cyan crosses).}
\label{fig:variability}
\end{figure}

These 19 variability classes are also listed as the labels\footnote{Note that the \texttt{ACV|CP|MCP|ROAM|ROAP|SXARI} class is abbreviated to \texttt{ACV|CP|...|SXARI} in Figure~\ref{fig:variability} for plotting purposes (following Table~1 in \citealt{rimoldini23}).} of the bar chart in the top panel of Figure~\ref{fig:variability}. The categories are sorted in decreasing order from the top, based on the number of stars recovered in each, as indicated by the text inside each bar (followed by the number of those that pass the CMD quality cuts in parenthesis). The CMD projection of the sources flagged as variables (e.g., see \citealt{gaia19}) is shown in the middle panel of Figure~\ref{fig:variability}. The color-coding is identical to that of their variability class in the top panel, and the marker sizes are larger for smaller samples. Note that, for several categories (e.g., \texttt{S}, \texttt{CV}), only a small fraction (or even none) of the targets are shown on the CMD. Since our CMD analysis of Sect.~\ref{sec:characterization_CMD} has been based on the mean {\gaia} DR3 magnitudes (see Sect.~\ref{subsec:characterization_CMD_caveats}), photometrically variable stars are naturally more prone to failing the CMD quality cuts. Additionally, the CMD placement can only be done for targets that pass the parallax quality cut (see Sect.~\ref{subsec:characterization_CMD_qualitycuts}), which as expected prevents extragalactic sources from appearing on the CMD (e.g., \texttt{AGN}).

The predominant variability class is \texttt{SOLAR\_LIKE} (8,441 targets), followed by \texttt{DSCT|GDOR|SXPHE} (2,851 targets). Moreover, the CMD of Figure~\ref{fig:variability} shows that most of the variable targets are located along the MS. A fraction of them extend to evolved regions, such as the \texttt{LPV}, \texttt{BE|GCAS|SDOR|WR}, and \texttt{CEP} classes. Additionally, many stars that may have appeared as CMD outliers, can now be explained by their photometric variability classification. Although only including a few targets, the \texttt{SDB}, \texttt{YSO}, and \texttt{CV} categories occupy CMD regions that are expected given their variability class. We also note that the targets classified as \texttt{ECL}, \texttt{CV}, \texttt{RS}, and \texttt{SYST} are the same binary candidates previously discussed in Sect.~\ref{subsec:characterization_binaries_eclipsing_gaia} and Sect.~\ref{subsec:characterization_binaries_othervariable_gaia}. Other interesting targets are those classified as \texttt{EP} and \texttt{AGN}, which we discuss in more detail in Appendix \ref{sec:app_variability_EP_AGN}.

We now turn to the 142 ($= 13,803 - 13,661$) targets flagged as \texttt{VARIABLE} in {\gaia} DR3 that lack variability classifications in \texttt{gaiadr3.vari\_classifier\_result} (and are thus absent from the top and middle panels of Figure~\ref{fig:variability}). For completeness we query the \texttt{gaiadr3.vari\_summary} table and find that all of them are also in the \texttt{gaiadr3.vari\_short\_timescale} table, which in turn reports variable candidates that show short-timescale phenomena ($<$ 0.5 to 1 day; see also \citealt{roelens17,roelens18}). These 142 targets do not appear in any of the other {\gaia} variability tables, and we thus refer to them as `{\gaia} short-timescale variable only'. We show their CMD positions as the black squares in the bottom panel of Figure~\ref{fig:variability} and inspect their CMD categories, finding 80 targets classified as \texttt{`Dwarf'}, 21 as \texttt{`Subgiant'}, 14 as \texttt{`Photometric Binary'}, 8 as \texttt{`Overlap Dwarf/Subgiant'}, 6 as \texttt{`Giant Branch'}, 4 as \texttt{`Uncertain~MS'}, and 9 that fail the CMD quality cuts. In terms of their median distance and apparent $G$-band magnitudes, these targets are more distant ($\Delta d \approx 500$ pc) and slightly brighter ($\Delta G \approx 0.5$ mag) than their counterparts with \texttt{gaiadr3.vari\_classifier\_result} classifications. Interestingly, 61 of the 142 targets are flagged as eclipsing binaries from the {\kepler} light curves (Sect.~\ref{subsec:characterization_binaries_eclipsing_kepler}). We show these as the overplotted cyan crosses in the bottom panel of Figure~\ref{fig:variability}.

Additional explorations of the over 13,000 photometrically variable targets are left as future work. We encourage the community to further investigate these variable sources by taking advantage of the different observing baselines of the {\kepler} and {\gaia} missions (e.g., \citealt{hey24,zhou24}).
\section{Conclusions} 
\label{sec:conclusions}

The stars observed by the {\kepler} mission remain a highly relevant sample for studies of stellar, exoplanetary, and Galactic astrophysics. In this paper, we characterize the $\sim$ 200,000 stars observed by {\kepler} in light of the recent {\gaia} DR3. These data provide unprecedented constraints regarding astrometry, photometry, spectroscopy, and stellar parameters.

We use the {\gaia} DR3 distances and magnitudes, in combination with a series of quality cuts that ensure reliable parameters, and place the {\kepler} stars on the absolute and de-reddened CMD (Figure~\ref{fig:characterization_CMD}). Regarding extinction, we consider a number of maps from the literature. We choose the map that maximizes homogeneity for our sample, and use the comparisons among the different maps to estimate extinction uncertainties. The photometry is de-reddened using the latest {\gaia} DR3 coefficients, and the photometric, astrometric, and extinction uncertainties are propagated to the reported CMD positions (Table~\ref{tab:summary_table}). From this, by comparing with stellar models and empirical analyses, we separate the sample into several categories based on CMD location. These include stars classified as: \texttt{`Dwarf'}, \texttt{`Photometric Binary'}, \texttt{`Subgiant'}, `\texttt{Giant Branch}', as well as other categories that represent more rare CMD regions (\texttt{`Overlap Dwarf/Subgiant'} and \texttt{`Uncertain~MS'}). We validate these categories by performing internal tests that consider the associated CMD uncertainties and the effects of metallicity, as well as external tests with asteroseismic catalogs and results from the {\gaia} DR3 FLAME module. The caveats involved in the classification are discussed.

We also report several categories of candidate binary systems (Figure~\ref{fig:characterization_binaries}). These are identified through a number of detection methods applied to the {\gaia} data, as well as by crossmatching with published {\gaia} DR3 binary tables and complementary literature data sets. The binary categories are RUWE, RV Variables, {\gaia} NSS, {\kepler} eclipsing, {\gaia} eclipsing, {\gaia} variables, spectroscopic from the SB9 catalog, multiples from the NASA Exoplanet Archive, {\hipparcos}-{\gaia} catalog of accelerations, and binaries from the Washington Double Star catalog. We report their respective sample sizes (Table~\ref{tab:summary_table}) and discuss the different degrees of overlap among them. For convenience to readers, we also report the `Binary Union' category, which identifies all the targets flagged by any of the binary detection methods.

Additionally, we leverage the {\gaia} data to carry out three further characterizations of the {\kepler} targets. First, we assess how the astrometric differences between {\gaia} DR3 and DR2 are translated into changes on CMD location (Figure~\ref{fig:comparisonDR3vsDR2}). Second, we revisit a sample of misclassified asteroseismic targets. We find that disagreements in distance estimates between {\gaia} versus photometric plus asteroseismic constraints are typically caused by mismatches due to bright, nearby neighbors that contaminate the {\kepler} light curves (Figure~\ref{fig:misclassifiedstars}). Third, we examine the photometric variability flags and classes reported in {\gaia} DR3. The {\kepler} targets span 19 different variability categories, and their classifications provide valuable complements to the CMD analysis (Figure~\ref{fig:variability}). Moreover, we find that $\sim$ 40\% of the `{\gaia} short-timescale variable only' targets correspond to eclipsing binaries as seen by {\kepler}.

The applications of this catalog are multiple. Our work will aid in the investigation of stellar properties and how they are influenced by stellar architectures and evolutionary stages. For instance, the catalog allows the straightforward selection of single stars or binary candidates, or of stars on the MS or in post-MS phases. Examples include the criteria to exclude binaries (`Flag Binary Union $=$ \texttt{FALSE}'), focus on MS stars (`Flag CMD $=$ \texttt{Dwarf'}), select evolved stars (`Flag CMD $=$ \texttt{Subgiant}' and/or `\texttt{Giant Branch}'), or target specific binary categories (e.g., `Flag RV Variable $=$ \texttt{TRUE}'). The same benefits extend to the selection of exoplanet hosts (e.g., \citealt{mcquillan13,berger23}), with a preliminary version of the catalog being used by \citet{garcia23} to identify single MS planet hosts. The parameters we report allow for valuable characterizations and comparisons, though it is important to consider the assumptions and caveats involved. We make our catalog publicly available (Table~\ref{tab:table_catalog}) and encourage the community to take full advantage of it as a useful resource in the continued exploration of the stars observed by the {\kepler} mission.
\section*{Data availability} 

Tables \ref{tab:table_catalog} and \ref{tab:coordinates_CMDregions} are available in electronic form on Zenodo (\url{https://zenodo.org/records/14774100}) and at the CDS via anonymous ftp to cdsarc.u-strasbg.fr (130.79.128.5) or via http://cdsweb.u-strasbg.fr/cgi-bin/qcat?J/A+A/.
\begin{acknowledgements}

We thank the referee for their insightful, detailed, and knowledgeable feedback, which has significantly improved the quality and applicability of our work.

We thank Phil R. Van-Lane for stimulating and valuable discussions.

We thank the organizers of the 11$^{\text{th}}$ Iberian Meeting on Asteroseismology for fostering valuable interactions that contributed to this work.

We thank Jamie Tayar for helpful exchanges about stellar evolution.

We thank the Stars Group at the Instituto de Astrofísica e Ciências do Espaço at Porto for their constructive feedback on our work.

We thank James Davenport for making the \texttt{KIC-to-TIC} catalog publicly available.\\ 

\indent DGR acknowledges support from the Juan de la Cierva program under contract JDC2022-049054-I. DGR and SM acknowledge support from the Spanish Ministry of Science and Innovation (MICINN) with the grant No. PID2019-107187GB-I00. SM acknowledges support by the Spanish Ministry of Science and Innovation with the Ramon y Cajal fellowship number RYC-2015-17697, the grant no. PID2019-107061GB-C66, and through AEI under the Severo Ochoa Centres of Excellence Programme 2020--2023 (CEX2019-000920-S). RAG acknowledges the support from the PLATO Centre National D'{\'{E}}tudes Spatiales grant. ARGS acknowledges the support by FCT through national funds and by FEDER through COMPETE2020 by these grants: UIDB/04434/2020, UIDP/04434/2020 \& 2022.03993.PTDC. ARGS is supported by FCT through the work contract No. 2020.02480.CEECIND/CP1631/CT0001. PGB acknowledges support by the Spanish Ministry of Science and Innovation with the \textit{Ram{\'o}n\,y\,Cajal} fellowship number RYC-2021-033137-I and the number MRR4032204. DHG received the support of a fellowship from ``la Caixa” Foundation (ID 100010434), with fellowship code LCF/BQ/DI23/11990068. PGB and DHG acknowledge support of the NAWI Graz consortium. JM acknowledges support from the Instituto de Astrofísica de Canarias (IAC) received through the IAC early-career visitor program and the support from the Erasmus+ programme of the European Union under grant number 2020-1-CZ01-KA203-078200. The research of JM was supported by the Czech Science Foundation (GACR) project no. 24-10608O. AE received the support of a fellowship from ”la Caixa” Foundation (ID 100010434) with fellowship code is LCF/BQ/PI23/11970031. DGR, RAG, MHP, PGB, DHG, and JM acknowledge support from the Spanish Ministry of Science and Innovation with the grant no. PID2023-146453NB-100 (\textit{PLAtoSOnG}).

This paper includes data collected by the {\kepler} mission and obtained from the MAST data archive at the Space Telescope Science Institute (STScI). Funding for the {\kepler} mission is provided by the NASA Science Mission Directorate. STScI is operated by the Association of Universities for Research in Astronomy, Inc., under NASA contract NAS 5–26555.

This work has made use of data from the European Space Agency (ESA) mission {\gaia} (\url{https://www.cosmos.esa.int/gaia}), processed by the {\gaia} Data Processing and Analysis Consortium (DPAC,
\url{https://www.cosmos.esa.int/web/gaia/dpac/consortium}). Funding for the DPAC has been provided by national institutions, in particular the institutions participating in the {\gaia} Multilateral Agreement.

This work made use of the \texttt{Gaia-Kepler.fun} crossmatch database created by Megan Bedell. 

This research has made use of the NASA Exoplanet Archive, which is operated by the California Institute of Technology, under contract with the National Aeronautics and Space Administration under the Exoplanet Exploration Program.

This work made extensive use of TOPCAT \citep{taylor05}.

This research has used data, tools or materials developed as part of the EXPLORE project that has received funding from the European Union’s Horizon 2020 research and innovation programme under grant agreement No 101004214.

This research has made use of the Washington Double Star Catalog maintained at the U.S. Naval Observatory.

\end{acknowledgements}

\bibliographystyle{./aa} 
\bibliography{./keplerfieldcharacterization}

\begin{appendix}
\section{{\kepler}-{\gaia} DR3 catalog} 
\label{sec:app_catalog_table}

We report our catalog in Table~\ref{tab:table_catalog}. We include the main {\gaia} DR3 astrometric, photometric, and spectroscopic data, as well as the several flags and parameters defined throughout the paper.

\begin{table}
\caption{Column descriptions of the {\kepler}-{\gaia} DR3 catalog.}
\label{tab:table_catalog}
\centering                         
\scriptsize 
\begin{tabular}{ll}
\hline
\hline
Column & Description\\
\hline 
KIC & KIC ID\\
{\gaia} DR3 & {\gaia} DR3  source ID\\
TIC & TIC ID\\
2MASS & 2MASS ID\\
$\varpi_{\text{Corr}}$ & Parallax, zero point corrected in mas\\
$\sigma_{\varpi,\text{Corr}}$ & Parallax error, inflation factor corrected in mas\\
Distance & Distance in pc\\
$\sigma_{\text{Distance}}$ & Distance error in pc\\
$T_{\text{eff } \texttt{gspspec}}$ & \texttt{gspspec} $T_{\text{eff}}$ in K\\
$T_{\text{eff } \texttt{gspspec}_\text{16th}}$ & 16th pct of \texttt{gspspec} $T_{\text{eff}}$ in K\\
$T_{\text{eff } \texttt{gspspec}_\text{84th}}$ & 84th pct of \texttt{gspspec} $T_{\text{eff}}$ in K\\
$\log(g)_{\texttt{gspspec},\text{calibrated}}$ & Calibrated \texttt{gspspec} $\log(g)$ in dex\\
$\log(g)_{\texttt{gspspec},\text{16th}}$ & 16th pct of \texttt{gspspec} $\log(g)$ in dex\\
$\log(g)_{\texttt{gspspec},\text{84th}}$ & 84th pct of \texttt{gspspec} $\log(g)$ in dex\\
$[\text{M/H}]_{\texttt{gspspec},\text{calibrated}}$ & Calibrated \texttt{gspspec} metallicity in dex\\
$[\text{M/H}]_{\texttt{gspspec},\text{16th}}$ & 16th pct of \texttt{gspspec} metallicity in dex\\
$[\text{M/H}]_{\texttt{gspspec},\text{84th}}$ & 84th pct of \texttt{gspspec} metallicity in dex\\
Quality \texttt{gspspec} & \texttt{gspspec} quality sample(s) if any\\
Flag Quality Cuts & Identifies failed quality criteria if any\\
$G$ & Apparent $G$-band magnitude (raw)\\
$BP-RP$ & $BP-RP$ color (raw)\\
$M_{G_0}$ & Absolute de-reddened $G$-band magnitude\\
$(BP-RP)_0$ & De-reddened $(BP-RP)$ color\\
$\sigma_{M_{G_0}}$ & Error in $M_{G_0}$\\
$\sigma_{(BP-RP)_0}$ & Error in $(BP-RP)_0$\\
$A_{0}$ & Monochromatic extinction\\
$\sigma_{A_{0}}$ & Error in monochromatic extinction\\
Flag CMD & Identifies CMD category\\
$P_{\text{CMD}}$ & Probability of CMD category\\
$P_{\text{CMD,[M/H]}}$ & Probability of CMD category including $\sigma_{\text{[M/H]}}$\\
Flag Metal-Poor Tail & Identifies targets in the metal-poor distribution tail\\
Flag Metal-Rich Tail & Identifies targets in the metal-rich distribution tail\\
$\Delta M_{G_0}$ & Magnitude difference to reference isochrone\\
RUWE & Renormalised Unit Weight Error\\
Flag RUWE & Identifies targets with RUWE$\geq1.4$\\
Flag RV Variable & Identifies targets classified as RV Variable\\
Flag NSS & Identifies targets in NSS tables\\
NSS Type & Binary type from NSS\\
NSS Tables & Union of acronyms of NSS tables\\
Flag EB {\kepler} & Identifies {\kepler} eclipsing binaries\\
Flag EB {\gaia} & Identifies {\gaia} eclipsing binaries\\
Flag {\gaia} Var. Binary & Identifies {\gaia} variable binaries\\
Flag SB9 & Identifies targets in the SB9 catalog\\
Flag NEA \texttt{sy\_snum} & Identifies multiple systems from NEA\\
NEA \texttt{sy\_snum} & \texttt{sy\_snum} value from NEA database\\
Flag HGCA High $\chi^2$ & Identifies targets with Hipparcos-{\gaia} $\chi^2>11.8$\\
Flag WDS & Identifies targets in the WDS catalog\\
WDS & Name in the WDS catalog\\
Flag Binary Union & Union of the eight non-CMD binary flags\\
$\Delta$DM & Distance modulus difference (DR3-DR2)\\
Flag ${\Delta \text{DM}}$ & Identifies changes in Flag CMD due to $\Delta$DM\\
Phot. Var. Flag & \texttt{phot\_variable\_flag} from {\gaia} DR3\\
Phot. Var. Class & \texttt{best\_class\_name} from {\gaia} DR3\\
\hline
\\
\end{tabular}
\tablefoot{Catalog of the 196,762 stars characterized in this work. (The full table is available in Section `Data availability'.)}
\end{table}

\section{\texttt{gspspec} parameters} 
\label{sec:app_calibration_quality_gspspec}

Our query for \texttt{gspspec} metallicities, surface gravities, and effective temperatures is described in Sect.~\ref{sec:data}. As reported in \citet{recioblanco23}, these {\gaia} DR3 parameters are deliberately uncalibrated, and thus some corrections are necessary to place them in a common scale with other spectroscopic surveys. Additionally, in order to select reliable parameters, a number of quality flags need to be considered. For our purposes, the above can become important when examining the metallicity distribution and Kiel diagram.

We follow the calibration recipe provided by \citet{recioblanco23} (see their Section 9.1.1). On the one hand, the temperatures do not need to be calibrated, as no significant offset was found when compared with the literature. On the other hand, the surface gravities and metallicities do need to be calibrated to account for underestimated gravities and $\log(g)$-dependent composition trends, respectively. We calibrate them using the proposed polynomials\footnote{For metallicity, we adopt the calibration with respect to the literature sample over the open cluster sample, as the latter is restricted to a more metal-rich regime.}, based on comparisons with high-quality literature data. With this, in Table \ref{tab:table_catalog} we report the raw temperatures, calibrated surface gravities, and calibrated metallicities (independently of the quality flags that we now analyze).

\begin{figure}
\centering
\includegraphics[width=8.0cm]{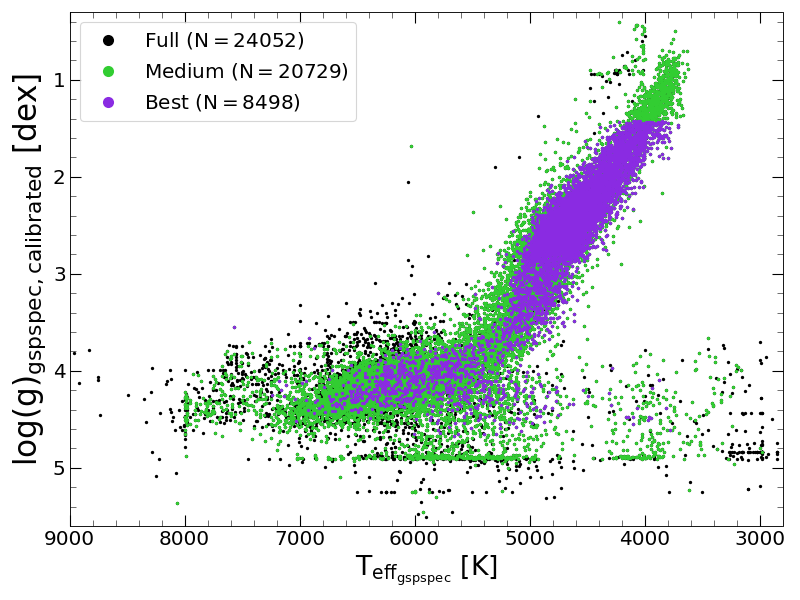}\\
\includegraphics[width=8.0cm]{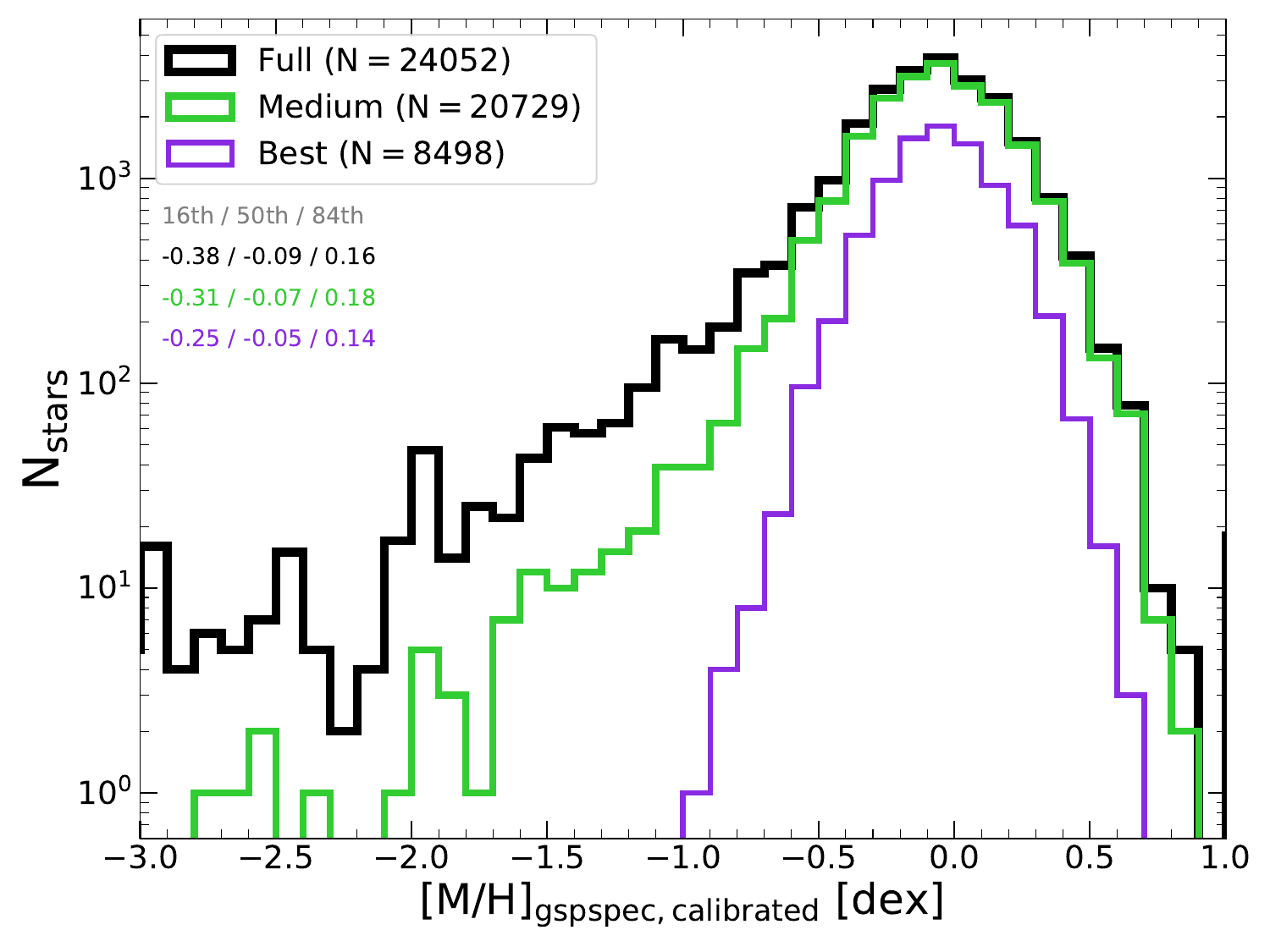}\\
\includegraphics[width=8.0cm]{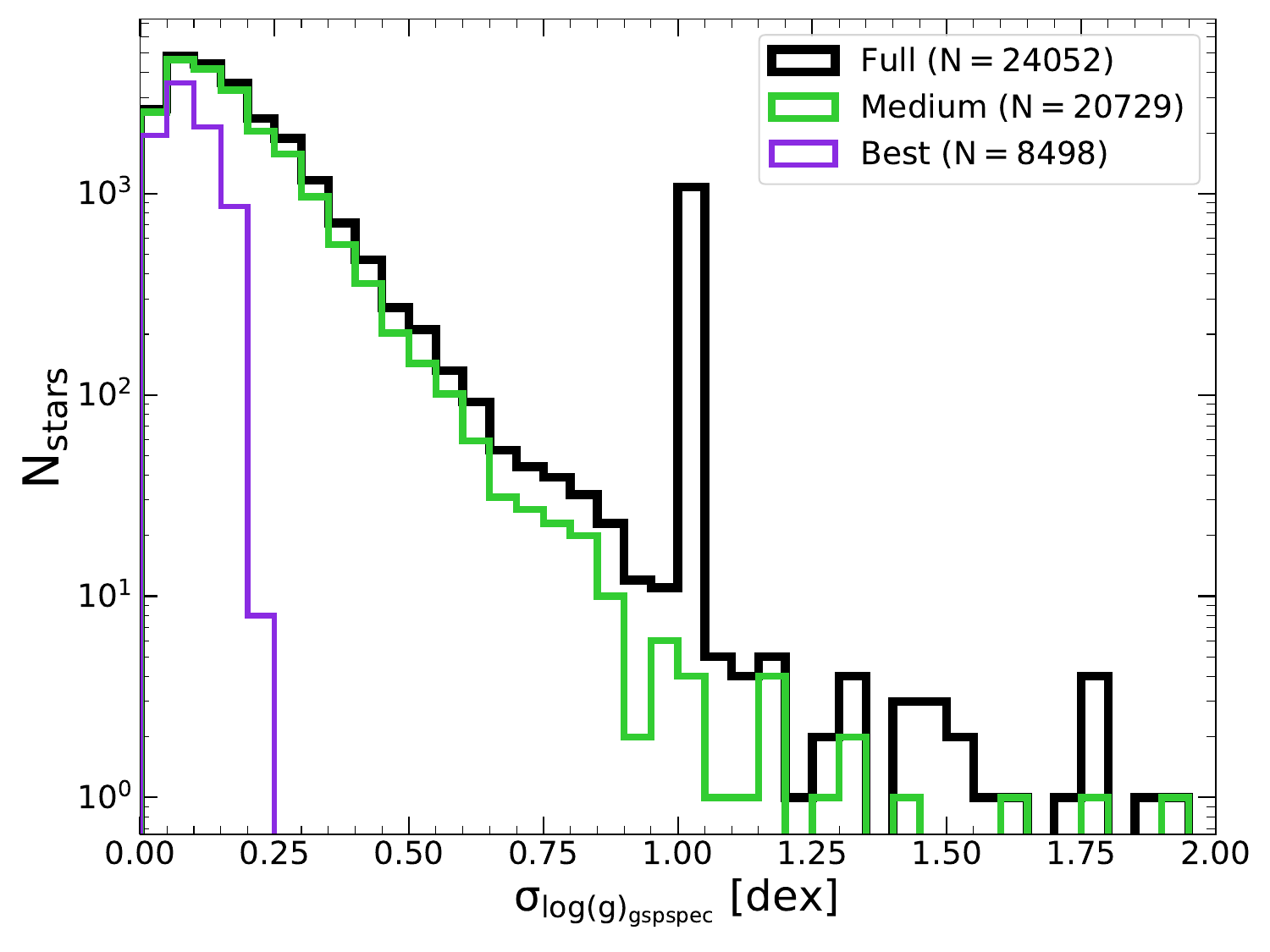}\\
\caption{Impact of quality flags on the \texttt{gspspec} parameters. The data sets shown are the full (black), medium-quality (green), and best-quality (purple) samples. Top: Kiel diagram. Artifacts are progressively removed for more stringent selections. Middle: Calibrated metallicity distribution. The 16th, 50th, and 84th percentiles of each sample are shown in the colored text. The metal-poor tail is more heavily affected by the flags. Bottom: Distribution of surface gravity uncertainties. Only targets with small uncertainties survive the most stringent quality cuts.}
\label{fig:gspspec_quality_samples}
\end{figure}

Regarding the quality flags, we test the impact of different criteria on the \texttt{flags\_gspspec} values following the selections from \citet{recioblanco23} (see their Section 9.1.1). From lowest to highest quality data, these are the full \texttt{gspspec} sample (24,052 stars or 12.2\% of the {\kepler} target list), the medium-quality sample (20,729 stars or 10.5\%), and the best-quality sample (8,498 stars or 4.3\%). We show the projections of these in the \texttt{gspspec} parameter space in Figure \ref{fig:gspspec_quality_samples}. Artifacts are clearly present in the full sample (e.g., horizontal line in the Kiel diagram at $\log(g) \approx$ 5 dex, or spike at $\sigma_{\log(g)} \approx 1$ dex). Some of these are removed from the medium-quality sample, and they are entirely removed from the best-quality sample. We find the metallicity distribution to only moderately change as a function of the quality criteria (note the logarithmic $y$-axis in the middle panel of Figure \ref{fig:gspspec_quality_samples}), with the metal-poor tail being progressively diminished with more restrictive quality cuts. Regarding the parameter uncertainties, the quality flags have a strong impact on their overall distribution (see also Appendix C in \citealt{recioblanco23}). For instance, targets with $\sigma_{\log(g)} > 0.2$ dex are present in the full and medium-quality samples but absent in the best-quality sample. Similar behaviors are seen for the distribution temperature and metallicity uncertainties (e.g., there are no stars with $\sigma_{T_{\text{eff}}}>$ 100 K or $\sigma_{[\text{M/H}]}>$ 0.1 dex in the best-quality sample.)

To facilitate the selection of {\kepler} stars belonging to the different quality samples, we report the `Quality \texttt{gspspec}' flag in Table \ref{tab:table_catalog}. This flag is a string composed of anywhere from 0 to 3 characters. The presence of a star in a given \texttt{gspspec} sample is indicated by one letter, namely: `f' for full, `m' for medium-quality, and `b' for best-quality. For instance, a star that is present in the full and medium-quality samples (but not in the best-quality sample) will have a value of `fm'. A star present in all the samples will have a value of `fmb'. The entry for a star without \texttt{gspspec} data will be empty. To maximize reliability, the metallicity distribution of Figure \ref{fig:characterization_Gmag_distance_RUWE_MHgspspec} is that of the best-quality sample. 

\section{Specifics of the quality cuts} 
\label{sec:app_specifics_quality_cuts}

The $G$/$BP$/$RP$ flux SNR parameters we use are reported in the \texttt{gaiadr3.gaia\_source} table as \texttt{phot\_{g/bp/rp}\_mean\_flux\_over\_error}. The values of the $BP$ and $RP$ flux excess factor (\texttt{phot\_bp\_rp\_excess\_factor}) are corrected by the color-dependent fit from Table 2 of \citet{riello21}. We only keep the targets with corrected excess factors within $\pm$3 times the typical scatter given their apparent $G$-band magnitudes (e.g., \citealt{mikkola23}). Finally, while the \texttt{visibility\_periods\_used} ($N_{\text{vpu}}$) criterion is not explicitly included in Sect.~\ref{subsec:characterization_CMD_qualitycuts} (e.g., \citealt{gaia18b}), stars with parallax values already incorporate a $N_{\text{vpu}} \geq 9$ cut \citep{lindegren21a}. The stars excluded by the quality cuts, and the reason for their exclusion, are identified as such in Table~\ref{tab:table_catalog} via the column `Flag Quality Cuts'.
\section{Comparing extinction maps} 
\label{sec:app_comparing_extinction_maps}

\begin{figure*}
\centering
\includegraphics[width=5.5cm]{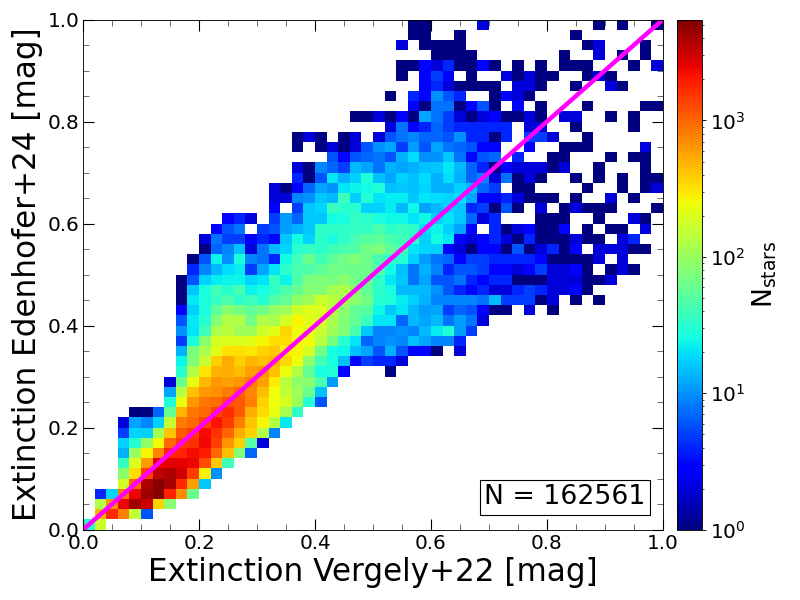}
\includegraphics[width=5.5cm]{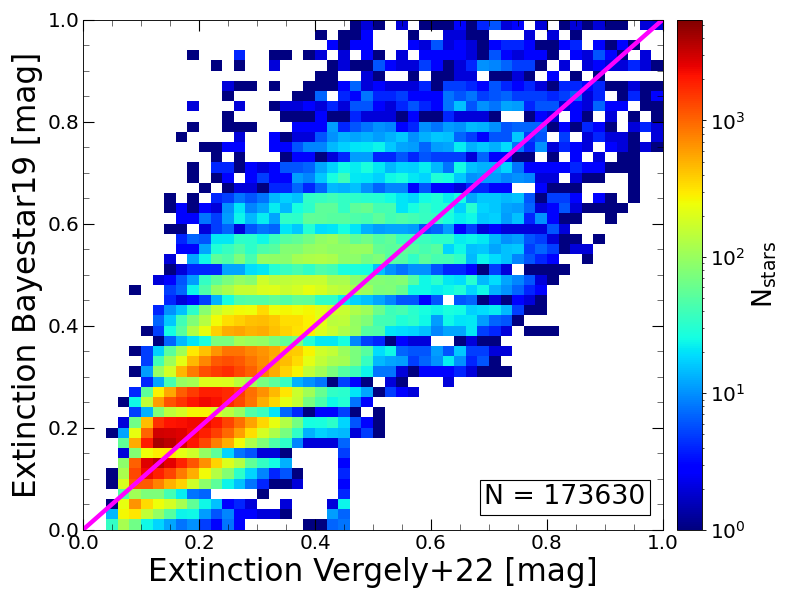}
\includegraphics[width=5.5cm]{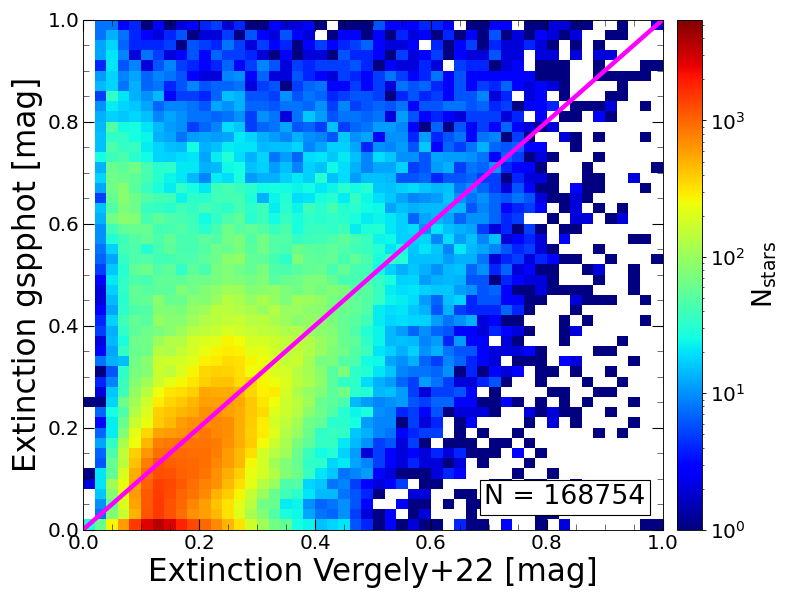}\\
\includegraphics[width=5.5cm]{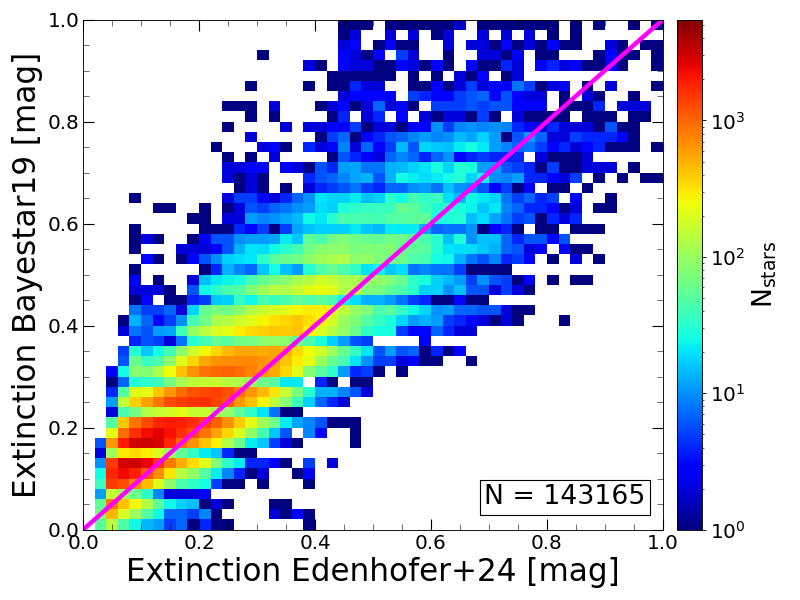}
\includegraphics[width=5.5cm]{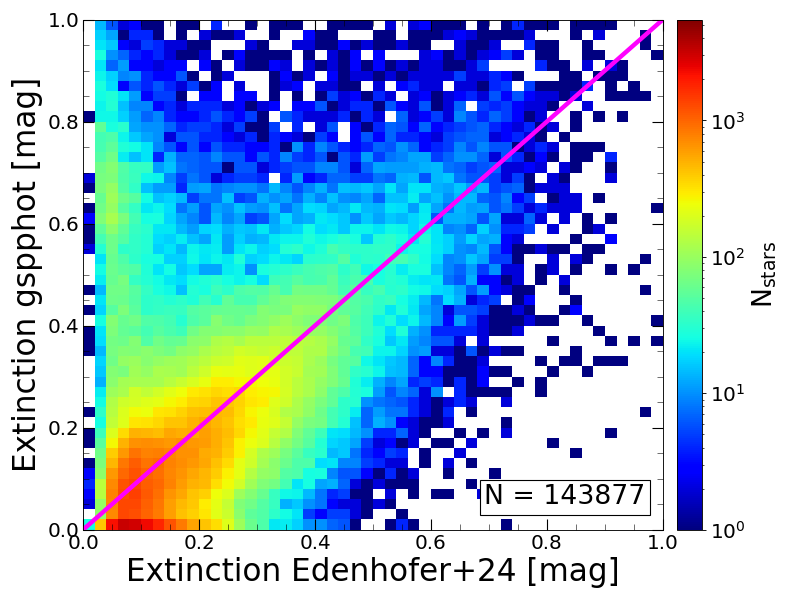}
\includegraphics[width=5.5cm]{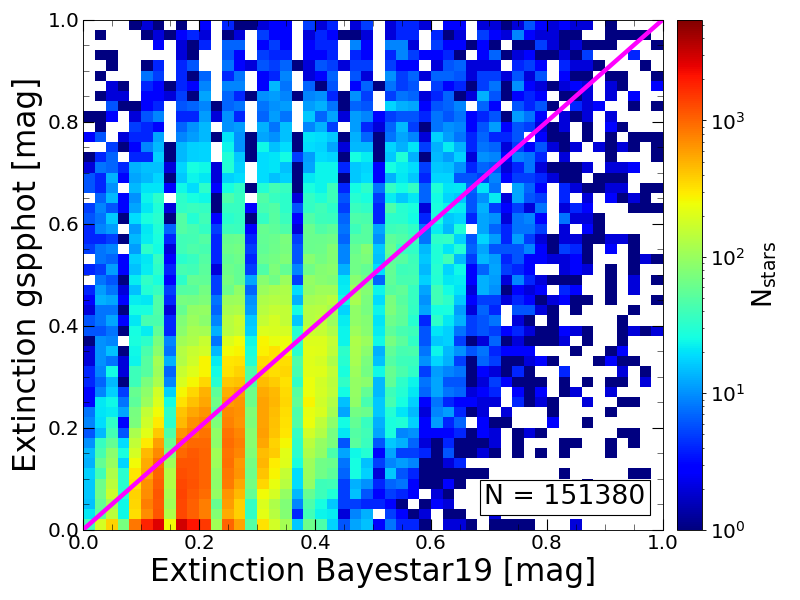}\\
\caption{Comparison of the different extinction references considered in this work: \citet{vergely22}, \citet{edenhofer24}, \texttt{Bayestar19}, and \texttt{gspphot}. Each panel shows a 2D histogram of the extinction comparison between two references. The color bars indicate logarithmic density (with the same range for all panels), and the 1:1 line is shown in magenta. The number of stars in each plot is indicated in the bottom-right corner. Given that the \texttt{Bayestar19} values are reported to fewer decimal places than the other references, the panels involving this map display horizontal/vertical stripes. The \texttt{gspphot} extinctions show the largest disagreement of all the references. For the CMD analysis, we adopt the \citet{vergely22} map as the source of extinctions.}
\label{fig:extinction_maps_comparisons}
\end{figure*}

Several extinction maps and catalogs have become available in the literature in recent years. In this section, we examine a number of them to decide which is more appropriate for our CMD analysis (Sect.~\ref{sec:characterization_CMD}). We limit our comparison to references that take into account the distances of stars, and thus do not include 2D maps such as the {\gaia} DR3 \texttt{Total Galactic Extinction} map \citep{delchambre23}, or the classical \citet{schlegel98} map (see also \citealt{schlafly11}).

With this, the four references we consider are: 
\begin{enumerate}
\item \citet{vergely22}: we query the extinction values using the EXPLORE website\footnote{\url{https://explore-platform.eu/}} (see also \citealt{lallement22}), giving as input the Galactic longitudes, latitudes, and distances. This reference provides different maps with varying spatial extents and resolutions (namely 50pc, 25pc, and 10pc). We adopt the 50 pc resolution map, as the distances returned by its query do not saturate past 3 kpc (as is the case for the 25 pc and 10 pc maps). This decision does not impact our results, as the three maps provide extinctions in good agreement with each other for our targets (fractional differences of $\lesssim$ 5\%). The extinction is reported in units of monochromatic extinction $A_0$ ($\lambda \approx 550$ nm). This map provides extinction values for 98.7\% of our sample (194,299 stars), which includes every star in the CMD sample.
\item \texttt{Bayestar19} \citep{green19}: we query this map via the \texttt{dustmaps} package \citep{green18}, using the same inputs as above. We only consider targets with the quality flags \texttt{converged} and \texttt{reliable\_dist} set to \texttt{True}. The map reports reddening in units similar but not equal to $E(B-V)$\footnote{\url{http://argonaut.skymaps.info/usage}}. We follow the \texttt{dustmaps} documentation\footnote{\url{https://dustmaps.readthedocs.io/en/latest/examples.html}} and convert these reddenings to extinction $A_{V}$ by multiplying the \texttt{Bayestar19} values by 2.742 (see also Table 6 in \citet{schlafly11} with $R_{V}=3.1$). This map provides extinction values for 88.2\% of our sample (173,630 stars).
\item \citet{edenhofer24}: we query this map using \texttt{dustmaps} and the same inputs as above. This reference also reports maps with varying resolutions and spatial extents (namely 1.25 kpc and 2 kpc). Both maps are in excellent agreement with each other (fractional difference of $\lesssim$ 1\%). Given the distance distribution of our targets (see Sect.~\ref{sec:data}), we choose the map that extends out to 2 kpc. The map reports reddening in the units of \citet{zhang23}, who in turn report their values in the same units as \texttt{Bayestar19}, and we thus apply the same multiplicative factor as above to convert them to $A_{V}$. This map provides extinction values for 82.6\% of our sample (162,561 stars).
\item {\gaia} DR3 \texttt{gspphot} \citep{creevey23}: we query this catalog using the \texttt{gaiadr3.astrophysical\_parameters} table. The extinction is reported in units of monochromatic extinction $A_0$ (\texttt{azero\_gspphot}), as well as in the {\gaia} bands. This catalog provides extinction values for 85.9\% of our sample (169,047 stars).
\end{enumerate}

We compare these references with each other in Figure \ref{fig:extinction_maps_comparisons}. We note that some of them are in slightly different units ($A_{V}$ vs. $A_{0}$), but their differences are very small (e.g., \citealt{yamaguchi24}). Figure~\ref{fig:extinction_maps_comparisons} provides valuable insights:
\begin{itemize}
\item The \texttt{Bayestar19} values exhibits periodic gaps. This is due to this map being reported down to fewer decimal places than the others, hence producing quantized extinctions (appearing as horizontal/vertical stripes).
\item The \texttt{gspphot} values show a strong concentration of points near extinction $\approx$ 0 mag. Additionally, \texttt{gspphot} is the only reference with a significant fraction of its values around high extinctions ($\gtrsim 0.6$ mag). Both of these features point to inaccuracies in the \texttt{gspphot} values, possibly due to the extinction-temperature degeneracy reported in \citet{andrae23a}.
\item We find a good overall agreement between the \citet{vergely22}, \citet{edenhofer24}, and \texttt{Bayestar19} maps, with all three of them showing a concentration of points around extinction $\sim 0.2$ mag. The comparison that more closely follows the 1:1 relation is that of \citet{vergely22} vs. \citet{edenhofer24}.
\end{itemize}

With this, we discard the \texttt{gspphot} values as the source of extinction for our CMD analysis. Regarding the other three maps, they all have similar distributions, with median (standard deviation) values of 0.20 (0.14) mag for \citet{vergely22}, 0.16 (0.13) mag for \citet{edenhofer24}, and 0.24 (0.15) mag for \texttt{Bayestar19}. Thus, the \citet{vergely22} median is located halfway between the \citet{edenhofer24} and \texttt{Bayestar19} values, with all three maps having almost identical standard deviations. For complementary discussions regarding extinction map comparisons, we refer to \citet{vanlane24} (see their Appendix G.1).

In this work, whenever possible, we aim for homogeneity in the stellar properties. In this regard, the \citet{vergely22} map is the only one that provides extinction values for all our CMD targets. Given that this map is also in good agreement with \citet{edenhofer24} and \texttt{Bayestar19}, and that its distribution is an intermediate point between the other two, we adopt \citet{vergely22} as the source of extinction values throughout this paper.
\section{Characterization of the adopted extinction values} 
\label{sec:app_characterization_extinction_values}

Following the discussion of Appendix \ref{sec:app_comparing_extinction_maps}, Figure \ref{fig:extinction_distribution} shows the distribution of extinction values from \citet{vergely22} for the CMD sample of Sect.~\ref{sec:characterization_CMD}. The histogram peaks around $A_0 \approx 0.15$ mag, with 16th and 84th percentiles of 0.12 to 0.32 mag.

To transform these monochromatic extinction values into the {\gaia} bands, we follow the (E)DR3 extinction law\footnote{\url{https://www.cosmos.esa.int/web/gaia/edr3-extinction-law}}. We mimic the implementation of \citet{godoyrivera21b}, updating the coefficients to those of \citet{fitzpatrick19}. In brief, this is an iterative process that solves for $k_m = A_m / A_0$ (where $m$ is one of the $G$, $BP$, or $RP$ {\gaia} bands), as a function of de-reddened $(BP-RP)_0$ color and $A_0$. We follow the {\gaia} documentation and use different coefficients for the MS and the top of the HRD (transition at $M_G = 5$ mag). We find $k_m$ values from this method to be in good agreement with the values that can be obtained from \texttt{gspphot} (calculated from their $A_{m}$ and $A_0$ values; see also \texttt{dustapprox}, \citealt{fouesneau_dustapprox_2022}). We also note that the coefficients are recommended to be used for colors inside the $-0.06 \leq (BP-RP)_0 \leq 2.5$ mag range. However, when applying the coefficients for the entire color range of our sample, we observe a continuous behavior around these limits in the CMD and Hess diagram of Figure \ref{fig:characterization_CMD}. Thus, given that the stars located outside the recommended color range are only 0.6\% of the {\kepler} sample (1,211 targets), we choose to keep them in our catalog and leave it to the reader to decide whether to use them.

With the above, we obtain star-by-star $k_m$ values and calculate the extinction coefficients in the {\gaia} bands (i.e., $A_{m} = k_m \times A_0$). The extinction errors are calculated with the analogous relation ($\sigma_{A_{m}} = k_m \times \sigma_{A_0}$), using the $\sigma_{A_0}$ values explained below. These $A_{G}$, $\sigma_{A_{G}}$, $A_{BP}$, $\sigma_{A_{BP}}$, $A_{RP}$, and $\sigma_{A_{RP}}$ values are used to de-redden the photometry and propagate the errors in Sect.~\ref{subsec:characterization_CMD_CMDsample}.

Regarding the error in the monochromatic extinction values, we consider both the random and systematic contributions. The errors reported by \citet{vergely22}, which we take as the random errors, have a median of $\sigma_{A_{0,\text{random}}}/A_{0} \approx 0.6\%$. These are significantly smaller than the values of the other extinction maps. The analogous values are $\approx$ 8.3\% for \texttt{Bayestar19}, $\approx 9.6\%$ for \texttt{gspphot} (or 17.3\% when considering the $\times$ 1.8 inflation factor from Table 4 of \citealt{andrae23a}), and $\approx$ 5.6\% for \citet{edenhofer24} (Phil Van-Lane, private communication). Thus, the random errors from \citet{vergely22} are probably underestimated.

\begin{figure}
\centering
\includegraphics[width=8.0cm]{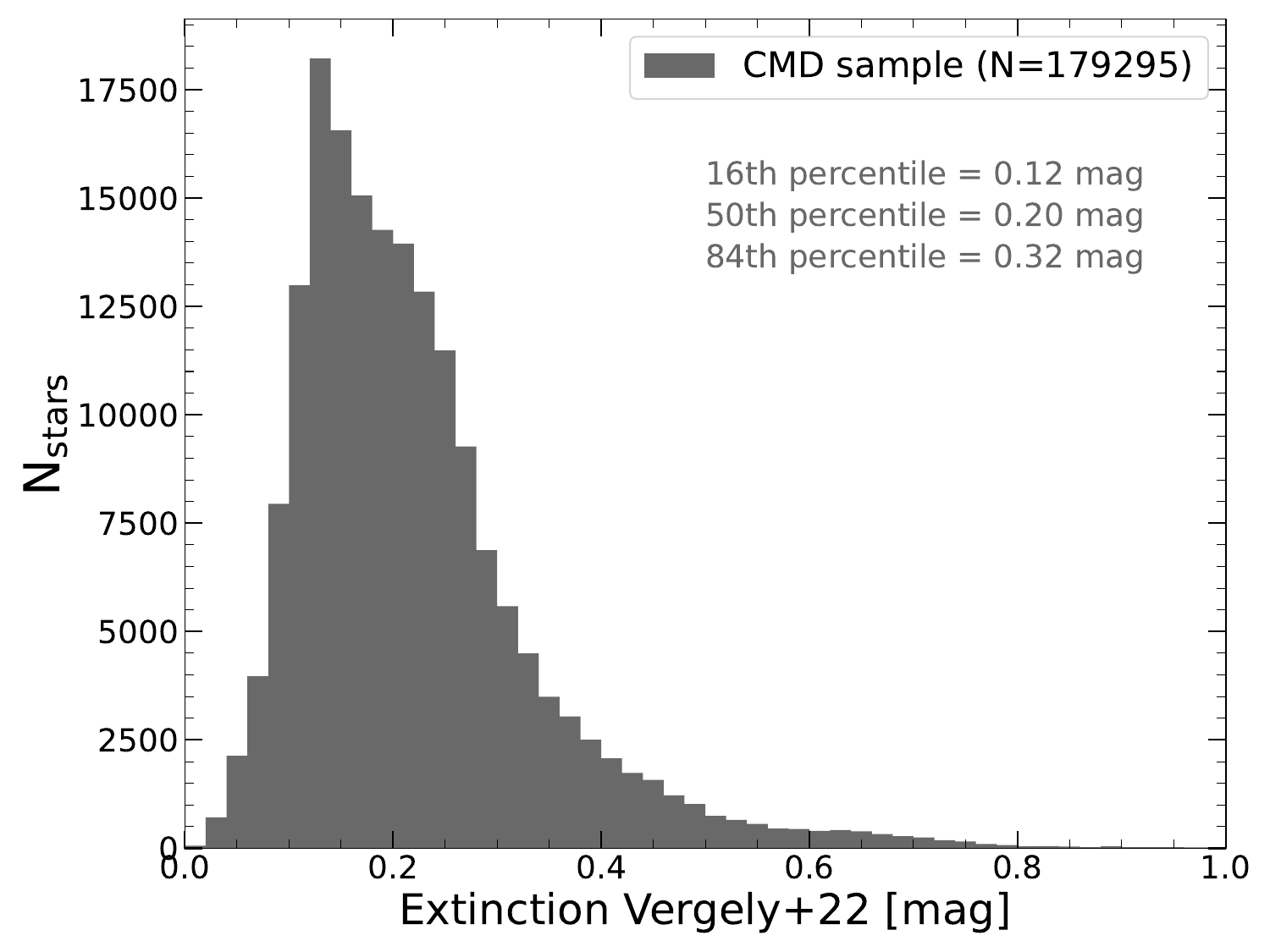}
\caption{Distribution of the \citet{vergely22} monochromatic extinction values ($A_0$) adopted for the CMD sample. The distribution peaks around $\approx 0.15$ mag, with a median of $\approx$ 0.20 mag.}
\label{fig:extinction_distribution}
\end{figure}

To provide more realistic monochromatic extinction errors that include systematics (e.g., due to our choice of extinction map), we calculate a combined error as
\begin{equation}
\sigma_{A_{0}} = \sqrt{(\sigma_{A_{0,\text{random}}})^2 + (\sigma_{A_{0,\text{syst}}})^2 }.
\end{equation}
We use the difference between the \citet{vergely22} versus the \citet{edenhofer24} and \texttt{Bayestar19} maps as a measure of this systematic error. The comparison yields a combined median fractional difference of $\approx 19.1\%$. We approximate this to $20\%$, which we adopt as a global value for $\sigma_{A_{0,\text{syst}}}$ (i.e., $\sigma_{A_{0,\text{syst}}}=0.2 \times A_0$). Both $A_0$ and $\sigma_{A_0}$ values are reported in Table~\ref{tab:table_catalog}.
\section{Coordinates of CMD regions} 
\label{sec:app_CMD_categories}

\begin{table}
\caption{CMD regions.}
\label{tab:coordinates_CMDregions}
\centering                         
\scriptsize 
    \begin{tabular}{lrr}
    	\hline
    	\hline
        Region & $(BP-RP)_0$ & $M_{G_0}$ \\        
        \hline
	Dwarf & -0.601 & -3.991 \\
	Dwarf & -0.210 & 0.600 \\
	Dwarf & -0.140 & 0.900 \\
	Dwarf & -0.090 & 1.200 \\
	Dwarf & -0.030 & 1.400 \\
       Dwarf & \ldots & \ldots \\
       
	Photometric Binary & 0.900 & 4.892 \\
	Photometric Binary & 0.920 & 5.042 \\
	Photometric Binary & 0.970 & 5.242 \\
	Photometric Binary & 1.030 & 5.442 \\
	Photometric Binary & 1.070 & 5.642 \\
       Photometric Binary & \ldots & \ldots \\
       
	Uncertain MS & -0.601 & -3.991 \\
	Uncertain MS & -0.601 & 6.657 \\
	Uncertain MS & 1.619 & 15.315 \\
	Uncertain MS & 4.591 & 15.315 \\
	Uncertain MS & 4.591 & 15.100 \\
	Uncertain MS & \ldots & \ldots \\		
       \ldots & \ldots & \ldots \\
       \hline
    \end{tabular}
\tablefoot{Color and magnitude coordinates of the CMD regions presented in Sect.~\ref{sec:characterization_CMD}. Note that these points define the borders of the polygons, and thus some of the regions share borders with each other. (The full table is available in Section `Data availability'.)}
\end{table}

\begin{figure*}
\centering
\includegraphics[width=8cm]{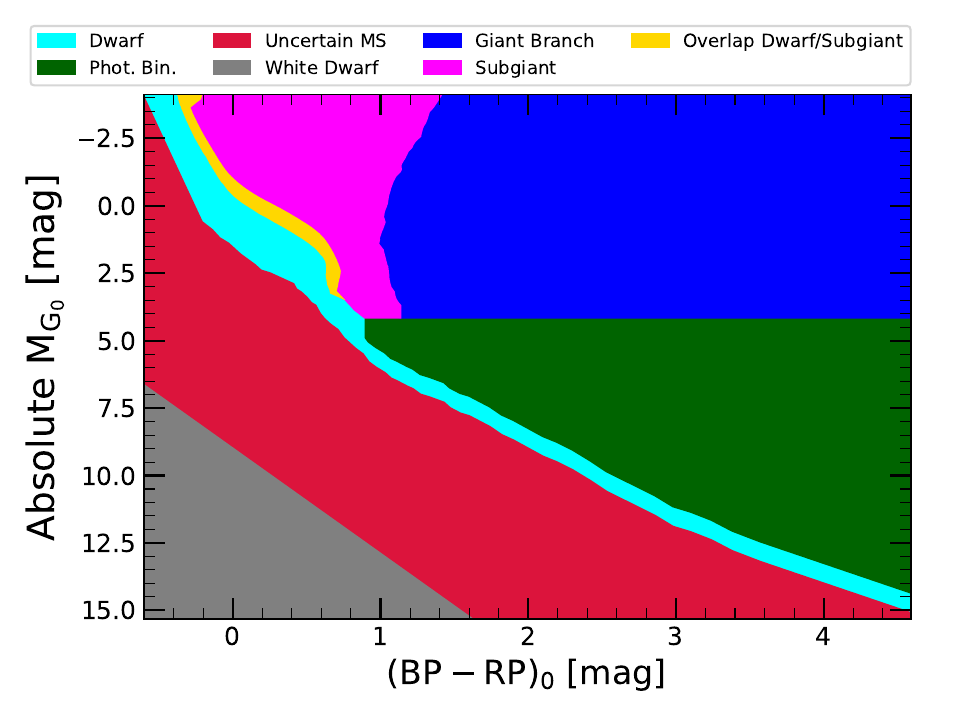}
\includegraphics[width=7.5cm]{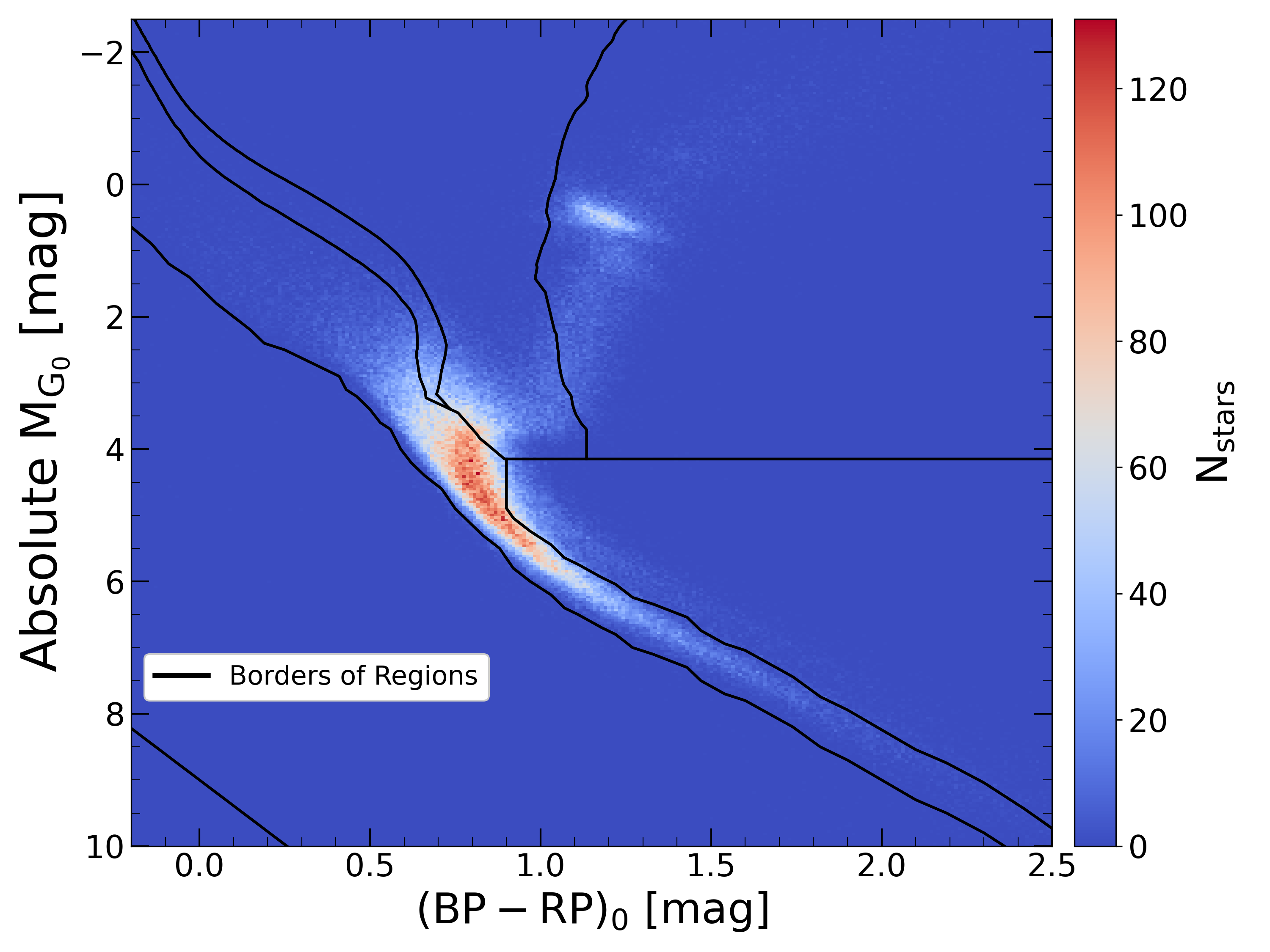}
\caption{Illustration of the CMD regions presented in Sect.~\ref{sec:characterization_CMD}. The left panel displays the coordinates of the regions, and the right panel shows the Hess diagram of the CMD sample with the borders of the CMD regions overplotted.}
\label{fig:cmd_categories_regions}
\end{figure*}

For reproducibility, we report the borders of the CMD regions defined throughout Sect.~\ref{subsec:characterization_CMD_categories} as (color, magnitude) data points in Table~\ref{tab:coordinates_CMDregions}. We summarize these graphically in the left panel of Figure~\ref{fig:cmd_categories_regions} by showing the extent of the regions. In the right panel, we project these onto the Hess diagram of the CMD sample. As expected from the selection function of the {\kepler} mission, the region of dwarf stars concentrates the highest density (see also Table~\ref{tab:summary_table}). Note that, for plotting purposes, the region plotted in the right panel is smaller than that of the left panel.
\section{Metallicity impact on suites of PARSEC models} 
\label{sec:app_PARSEC_metallicity}

\begin{figure*}
\centering
\includegraphics[width=5.5cm]{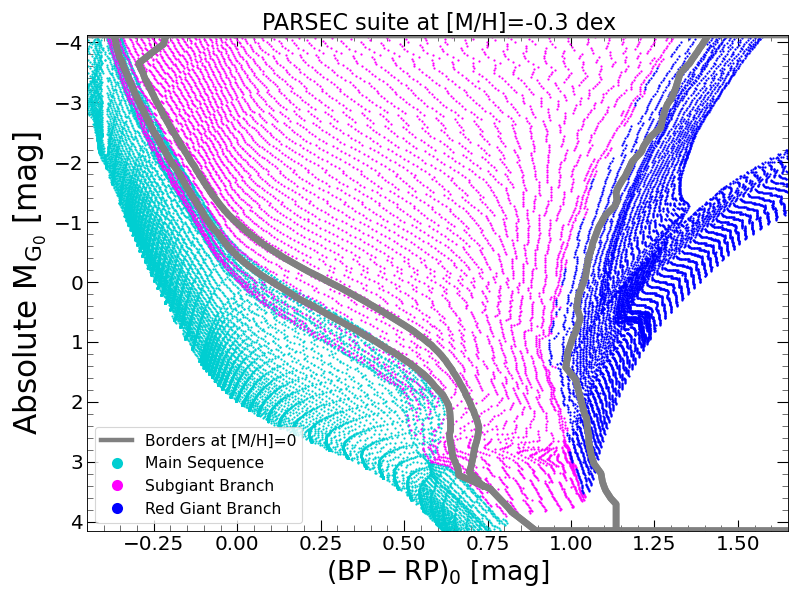}
\includegraphics[width=5.5cm]{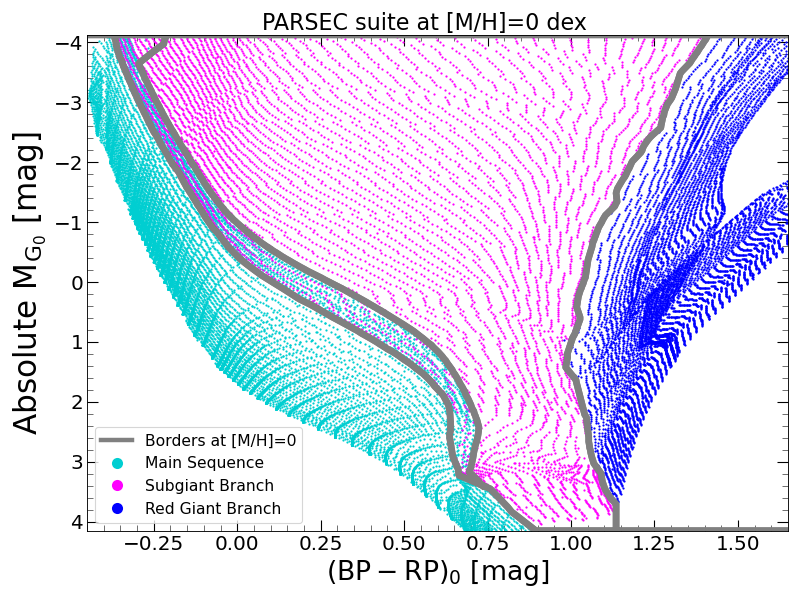}
\includegraphics[width=5.5cm]{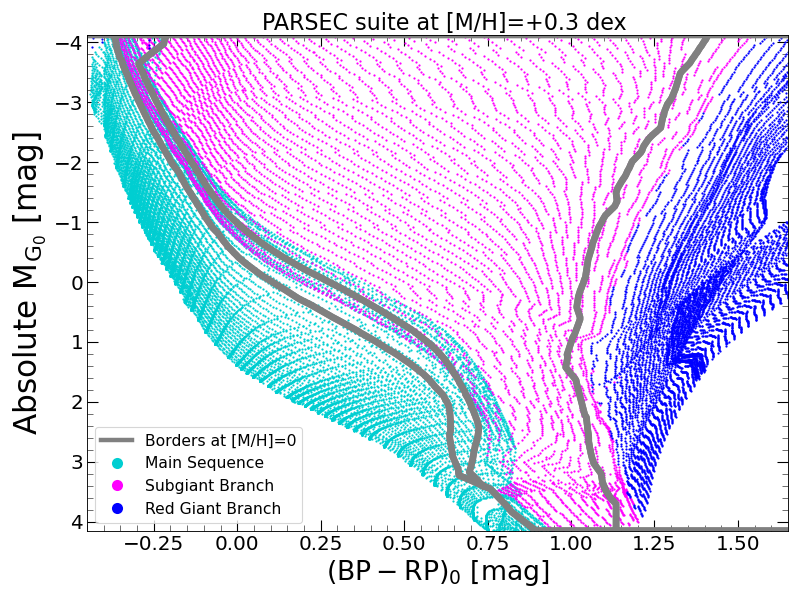}\\
\caption{Illustration of the metallicity-dependence of the evolved CMD regions, analogous to the top-left panel of Figure \ref{fig:characterization_CMDregions_upper_lower}. From left to right, the panels show suites of PARSEC models at [M/H]=$-0.3$, 0, and $+0.3$ dex. The colors indicate the evolutionary stage for each metallicity bin. The borders used in the CMD classification of Sect.~\ref{sec:characterization_CMD}, calculated at [M/H]=0 dex, are shown as the thick grey lines in all panels.}
\label{fig:suitePARSEC_vs_metallicity}
\end{figure*}

Following Sect.~\ref{subsubsec:characterization_CMD_validation_modelsuites}, in Figure \ref{fig:suitePARSEC_vs_metallicity} we replicate the suite of PARSEC models used to define the borders of the evolved CMD regions. We do this for three metallicity bins, namely [M/H]=$-0.3$, 0, and $+0.3$ dex, but keeping the borders of the fiducial CMD regions at [M/H]=0 dex identical in all panels for visual reference. Although they vary as a function of CMD location, typical color and magnitude changes are of order $\frac{\text{d}(BP-RP)_0}{\text{d}\text{[M/H]}} \sim 0.3$ and $\frac{\text{d}M_{G_0}}{\text{d}\text{[M/H]}} \sim 0.9$ mag/dex (or 0.03 and 0.09 mag per +0.1 dex).

\section{{\gaia} NSS binary types and tables} 
\label{sec:app_NSS_tables}

The NSS binary types are encoded in a bit-condensed flag in the \texttt{non\_single\_star} column in the \texttt{gaiadr3.gaia\_source} table. We have translated these to their plain meanings (astrometric, spectroscopic, eclipsing, and combinations thereof), and report them in the column `NSS Binary Type' of Table~\ref{tab:table_catalog}. For example, a value of $\texttt{non\_single\_star}=3$ is translated to 011 in binary notation, which in turn implies a `spectroscopic+astrometric'  binary type.

Additionally, the {\gaia} DR3 NSS systems are reported in four tables, namely \texttt{nss\_two\_body\_orbit} (TBO; systems with orbital parameters), \texttt{nss\_acceleration\_astro} (ACA; systems with acceleration solutions due to non-linear proper motions), \texttt{nss\_non\_linear\_spectro} (NLS; spectroscopic binaries with long-period trend solutions), and \texttt{nss\_vim\_fl} (VIM; sources with variability induced movers). For further details, we refer to \citet{gaia23c}. For completeness, we investigate the provenance of the 4,005 {\kepler} targets flagged as NSS in Sect.~\ref{subsec:characterization_binaries_nss} by crossmatching with each of the four tables individually. We find 1,958 targets in the TBO, 1,866 targets in the ACA, 294 targets in the NLS, and 7 targets in the VIM. Some degree of overlap exists among them, with 71 targets appearing in both TBO and ACA, 1 target in both TBO and NLS, 3 targets in both TBO and VIM, and 45 targets in both ACA and NLS (with no other intersections being found). For the targets classified as NSS by {\gaia}, we join the 3-letter acronym of the tables they appear in, and create the column `NSS Tables' of Table~\ref{tab:table_catalog}. For example, a star found in both TBO and ACA will have `NSS Tables = TboAca'.
\section{Additional scrutiny of photometrically variable targets} 
\label{sec:app_variability_EP_AGN}

We now expand on two of the variability classes discussed in Sect.~\ref{sec:variability}, namely the stars with exoplanet transits (\texttt{EP}) and active galactic nuclei (\texttt{AGN}).

The {\gaia} photometry allows for the detection of exoplanet candidates via the transit method \citep{panahi22,rimoldini23}. We find 10 {\kepler} targets classified as \texttt{EP}. These correspond to KIC 4150804, KIC 5780885, KIC 9651668, KIC 9818381, KIC 10019708, KIC 10666592, KIC 10874614, KIC 11517719, KIC 11804465, and KIC 12019440. As per the Simbad database, all of these are reported in literature studies of exoplanet hosts (e.g., \citealt{su22,maxted23}).

Additionally, we find 21 {\kepler} targets classified as \texttt{AGN} by {\gaia} DR3 \citep{carnenero23}. None of them pass our parallax SNR quality cut, as expected for extragalactic objects (e.g., \citealt{luri18}). We query the Simbad database and find that 13 of these targets are reported in literature studies of {\kepler} AGNs and quasars (e.g., \citealt{scaringi13,dobrotka17,smith18}). For the remaining 8 targets, however, Simbad does not return any results. We list their IDs for interested readers: KIC 3730597, KIC 4356027, KIC 7729019, KIC 8160685, KIC 9339957, KIC 10070645, KIC 11862867, and KIC 11913354.

\end{appendix}

\end{document}